\documentclass[review]{elsarticle}

\usepackage{sphinx}
\usepackage[font=small,labelfont=bf]{caption}
\usepackage{subcaption}
\usepackage[english]{babel}

\usepackage[version=4]{mhchem}
\usepackage{amsmath}
\usepackage{graphicx}
\usepackage{listings}
\usepackage{color}
\usepackage{threeparttable}
\graphicspath{{./images/}}
\usepackage[colorlinks=true, allcolors=blue, draft=false]{hyperref}
\usepackage{hypcap}
\usepackage{siunitx}
\usepackage{easyReview}
\usepackage{ulem}

\definecolor{dkgreen}{rgb}{0,0.6,0}
\definecolor{gray}{rgb}{0.5,0.5,0.5}
\definecolor{mauve}{rgb}{0.58,0,0.82}
\definecolor{gray}{rgb}{0.4,0.4,0.4}
\definecolor{darkblue}{rgb}{0.0,0.0,0.6}
\definecolor{lightblue}{rgb}{0.0,0.0,0.9}
\definecolor{cyan}{rgb}{0.0,0.6,0.6}
\definecolor{darkred}{rgb}{0.6,0.0,0.0}

\lstdefinelanguage{XML}
{
  morestring=[s][\color{dkgreen}]{"}{"},
  morecomment=[s][\color{red}]{<!--}{-->},
  stringstyle=\color{black},
  keywordstyle=\color{lightblue},
  identifierstyle=\color{darkblue},
  morekeywords={auxtype, auxvalue, name, value, lunit, x, y, z, ref, version}
}

\lstdefinelanguage{bash}
{
  basicstyle=\ttfamily,
  stringstyle=\color{black},
  keywordstyle=\color{blue},
  identifierstyle=\color{black},
  morekeywords={prompt, python},
  comment=[l]{\#!},
}

\title{Prompt: Probability-Conserved Cross Section Biasing
Monte Carlo Particle Transport System}

\author[1,2]{Zi-Yi Pan}
\author[1,2,3]{Ni Yang}
\author[1,2]{Ming Tang}
\author[1,2]{Peixun Shen}
\author[1,2]{Xiao-Xiao Cai\corref{cor1}}
\ead{caixx@ihep.ac.cn}
\cortext[cor1]{Corresponding author: caixx@ihep.ac.cn}

\address[1]{Institute of High Energy Physics, Chinese Academy of Sciences, China}
\address[2]{Spallation Neutron Source Science Center, China}
\address[3]{University of Chinese Academy of Sciences, China}

\DeclareSIUnit\angstrom{\text {Å}}

\newcolumntype{L}[1]{>{\raggedright\arraybackslash}p{#1}}

\newcolumntype{C}[1]{>{\centering\arraybackslash}p{#1}}

\newcolumntype{R}[1]{>{\raggedleft\arraybackslash}p{#1}}

\begin{document}

\begin{abstract}
An open source software package for simulating thermal neutron propagation in geometry is presented.  
In this system, neutron propagation can be treated by either the particle transport method or the ray-tracing method. 
Supported by an accurate backend scattering physics engine, 
this system is capable of reproducing neutron scattering experiments in complex geometries and is expected to be used in the areas of instrument characterisation, optimisation and data analysis.

In this paper, the relevant theories are briefly introduced.
The simulation flow and the user input syntax to control it are provided in detail.
Five benchmarking simulations, focusing on different aspects of simulation and scattering techniques, are given to demonstrate the applications of this simulation system. 
They include an idealised total scattering instrument, a monochromatic powder diffractometer, a neutron guide, a chopper and an imaging setup for complex geometries.
Simulated results are benchmarked against experimental data or well-established software packages when appropriate.  Good agreements are observed.

\end{abstract}
\maketitle

{\bf PROGRAM SUMMARY}

\begin{small}
\noindent
{\em Program Title: }
\texttt{Prompt} \\
{\em Developer's respository link:  }       
\texttt{https://gitlab.com/cinema-developers/prompt}    
\\
{\em Programming language:}
\texttt{C++}, \texttt{C} and \texttt{Python}\\                           
{\em External routines/libraries:}       
  \texttt{NCrystal}, \texttt{VecGeom}, \texttt{MCPL}, \texttt{PyVista}
  \\
{\em Licensing provisions:  }       
\texttt{Apache License, 2.0 (Apache-2.0)}    
\\
{\em Nature of problem:}\\
  It is challenging for traditional neutron ray-tracing simulation methods to consider multiple scatterings in complex geometries, hence the accuracy and precision of the simulated neutron scattering experiments are limited. 
   \\
{\em Solution method:}
  A hybrid system of Monte Carlo particle transport method and ray-tracing method is proposed.
  \\
\end{small}
\label{sIntro}


\newpage
\section{Introduction}
\label{sIntroduction}

Along with the instrument design activities at the powerful neutron spallation sources over the last decades~\cite{TAYLOR2006728,MASON2006955,KAJIMOTO2019148,lindroos2011european,wei2009china}, Monte Carlo ray-tracing codes emerged as the de facto standard  for neutron instrument characterisation and optimisation.
There are many packages currently available, such as McStas~\cite{lefmann1999mcstas}, VITESS~\cite{zsigmond2002monte}, and RESTRAX/SIMRES~\cite{vsaroun1997restrax,saroun2004neutron}. 
Most of them employ the linear chain method, in which an instrument is decomposed into a series of components and the neutron beam is only allowed to transmit from an upstream component, to the adjoining downstream component. 
This method is computationally efficient and proven to be highly valuable in estimating neutron instrument flux on the sample~\cite{LEFMANN20061083,CHEN2018370,Santisteban:ks5105,ohl2004high} and resolution functions~\cite{doi:10.1063/1.3680104,doi:10.1143/JPSJS.80SB.SB025,doi:10.1063/1.3626935}. 
A recent review on the state-of-the-art of these packages can be found in~\cite{Willendrup2020}. 

There are cases when a neutron stream is split and goes into multiple downstream components. To address these cases, McStas offers a flexible ``GROUP'' keyword for users to specify the paths.
On the other hand, VITESS implements similar approach in several special components, such as the Fermi choppers, supermirrors and crystal analysers~\cite{zsigmond2004survey}.
However, the linear chain method becomes more cumbersome for complex geometries, where the internal structures cannot be arranged in an upstream-downstream manner.
For instance, in an assembly of a sample environment vessel containing a sample, the unwanted scatterings from the vessel can occur before or after neutrons passing by the sample.
To address this problem, McStas introduced the Union component~\cite{bertelsen2017software} to allow the definition of complex internal structures. The neutron path within this component depends only on the geometry and scattering physics.  

It is worth noting that the linear chain method is not enforced in NISP~\cite{doi:10.1063/1.59486,seeger2004neutron}, a package developed in the 1990s.
This package inherits a part of the Monte Carlo engine from an early version of a particle transport code~\cite{Willendrup2020}.
Due to the absence of the linear assumption, neutrons in this package are not travelled in a pre-defined chain, hence the simulation is more realistic.

Despite fewer assumptions being made, NISP is not currently  among the most popular packages. 
Apart from the computational efficiency, license  and manpower shortage problems outlined in~\cite{Willendrup2020}, a limited number of scattering models in bulk materials could also be a major disadvantage.
The departure between the model and actual physics introduces errors in the cross section and leads to inaccurate results. 
Since 2015, NCrystal~\cite{Cai2020},  a general purpose open-source engine for neutron scattering has been under active development.  
In this package, microscopic cross sections are calculated based on material basic properties~\cite{Cai2019,Kittelmann2021,NCWiki}.
Thanks to the implemented robust numerical methods, neutron scattering can be modelled in a wide energy range, from ultracold to tens of electronvolts. 

The presented software package, probability-conserved cross section biasing Monte Carlo particle transport system (\texttt{Prompt}), is our first attempt to tackle the cross section challenge. 
Available neutron scattering models and the probability-conserved cross section biasing technique are briefly introduced in 
Section~\ref{sTheory}. 
Section~\ref{sProgram} explains the concepts of the system along with code examples and shows a minimal but complete simulation input script. Section~\ref{sBenchmark} demonstrates overall system performance by studying a few problems, including an idealised total scattering instrument, a monochromatic powder diffractometer, a neutron guide, an imaging setup for complex geometries and a disk chopper.
This work is discussed and concluded in section~\ref{sConclusion}.

\section{Theories}
\label{sTheory}

The objective of the system is outlined in section~\ref{ssTransport}.
Section~\ref{sScatteringLength} and~\ref{ssScatteringType} 
introduce the neutron scattering length and scattering types, respectively. As the theory of neutron scattering is given elsewhere in more detail, e.g.~\cite{Lovesey,Squires,Schober2014}, 
these sections are dedicated to introducing the physical meanings of the relevant parameters and the underneath cross section models.
Section~\ref{ssBiasing} introduces the implemented  probability-conserved cross section biasing method that is able to effectively control the interaction number of neutrons in a finite volume. 
Section~\ref{ssMirrow} introduces the mirror and chopper models.  

\subsection{The scope of the system}
\label{ssTransport}

\begin{figure}
\centering
\includegraphics[width=\linewidth]{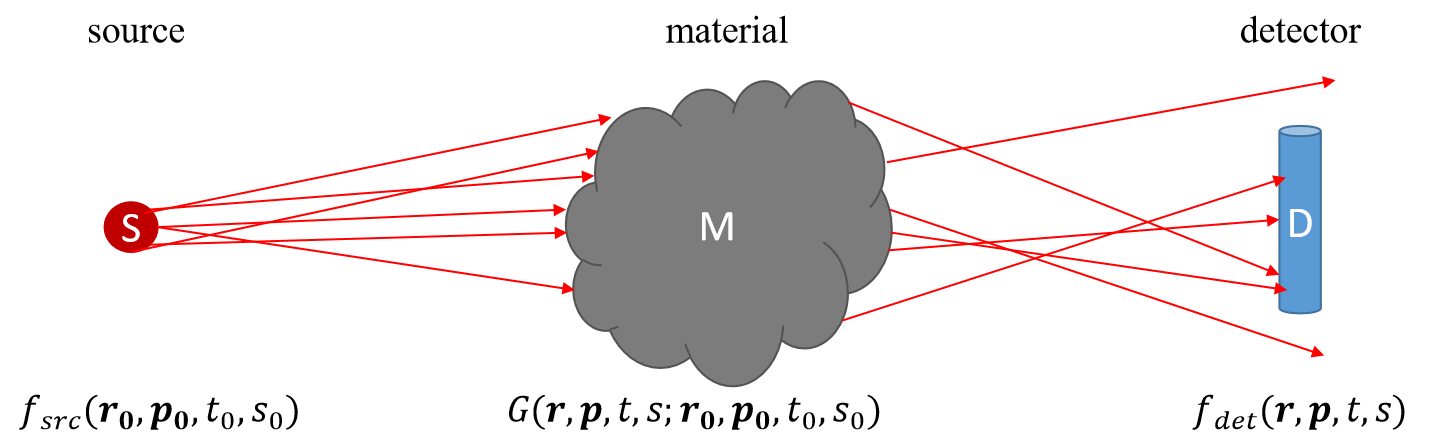}
\caption{A simplified sketch of particle detection after a volume of material}\label{fTransport}
\end{figure}

Fig.~\ref{fTransport} illustrates that particles that are originated
from the source interact with a volume of material. A part of the particles is detected by a detector at the end.
The problem of interest here is the probability of detection, $f_{det}(\boldsymbol{r}, \boldsymbol{p}, t, s)$. 

From the viewpoint of particle transport, it can be expressed as
\begin{equation}
\label{eTransport}
\begin{split}f_{det}(\boldsymbol{r}, \boldsymbol{p}, t, s)= \int {\mathop{}\!\mathrm{d}}^3
\boldsymbol{r}_0 \int {\mathop{}\!\mathrm{d}}^3 \boldsymbol{p}_0 \int {\mathop{}\!\mathrm{d}} t_0
\int {\mathop{}\!\mathrm{d}} s_0 \,\,
G(\boldsymbol{r}, \boldsymbol{p}, t, s; \boldsymbol{r}_0, \boldsymbol{p}_0, t_0, s_0) f_{src}(\boldsymbol{r}_0, \boldsymbol{p}_0, t_0, s_0)\end{split}
\end{equation}
where \(\boldsymbol{r}\) is the position, \(\boldsymbol{p}\) is the
momentum, \(t\) is the time, \(s\) is the spin, and \(\textbf{$G$}\) function is the propagator.  

This high-dimensional integral is typically evaluated using 
Monte Carlo particle transport systems, in which the propagator is decomposed into geometry and physics modules in general. In such systems, the physics module predicts the particle states by sampling the physical distributions for given prior states, materials and any applied external fields;
while the geometry model traces individual particles in volumes along their trajectories.

On the other hand, the ray-tracing method simplifies the integration by a scaling function $\epsilon$, which represents the weight adjustment of the particles. In addition, when the linear chain approximation is applied, the $\epsilon$ function is set to zero if the leaving particles do not intersect with the detector. 
\begin{equation}
\label{eRayTracing}
\begin{split}f_{det}(\boldsymbol{r}, \boldsymbol{p}, t, s)= 
\epsilon (\boldsymbol{r}, \boldsymbol{p}, t, s; \boldsymbol{r}_0, \boldsymbol{p}_0, t_0, s_0) f_{src}(\boldsymbol{r}_0, \boldsymbol{p}_0, t_0, s_0)\end{split}
\end{equation}
This method describes neutron transmission behaviour reasonably well in many optical components. One of the simplest examples is the rotating neutron disk chopper. 
The $\epsilon$ function is either unity or zero, representing respectively the event of neutron transmission or absorption. 

The objective of~\texttt{Prompt} is to solve Eq.~\ref{eTransport} in bulk materials and  Eq.~\ref{eRayTracing} to model moving optical components, i.e. choppers. The linear chain approximation is relaxed in the system.  
Only thermal neutrons are supported in this version of code. 
Notice that the gravitational field that may be important when simulating cold neutrons in a long beam guide. By limiting the particle step size, it is possible to simulate the gravitational effects. However, such implementation is computational demanding and outside the scope of current version of the code. It is not yet supported.

\subsection{Neutron scattering length}
\label{sScatteringLength}

The introduction starts with a limiting case, where neutrons scatter with a stationary nucleus that is fixed at a position throughout the scattering process.
If the target nucleus remains in the ground electronic state,  and the neutron wavelengths are much longer than the range of neutron-nucleus interaction, the scatterings are contributed only by the \textit{s}-wave components in terms of partial waves~\cite{Schober2014}.
The scatterings in that case are essentially elastic isotropic and can be characterized by a single parameter $b$, the so-called \textit{bound} scattering length~\cite{Lovesey}.  
The cross section for this hypothetical~\footnote{Fixing a stationary nucleus is only sound classically, but violates the uncertainty principle. } case is simply $4\pi|b|^2$.

In the opposite limiting case, where the motion of a nucleus is not hindered at all, the process can be viewed as elastic scattering in the centre of mass frame of reference.
The so-called \textit{free} scattering length for this case can be derived~\cite{Squires} to be  $b_f=bM/(m+M)$, where $m$ and $M$ are the mass of the neutron and target nucleus, respectively. When the incident neutrons are far more energetic than the chemical bonds in the system, say above a few electronvolts, the scattering cross section is approaching the limiting value $4\pi|b_f|^2$, for instance, see ref.~\cite{CAB2014}.

The angular distributions in the laboratory and centre-of-mass frames can be correlated as Eq.~\ref{eMassofCentre}~\cite{Schober2014}. From the viewpoint of this equation, the angular distribution of the \textit{bound} situation is a special case for the \textit{free} situation when $\gamma$ is approaching zero, i.e. the mass of the target nucleus is infinity in comparison with a neutron.     

\begin{align}
\nonumber
P(\theta_{lab}) =& \frac{(1+\gamma^2 + 2\gamma\cos\theta_{c.m.})^{3/2}}{|1+\gamma\cos\theta_{c.m.}|}  P(\theta_{c.m.}) \\ &\mathrm{ with} \, \,\,\,\,\gamma=\frac{m}{M}
\label{eMassofCentre}
\end{align}

However, neither the \textit{bound} nor the \textit{free} cases can describe the neutron scattering process adequately accurately in a finite and correlated system, especially when incident neutron wavelengths are comparable with the shortest inter-atomic distances.

Notice that the scattering length $b$ is complex in general.
The amplitude of absorption is the imaginary part of $b$.
For thermal neutrons, it is generally adequately accurate to decouple the absorption and scattering processes, see more discussions in section 2.2 of~\cite{Cai2019}. 
As a result, the scattering lengths are typically real-valued for cross section calculations.  
The absorption cross sections, without considering nuclear resonances, are often taken to be inversely proportional to the neutron velocity, and insensitive to atomic structure and dynamics.

\subsection{Neutron scattering cross section}
\label{ssScatteringType}

Neutral particles exhibit step-like behaviours in materials. 
In the Lambert-Beer law, the probability of observing a free particle at length $x$ in a homogeneous material can be described as 
\begin{equation} \label{eDistribition}
P(x) = \exp(-\rho \sigma x )
\end{equation}
where $\rho$ is the scatter number density and $\sigma$ is the cross section. 
The cross section can be calculated from the double differential cross section.
\begin{align}
\sigma =& \int \int \frac{\mathrm{d}^2\sigma}{\mathrm{d}\Omega \mathrm{d}E'} \mathrm{d}\Omega \mathrm{d}E' 
\label{eDiff2Total}
\end{align}

Under the Born approximation and Fermi's golden rule, the thermal neutron scattering cross section is the double integral of the scattering function scaled by the ratio of the scattered and incident wavenumbers~\cite{CAI2020106851}, 
\begin{align}
{\sigma}(\vec k_i) 
=& \int \int \frac{|\vec{k_f}|}{|\vec{k_i}|} S(\vec Q,\omega) \mathrm{d}\Omega \mathrm{d}E'
\label{eSqw2Xs}
\end{align}
with $\vec{Q} = \vec{k_i}-\vec{k_f}$ and $\hbar \omega = \hbar^2 (k^2_i-k^2_f)/2m $, to obey the conservation law of momentum and energy. Here $\vec{Q}$ and $\omega$ are the reduced momentum and energy transfers, respectively.

The evaluation and implementation of Eq.~\ref{eSqw2Xs} depend on the orientation distributions of the statistically equivalent subsystems with respect to the laboratory frame of the simulation system~\cite{Cai2020,Kittelmann2021}. 
For oriented materials, i.e. single and layered crystals, the relevant crystal axes should be aligned with the simulated laboratory frame of reference. So that the moment transfer in the predefined scattering function is identical with that in the simulation world. See more discussions in section~\ref{ssMaterial} on how to configure the alignment.

The neutron scattering function can be decomposed into a series of $S_{j,j'}(\vec Q,\omega)$, the partial scattering function, contributed by the $j_\mathrm{th}$ and $j'_\mathrm{th}$ nuclei pairs~\cite{VanHove1954}.

\begin{align}
  S(\vec Q,\omega) =
\sum_{j,j'=1}^N{\overline{b_jb_{j'}}}
S_{j,j'}(\vec Q,\omega)
\label{eSqw}
\end{align}
Here $N$ is the number of nuclei, 
$\overline{b_jb_{j'}}$ emphasizes the averaging is performed over the isotopes and spins of all the subsystems.


For a more realistic treatment of the scattering, in the case that the value $b$ for the $j_{th}$ nucleus has no correlation with the isotope and spin, the averaging is 
\begin{align}
  \overline{b_j b_{j'}} = \begin{cases}
      \overline{b_j}\cdot\overline{b_{j'}}, & \text{for}\ j\ne j' \\
      \overline{b^2_j}, & \text{for}\ j=j'
    \end{cases}
\label{eScatlength}
\end{align}
Therefore,
\begin{align}
  S(\vec Q,\omega) 
=& 
\sum_{\substack{j,j'=1 \\ j\ne j'}}^N \overline{b_j}  \,\overline{b_{j'}}
S_{j,j'}(\vec Q,\omega) + 
\sum_{j}^N{\overline{b_j^2}}
S_{j,j}(\vec Q,\omega) \label{eSqwDS}
\\
=& S_{\mathrm{distinct}}(\vec Q,\omega) + S_{\mathrm{self}}(\vec Q,\omega)
\label{eSqw2}
\end{align}
In Eq.~\ref{eSqw2}, the first term is originated from the pairs that are built by different nuclei and is known as the \textit{distinct} scattering. The second term is contributed from a single nucleus solely, hence called the \textit{self} scattering. 

Inserting $\overline{b^2}=\overline{(b-\overline{b})^2}+(\overline{b})^2$ into Eq.~\ref{eSqwDS}, one can obtain
\begin{align}
  S(\vec Q,\omega) =&
\sum_{\substack{j,j'=1}}^N{\overline{b_j} \, \overline{b_{j'}}}
S_{j,j'}(\vec Q,\omega) + 
\sum_{j}^N{[\overline{b_j^2}-(\overline{b_j})^2]}
S_{j,j}(\vec Q,\omega)  \\
=& S_{\mathrm{coh}}(\vec Q,\omega) + S_{\mathrm{inc}}(\vec Q,\omega)
\label{eSqw3}
\end{align}

where $S_{\mathrm{coh}}$ and $S_{\mathrm{inc}}$ are the coherent and incoherent scatterings, respectively. 
Coherent scattering can be physically interpreted as the scattering that would arise if all the scattering lengths equal to their mean value.
While the incoherent scattering, proportional to the self scattering, represents the deviations of the scattering lengths from their mean value. 
It is worth emphasising that the coherent scattering is contributed by distinct scattering as well as a part of the self scattering.
\begin{align}
S_{\mathrm{coh}}(\vec Q,\omega) = S_{\mathrm{distinct}}(\vec Q,\omega) + \sum_{j}^N{\overline{b_j}^2}
S_{j,j}(\vec Q,\omega) 
\label{eDistic}
\end{align}


Coherent and incoherent scatterings can be further categorised by whether the neutron exchanges energy with the system, i.e. elastic and inelastic with respect to the laboratory frame of reference. 
\begin{align}
  S(\vec Q,\omega)
=& S_{\mathrm{coh}}^{\mathrm{el}}(\vec Q,\omega) + S_{\mathrm{coh}}^{\mathrm{inel}}(\vec Q,\omega)  + S_{\mathrm{inc}}^{\mathrm{el}}(\vec Q,\omega) +
S_{\mathrm{inc}}^{\mathrm{inel}}(\vec Q,\omega)
\label{eSqw4}
\end{align}

\subsubsection{Elastic scattering}

Homogeneous liquids and gases do not exhibit elastic scattering~\cite{Squires,Cremer2013}~\footnote{Indeed, measuring these materials in an elastic type neutron instrument does produce event data. However, this type of instrument generally guarantees that the event data are analysed without considering energy exchange in scatterings, but not what the actual physical processes are taking places. }. 
In the backend scattering engine, elastic scatterings are originated explicitly from solid type materials.  

The scattering functions of the incoherent and coherent elastic scattering are shown in Eq.~\ref{eIncEla} and~\ref{eCohEla}, respectively. 

\begin{align}
  S_{\mathrm{inc}}^{\mathrm{el}}(\vec Q,\omega) =&
  N \sum_{j}^N{[\overline{b_j^2}-(\overline{b_j})^2]} \exp(-2W_j)
\label{eIncEla}
\end{align}

\begin{align}
  S_{\mathrm{coh}}^{\mathrm{el}}(\vec Q,\omega) =&
  N  \frac{(2\pi)^3}{v_0} \sum_{\vec{\tau}} \delta(\vec Q - \vec \tau) 
  \left| \sum_{j}^N \overline{b_j}\exp(i\vec{Q} \cdot \vec{R_j}) \exp(-W_j) )\right|^2
\label{eCohEla}
\end{align}
$W$ is the Debye-Waller factor,  $v_0$ is the unit cell volume, $\vec{\tau}$ is the lattice point in the reciprocal space, and $\vec{R}$ is the equilibrium position. 

For idealised materials, the orientations of crystal unit cells are described by distribution functions. Numerical procedures are  implemented to compute the effective scattering functions averaged over the given distribution. 
The averaging is rather sophisticated and outside the scope of this paper, refer to Ref.~\cite{Kittelmann2021} for the detailed methods. The backend scattering engine supports the elastic scatterings in isotropic powders, mosaic single crystals and mosaic layer crystals.  

\subsubsection{Inelastic scattering}
\label{sssInelastic}

Limited by the memory footprint and initialisation time of typical Monte Carlo simulations, inelastic scattering is only supported in isotropic materials, e.g. powder and liquid. 

It has been shown that the incoherent part of the inelastic scattering for ideal crystal powders in the harmonic approximation~\cite{sjolander1958} and 
liquids in the Gaussian approximation under the fluctuation-dissipation theorem~\cite{Rahman1962} can  arrive independently to the identical form,

\begin{align}
	S_{\mathrm{inc}}^{\mathrm{inel}}(Q,\omega)	&= \frac{1}{2\pi} \int e^{-i \omega t}  \exp\left[- \frac {Q^2}{2}\Gamma(t)\right]  \mathrm{d} t \label{fGaussian}
\end{align}
Here the scattering function depends on the modulus of $\vec Q$ due to the isotropic nature of the material, and $\Gamma(t)$ is width function.  
\begin{equation}\label{gamma}
	\Gamma(t) = \frac{\hbar}{M} \int_0^{\infty} \mathrm{d}\omega \frac{\rho(\omega)}{\omega}
	\left[\coth\left(\frac{\hbar \omega}{2k_B T}\right)
	(1-\cos {\omega t})-i \sin{\omega t}\right]
\end{equation}
where $\rho(\omega)$ is the density of states, $k_B$ is the Boltzmann constant and $T$ is the temperature.
The expressions indicate that it is adequate to calculate the incoherent cross section in isotropic materials using only the density of states.

The numerical evaluation of Eq.~\ref{fGaussian} for crystals is computationally efficient when applying the phonon expansion method~\cite{sjolander1958}, therefore it is suitable to evaluate on-the-fly.
On the other hand, for liquids, $\Gamma$ becomes divergent when $\omega$ approaches zero, because of the diffusive motions. 
The scattering function, in that case, is typically pre-computed based on model-assisted evaluation methods~\cite{CAB2014} or more expansive direct convolutions~\cite{Du2022}.

The coherent inelastic scattering function may also be included in the pre-computed input cross section. In the case of liquid, the Sk\"old approximation~\cite{Skold1967}, i.e. Eq.~\ref{eSkoldPartial}, is often applied. 
\begin{align}
\begin{split}
\label{eSkoldPartial}
S_{\mathrm{coh}}^{\mathrm{inel}}(Q, \omega)
&= \sum_{jj'=1}^N \overline{b_j}\,\overline{b_{j'}}  {S_{jj'}}(Q)  {S_{jj}}\left(\frac{Q}{\sqrt{{S_{jj'}}(Q)}}, \omega\right) 
\end{split}
\end{align}
${S_{jj'}}(Q)$ is the partial static structure factor and equals to $\int {S_{jj'}}(Q, \omega) \mathrm{d}\omega$. As suggested by the zeroth order sum rule of the scattering functions, it is unity when $j=j'$, i.e. the case of self scattering function. 

For numerical simplification, the isotropic incoherent approximation, or otherwise known as the Gaussian approximation~\cite{Rahman1962}, is often employed to skip the computation of the coherent inelastic scattering function.
In that case, the distinct scattering functions are vanished,
hence the overall inelastic scattering is approximated entirely by the inelastic part of the self scattering function (see Eq.~\ref{eSqw2}).
The absence of distinct scattering results in a smoothened inelastic scattering function with respect to $Q$. 
Consequently, the calculated angular distributions are not expected to be directly comparable with those from measurements in every detail. 
Nonetheless, in practice, the incoherent approximation is known to be highly useful when the angular distribution is less of concern, in the sense that the final results of interest are the averaged values in a large angular range, such as when measuring phonon density of states  (see section 9.16 of~\cite{Schober2014}), characterising  neutron moderation processes~\cite{CAB2014} and estimating the intensities of multi-phonon scattering background (see the section 3.10 of~\cite{Squires}).

In additional to the incoherent approximation, a Debye model can be used for crystals, when only the structural information is mainly concerned, e.g. powder diffraction of a strong coherent sample. The density of states in that case is proportional to the power of phonon energy and the upper cut-off is defined by the Debye temperature~\cite{Sears1995PHONONDO}.     

In the situations where the correlations of the systems are relatively insignificant, e.g. in strong absorbing or gaseous materials, the short-collision-time approximation~\cite{MacFarlane2010,Cai2019}, can be used to describe the scattering processes, see section~\ref{ssMaterial} for more discussions.

\subsection{The cross section biasing mechanism}
\label{ssBiasing}
According to the probability expressed in Eq.~\ref{eDistribition},  the step length of a free particle can be sampled according to 
\begin{equation} \label{eStepLength}
    l=-\frac{\log_e(\xi)}{\rho \sigma}
\end{equation}
where $\xi$ is a number generated in a uniform distribution $[0,1)$.

If an arbitrary positive scaling factor $g$ is applied to the cross section to form an artificial cross section $\sigma_g = g\sigma(E)$, the sampled step length becomes
\begin{equation} \label{eStepLength2}
    l_g=-\frac{\log_e(\xi)}{g \rho \sigma}
\end{equation}

In comparison with $l$, $l_g$ is increased or decreased, when the factor is smaller or greater than unity, respectively. That may effectively control the particle interaction number in a finite volume. 

However, altering such an important parameter distorts the step length  distribution in Eq.~\ref{eDistribition} and produces wrong statistical results. 

It can be seen that by differentiating Eq.~\ref{eDistribition},  and then dividing the result by $P(x)$, one can obtain
\begin{equation} 
\frac{\mathrm{d} P(x)}{P(x)} = -\rho(x) \sigma \mathrm{d}x
\end{equation}

When applying an artificial cross section, this equation becomes
\begin{equation}\label{eBias} 
\frac{\mathrm{d}P_g(x)}{P_g(x)} = -\rho(x) \sigma_g \mathrm{d}x
\end{equation}

By the quotient rule of derivative, the impact of applying a biasing factor to the probability along the path can be quantified as
\begin{equation}\label{e} 
\frac{\mathrm{d}P_g(x)/P(x)}{\mathrm{d}P_g(x)/P(x)} = -(g-1) \rho(x) \sigma \mathrm{d}x
\end{equation}

It has been shown~\cite{MENDENHALL201238} that the distribution distortion can be eliminated when the particle weight is corrected by a compensation factor at the end of each interaction as 

\begin{equation} \label{eWeight} 
W_n = W_0 \prod_{l=1}^{n} \frac{1}{g_{m,l}} \prod_{m=1}^{n} \prod_{i=1}^{N_p } \exp[(g_{m,i}-1) x_m \rho \sigma_i] 
\end{equation}
Here, $W_o$ is the original weight of the particle;  $n$ is the number of interactions, the step length of the $m_{th}$ step is $x_m$,
$N_p$ is the number of reaction processes, the biasing factor for the $i_{th}$ process at the $m_{th}$ step is $g_{m,i}$, the $l_{th}$ process occurs at the end of the $m_{th}$ step, and the corresponding biasing factor is $b_{m,l}$.

The initial aim of the implementation is to facilitate the study of  scatterings between an incident neutron beam and a small sample of a scattering instrument.
In such a scenario, multiple scatterings are often considered to be a type of weak but important noise. Applying a biasing factor greater than unity, not only the statistics of the single scattering signal can be improved, but also the variances of the multiple scattering noises can be significantly reduced. Detailed validation and performance of this method are discussed in section~\ref{ssTotal} and~\ref{ssRabbit}.

\subsection{Neutron mirror and disk chopper}
\label{ssMirrow}
The empirical reflectivity model in \texttt{McStas}~\cite{McStasComp} is used for neutron mirrors. The reflection probability is a piecewise function, 

\begin{align}
  R(Q) = \begin{cases}
      R_0 , & \text{if } Q \leq Q_c\\
      \frac{1}{2} R_0 [1-\alpha(Q-Q_c)] \left[1-\tanh\left( \frac{Q-mQ_c}{W} \right ) \right] , & \text{if } Q > Q_c
    \end{cases}
\label{eMirrowRef}
\end{align}
where $R_0$ is the low-angle reflectivity, $Q_c$ is the critical scattering wavenumber, $\alpha$ is the reflectivity slope, $m$ is the m-value of the mirror, and $W$ is the cut-off width. The default values retaken from McStas are given in Table~\ref{tableMirror}.

When a neutron is reflected, its energy remains unchanged, so $Q$ only depends on the reflection angle. The reflected direction, $\Vec{n}_f$,  can be calculated using the neutron direction $\Vec{n}_i$, and the normal of the contact point $\Vec{n}_r$ as 

\begin{align}
\Vec{n}_f = \Vec{n}_i - 2\Vec{n}_r( \Vec{n}_i\cdot \Vec{n}_r)
\label{eMirrowNewDir}
\end{align}

\begin{table}
	\centering                       
	\caption{The parameters of the neutron mirror model and their values retaken from Mcstas \texttt{Guide} component. The m-value of mirror is required, while the others are fixed.}
	\label{tableMirror} 
	
    \begin{tabulary}{1.0\textwidth}{TTTT}
    \toprule
    Parameter & Value & Description & Unit
\\ 
    \midrule
    $m$ &  & Mirror m-value   & 1
\\
    $R_0$ & \num{0.99} & Low-angle reflectivity & 1
\\
    $Q_c$ & \num{0.0219} & Critical scattering wavenumber & \SI{}{\per\angstrom}
\\
    $\alpha$ & \num{6.07} & Slope of reflectivity & \SI{}{\angstrom}
\\
    $W$ & \num{0.003} & Width of cut-off & \SI{}{\per\angstrom}
\\
    \bottomrule
    \end{tabulary}
\end{table}

The neutron disk chopper is a neutron optics component that is modelled in ray-tracing method. The neutron weight is modified by the ray-tracing process instead of detailed scattering and propagation in the transport mode. 

When a neutron arrives at the disk chopper, the ray-tracing method is used. The corresponding $\epsilon$ function in Eq.~\ref{eRayTracing} is a mask of position and time. It is unity if a neutron hits a opening window, and is zero otherwise. In the implementation, the disk chopper is considered to be infinitely thin.

\section{The program}
\label{sProgram}




The executable, \texttt{prompt}, reads user input files in the Geometry Description Markup Language (\texttt{GDML}) format~\cite{chytracek2006geometry} and launches the simulation. 
The language schema is introduced in the  manual of the standard in detail~\cite{gdml2022}.
The available options are summarized in Table~\ref{tableLaunch}.  

\begin{table}
	\centering                       
	\caption{Options of the \texttt{Prompt} launcher. Parameters without
default values are required, the others are optional.}     
	\label{tableLaunch} 
	
    \begin{tabular}{p{0.2\textwidth}p{0.2\textwidth}p{0.6\textwidth}} 
    \toprule
    Option & Default & Description 
\\ 
    \midrule
    \texttt{-g [\textit{file}]} & & Set \texttt{GDML} input file. 
\\
    \texttt{-s [\textit{seed}]} & \num{4096} & Set the seed for the random generator. 
\\
    \texttt{-n [\textit{number}]} & \num{100} & Set the number of primary events. 
\\
    \texttt{-v} &  & The flag to activate the visualisation of geometries and trajectories.
\\
    \bottomrule
    \end{tabular}
\end{table}

In a typical  simulation, statistical distributions of particle histories are iteratively accumulated in a Monte Carlo loop of events. 
Each event begins with the generation of the particle, and ends with the absorption of the particle in materials or escape of the particle from all the volume boundaries. 
The accumulators are referred to as scorers in this system and are realised by histograms that are attached to volumes. 
The accumulation processes are triggered by the concept of particle tracing states (PTS) on the volume.

The simulation flow is illustrated in section~\ref{ssFlow}.
Available particle generators, i.e. particle guns, are presented in section~\ref{ssGun}.
The volume and scorers are introduced in section~\ref{ssGeometry} and~\ref{ssScorer}, respectively.
The syntax for initialising physics models is described in section~\ref{ssMaterial}. Finally, a short example of input script is listed and explained in section~\ref{sSample}.

\subsection{Simulation flow}
\label{ssFlow}
\begin{figure*}
    \centering
    \includegraphics[width=0.8\textwidth]{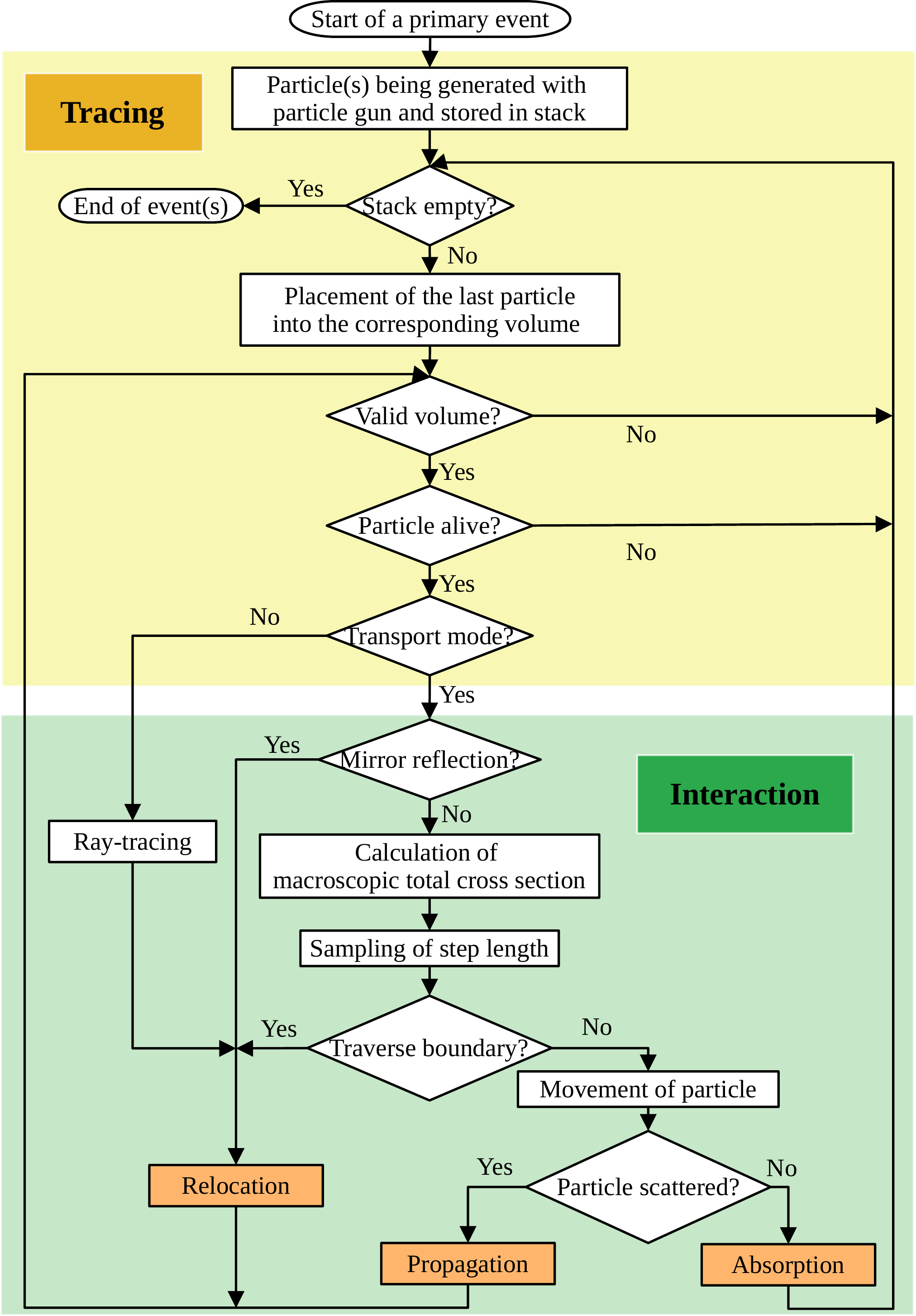}
    \caption{The main simulation flow of a primary event adopted in \texttt{Prompt}}
    \label{fig:sMainflow}
\end{figure*}

The flow of the simulation is divided into two phases, the tracing phase and the interaction phase, as illustrated in Fig.~\ref{fig:sMainflow}.

The system first looks for the volume that contains the last particle in the particle stack. The successful determination of a valid volume indicates that the particle is inside the world, i.e. the largest volume that contains all the child volumes in the simulation. Otherwise the simulation for this particle is terminated.
The program then asserts the particle alive status before picking the simulation mode for the interaction phase. 
If the particle is not just generated at a random position by a particle gun, it must  arrive at the boundary of a volume at this stage. 
In that case, the system checks whether a ray-tracing model is defined for the volume, before branching out into  the ray-tracing mode or the transport mode.  

In the transport mode, the particle is first treated by the mirror, if it is specified. 
Next, the macroscopic total cross section is evaluated, and the step length is sampled using Eq.~\ref{eStepLength2}. 
If the sampled step is longer than the maximum distance that allowed by the volume, the particle is moved to the geometry intersection point and enters the relocation subprocess. 
Otherwise, the particle is moved forward to the location where an interaction takes place. In this case, the particle enters the propagation subprocess or the absorption subprocess, depending on the sampling on the cross sections. 
In any case, whenever the particle is moved, the weight is adjusted according to Eq.~\ref{eWeight}, if the cross section is biased. 

In the ray-tracing mode, the particle is simulated in Eq.~\ref{eRayTracing}.
The main difference between this mode and the transport mode are that the particle states in the volume is completely manipulated by the implemented model. 
The system views the ray-tracing volume as a black box and has no knowledge about the particle states, until the model decides to eliminate the particle or sends the particle to the boundary of the volume.
Notice that, a volume containing child volumes is not supported by the ray-tracing mode, as the internal structure of the volume can not be modelled systematically. 

\begin{figure*}
    \centering
    \begin{subfigure}[b]{0.3\textwidth}
        \centering
        \includegraphics[width=\textwidth]{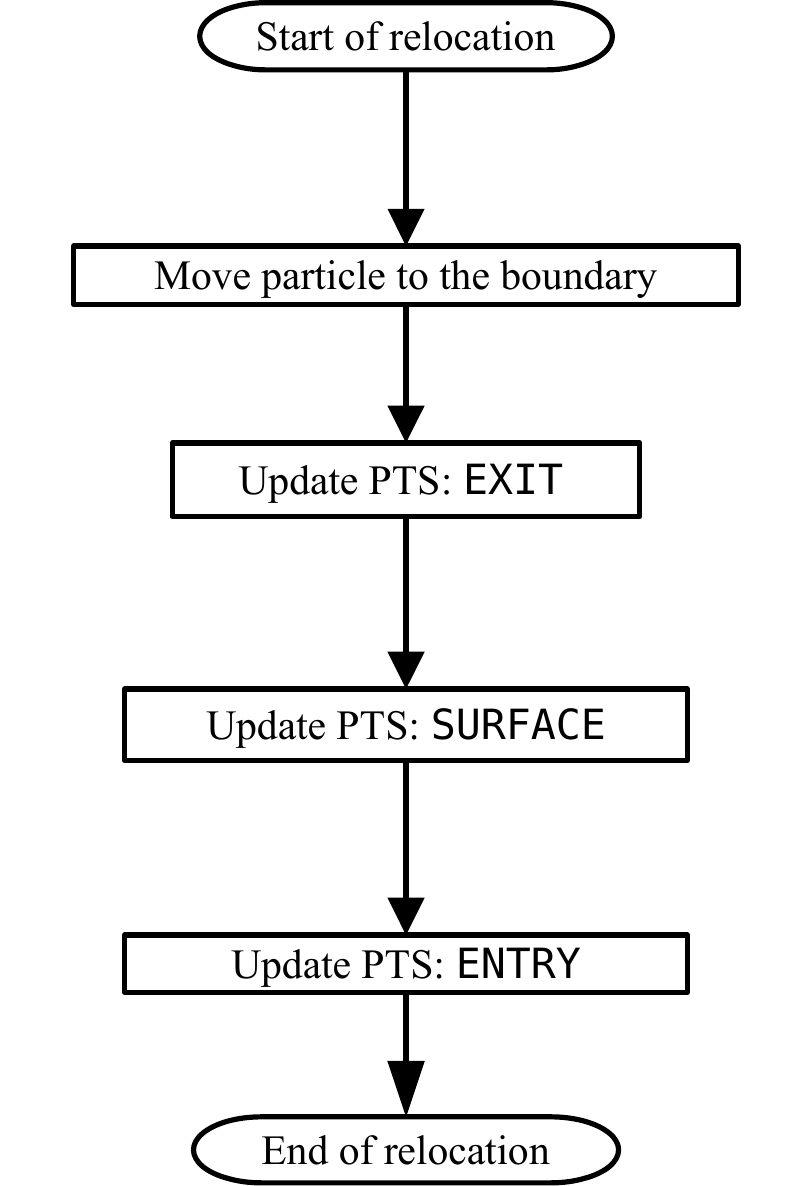}
        \caption{Relocation of particle between volumes}
        \label{fig:sRelocation} 
    \end{subfigure}
    \hfill
    \begin{subfigure}[b]{0.3\textwidth}
        \centering
        \includegraphics[width=\textwidth]{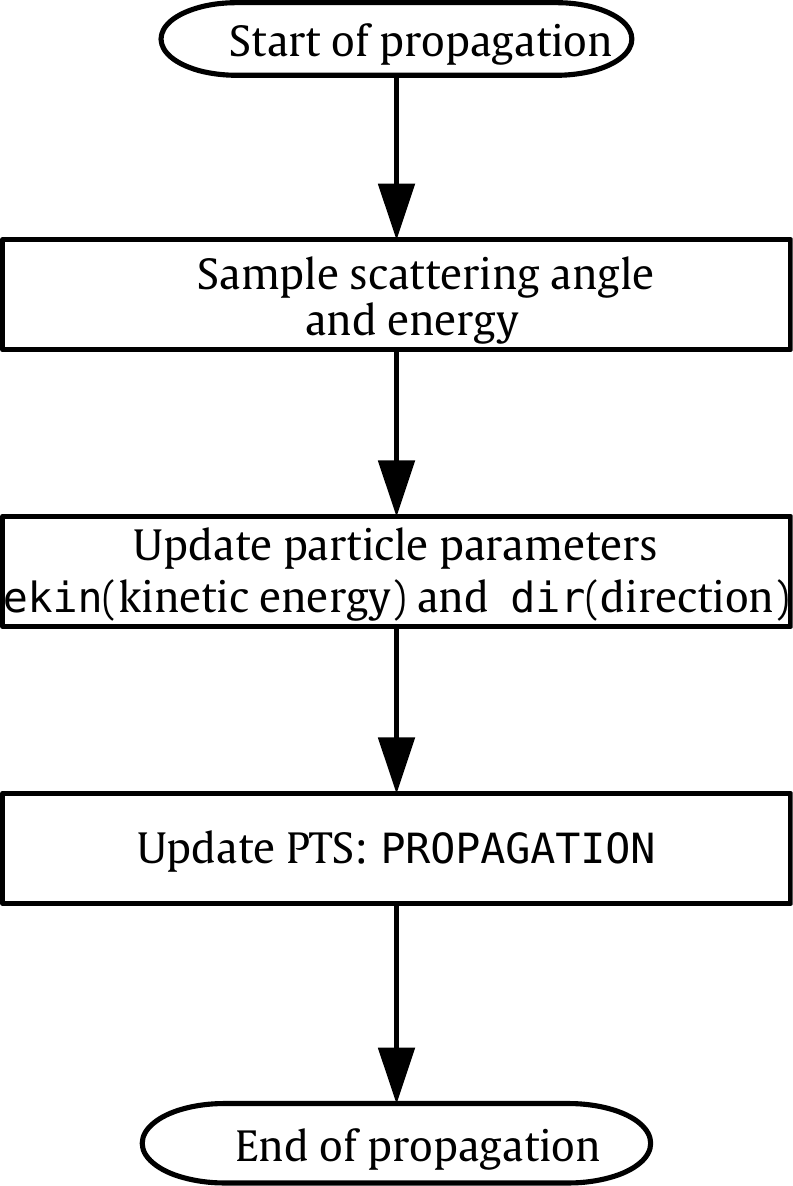}
        \caption{Propagation of particle within a volume}
        \label{fig:sPropagation}    
    \end{subfigure}
    \hfill
    \begin{subfigure}[b]{0.3\textwidth}
        \centering
        \includegraphics[width=\textwidth]{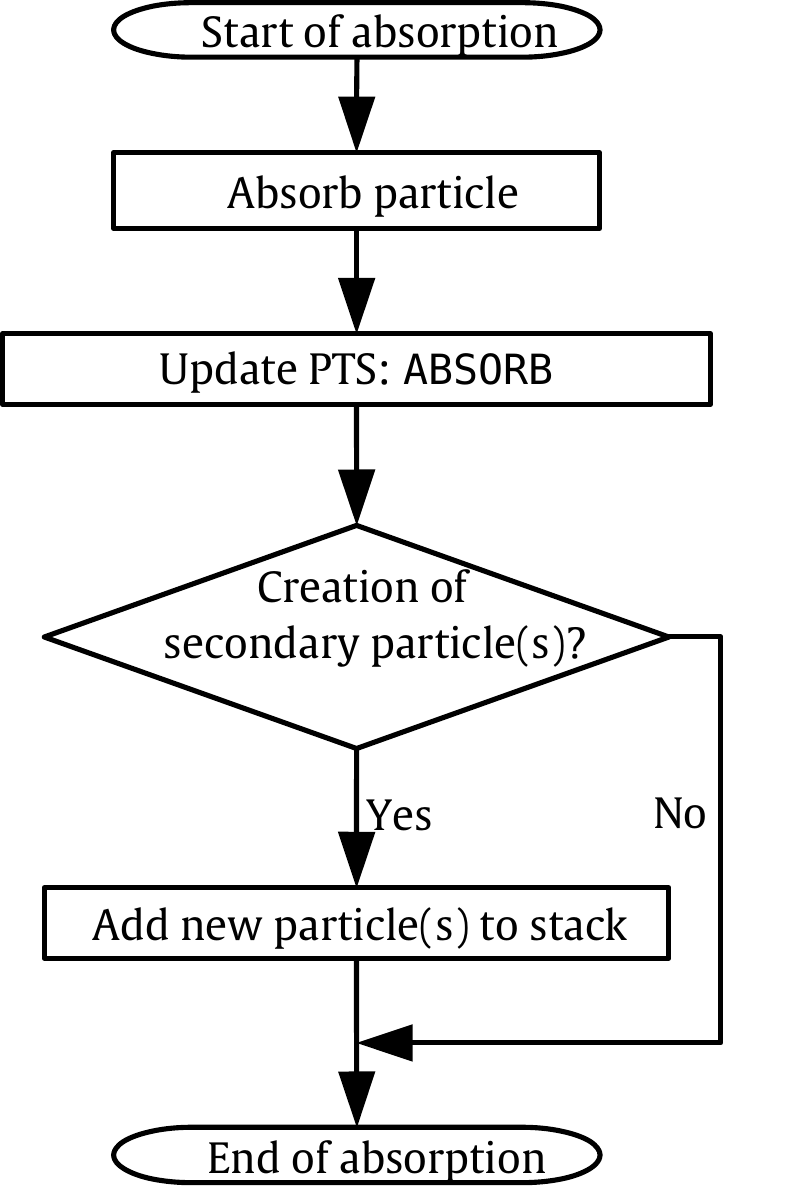}
        \caption{Absorption of particle within a volume}
        \label{fig:sAbsorption}
    \end{subfigure}
    \caption{Three types of sub-routines: relocation, propagation and absorption}
    \label{fig:sReaction}
\end{figure*}

The sub-routines  of the interaction phase are presented in Fig.~\ref{fig:sReaction}.
They can modify particle parameters such as location, energy and direction of motion, as well as changing the particle tracing states that can trigger available scorers. More details of the scorers are introduced in section~\ref{ssScorer}. 

In the relocation sub-routine shown in Fig.~\ref{fig:sRelocation}, the particle is moved to the boundary of the next volume, with the momentum unchanged. 
In the propagation sub-routine shown in Fig.~\ref{fig:sPropagation}, the scattering angle and energy of the particle are sampled.
In the absorption sub-routine shown in Fig.~\ref{fig:sAbsorption}, the alive status of the particle is set to false. 

\subsection{Particle gun}
\label{ssGun}
The initial neutron position, momentum direction, energy and time are generated by particle guns. 
A few types of guns are initially provided in this version of \texttt{Prompt} to facilitate the modelling of point sources, surface sources and particle histories in the standard Monte Carlo Particle Lists (\texttt{MCPL}) format from other software packages. 

Two basic particle guns, which can create neutron point sources, are provided, namely, \texttt{SimpleThermalGun} and 
\texttt{IsotropicGun}. 
There are three ways to specify neutron energies in these two guns, using a single value, sampling from uniform distribution or a Maxwell–Boltzmann distribution of coefficient $kT$. The \texttt{SimpleThermalGun} can produce neutrons aiming in a single direction, as shown in Listing~\ref{LisSimpleThermalGun}. While the \texttt{IsotropicGun} can produce isotropic neutrons, in which the momentum direction is equiprobable in all solid angles, see Listing~\ref{LisIsotropicGun}. 

\begin{lstlisting}[language=XML, 
label=LisSimpleThermalGun,
caption={\texttt{SimpleThermalGun} that produces neutrons with energy of \SI{0.0253}{\eV}, position of $(0,0,0)$ and direction along x axis.},
frame=single
]
<userinfo>
  <auxiliary auxtype="PrimaryGun" auxvalue="gun=SimpleThermalGun; energy=0.0253; position=0,0,0; direction=1,0,0" />
</userinfo>
\end{lstlisting}

\begin{lstlisting}[language=XML, 
label=LisIsotropicGun,
caption={\texttt{IsotropicGun} that produces radial neutron beams with energy of \SI{0.0253}{\eV}, position of $(0,0,0)$.},
frame=single
]
<userinfo>
  <auxiliary auxtype="PrimaryGun" auxvalue="gun=IsotropicGun; energy=0.0253; position=0,0,0" />
</userinfo>
\end{lstlisting}

Inspired by McStas, \texttt{Prompt} also provides \texttt{UniModeratorGun} and \texttt{MaxwellianGun} to generate neutrons from a moderator surface.
The particle positions are sampled randomly from a rectangular surface representing a moderator surface. The direction of each is determined by sampling another point in another opening representing a slit as the entrance of neutrons to a beam channel.  

The \texttt{UniModeratorGun} samples randomly the kinetic energy of neutrons from the uniform distribution formed by the wavelength and the range of wavelength taken as inputs, as shown in Listing~\ref{LisUniModeratorGun}. 

The \texttt{MaxwellianGun} takes a temperature as input, and samples randomly the kinetic energy of neutrons from a corresponding Maxwell–Boltzmann distribution, as presented in Listing~\ref{LisMaxwellianGun}.

\begin{lstlisting}[language=XML, 
label=LisUniModeratorGun,
caption={A \texttt{UniModeratorGun}, producing neutrons with energy randomly selected between $\SI{3.09}{\angstrom}$ and $\SI{3.69}{\angstrom}$, with position randomly selected in an opening of \SI{100}{\mm}$\times$\SI{50}{\mm} located at $(0,0,-400)$ (in \SI{}{\mm}), with directions determined by connecting two points of two openings},
frame=single
]
<userinfo>
  <auxiliary auxtype="PrimaryGun" auxvalue="gun=UniModeratorGun; mean_wl=3.39; range_wl=0.3; src_w=100; src_h=50; src_z=-400; slit_w=5; slit_h=10; slit_z=1" />
</userinfo>
\end{lstlisting}

\begin{lstlisting}[language=XML, 
label=LisMaxwellianGun,
caption={A \texttt{MaxwellianGun}, producing neutrons with energy randomly selected ($T=293.15\si{\kelvin}$), with position randomly selected in an opening of \SI{100}{\mm}$\times$\SI{50}{\mm} located at $(0,0,-400)$ (in \SI{}{\mm}), with directions determined by connecting two points of two openings},
frame=single
]
<userinfo>
  <auxiliary auxtype="PrimaryGun" auxvalue="gun=MaxwellianGun; src_w=100; src_h=50; src_z=-400; slit_w=5; slit_h=10; slit_z=1; temperature=293.15; " />
</userinfo>
\end{lstlisting}

An example of using an \texttt{MCPLGun} is shown in Listing~\ref{LisMCPLGun}. The input files are typically generated by  Monte Carlo packages, e.g. McStas, Geant4, MCNP and PHITS. Notice that when preparing the input files,  particles other than neutrons have to be filtered out. That can be easily done by using mcpltool in the \texttt{MCPL} toolkit as
\texttt{mcpltool --extract -p2112 example.mcpl justneutrons.mcpl
}

\begin{lstlisting}[language=XML, 
label=LisMCPLGun,
caption={A \texttt{MCPLGun} reads neutron histories from the specified MCPL files},
frame=single
]
<userinfo>
  <auxiliary auxtype="PrimaryGun" auxvalue="gun=MCPLGun; mcplfile=justneutrons.mcpl"  />
</userinfo>
\end{lstlisting}

\subsection{Geometry }
\label{ssGeometry}
\texttt{Prompt} uses \texttt{VecGeom}~\cite{amadio2005geantv} v1.2.0 as the geometry engine and \texttt{GDML}~\cite{chytracek2006geometry} as the language to define the geometry of a simulation. A geometry is a hierarchy of volumes, each of which a shape and a composition should be assigned to.
There are three essential parts to describe a volume, namely \texttt{solids} for the shape description, \texttt{structure} for the hierarchy relationship and \texttt{material} for the composition. In this section, the methods to define the \texttt{solids} and \texttt{structure} are given, and leaving the \texttt{materials} part in section~\ref{ssMaterial}.

The \texttt{solids} currently supported in \texttt{Prompt} are retaken from the \texttt{GDML} user's guide~\cite{gdml2022} and presented in~\ref{\detokenize{./prompt/usersmanual/geometry:geometry}}.
A \texttt{solid} defines solid type and the corresponding parameters. As an example, the lines presented in Listing~\ref{LisBox} define a box of the identical length each side. The length unit, \texttt{lunit}, is millimeter by default. It can be otherwise chosen to be nanometer (\texttt{nm}), micrometer (\texttt{um}), centimeter (\texttt{cm}),  meter (\texttt{m}) or kilometer (\texttt{km}). The default unit for angle, \texttt{aunit}, is degree (\texttt{deg}), and radian (\texttt{rad}) can also be used, alternatively.  

\begin{lstlisting}[
language=XML,
label=LisBox,
caption=A box named \textit{Cube} of \SI{10}{\mm} for each side length,
frame=single
]
<solids>
    <box lunit="mm" name="Cube" x="10.0" y="10.0" z="10.0" />
</solids>
\end{lstlisting}

The \texttt{structure} is a hierarchy of volumes. A volume is called logical when it refers to a \texttt{solid} and a \texttt{material}.
A logical volume becomes physical when it is positioned into a mother logical volume. Notice that any boundary of a daughter volume should be contained within its mother volume and boundary overlap test is not implemented in this version of the system.  
All volumes in a simulation are physical because they have to be positioned, except for the \texttt{world}, the largest volume on the top of the hierarchy. 
An example is presented in Listing~\ref{LisStructure}, where a hierarchy of two levels is defined. The first level corresponds to the \texttt{world} of this simulation, while the second is a volume named \texttt{phy\_2ndLevel} placed in the center of the \texttt{world}.

\begin{lstlisting}[
language=XML,
label=LisStructure,
caption=A hierarchy of two levels,
frame=single
]
<solids>
    <box lunit="mm" name="solWorld" x="10" y="10" z="10" />
    <orb lunit="mm" name="solSphere" r="5" />
</solids>
<structure>
    <volume name="2ndLevel">
        <solidref ref="solSphere" />
        <materialref ... />
    </volume>
    <volume name="World">
        <solidref ref="solWorld"/>
        <materialref ... />
        <physvol name="phy_2ndLevel">
            <volumeref ref="2ndLevel" />
            <position name="Position" unit="mm" x="0" y="0" z="0" />
        </physvol>
    </volume>
</structure>
\end{lstlisting}

\subsection{Scorer}
\label{ssScorer}
\begin{table}
	\centering                       
	\caption{Particle scorers in a volume. The statistical information is presented in form of one-dimensional histogram, if not otherwise noted. }
	\label{tableScorer} 
	
    \begin{tabular}{p{7.5em}p{25em}} 
    \toprule
    Scorer & Statistical Information   \\ 
    
    \midrule
    \texttt{ESpectrum} & Energy Spectrum\\
    \texttt{WlSpectrum} & Wavelength spectrum \\ 
    \texttt{TOF} & Time-of-flight spectrum of particles.\\ 
    \texttt{PSD} & Two-dimensional spatial spectrum on the projection of a volume\\
    \texttt{MultiScat} & Scattering number spectrum of scattered in a volume \\
    \texttt{DeltaMomentum} & Momentum transfer spectrum \\ 
    \texttt{Angular} & Angular spectrum  \\ 
    \texttt{VolFluence} & Particle flux spectrum in a volume \\ 
    
    \bottomrule
    \end{tabular}
\end{table}

Eight scorers that are designed to accumulate particle statistical information in  histograms are listed in Table~\ref{tableScorer}. 
The variables of interest are generally fixed to one of the attributes of the particles, such as energy, position, and time. 
For a \texttt{DeltaMomentum} scorer, the method used to calculate the variable should be chosen from elastic or inelastic scatterings. 
In the elastic case, the momentum transfer $Q$ is calculated as 
\begin{equation} \label{eQelastic}
Q = 2k_i\sin\theta
\end{equation}

While in the inelastic case, the momentum transfer is 
\begin{equation} \label{eQtrue}
Q^2 = {k_i}^2+{k_f}^2-2{k_i}{k_f}\cos2\theta
\end{equation}

To set up, a scorer is attached to a volume from which the statistical information is expected to be extracted. For instance, in Listing~\ref{LisPSD}, a \texttt{PSD} scorer is attached to the volume named Detector. The parameters of scorers are listed in~\ref{sScorers}.

\begin{lstlisting}[language=XML, 
label=LisPSD,
caption={A \texttt{PSD} scorer attached to the volume named Detector},
frame=single
]
<volume name="Detector">
  <materialref ref="Vacuum"/>
  <solidref ref="DetectorSolid"/>
  <auxiliary auxtype="Scorer" auxvalue= "Scorer=PSD; name=NeutronHistMap; xmin=-250; xmax=250; numBins_x=100; ymin=-250; ymax=250; numBins_y=100; ptstate=SURFACE; type=XY"/>
</volume>
\end{lstlisting}

The \texttt{ptstate} in Listing~\ref{LisPSD} denotes the particle tracing state. The particle tracing states reflect the relationships between the particles and volumes, and also act as triggers allowing scorers to accumulate particle weights.
Each scorer should be assigned to a particle tracing state to listen to.
In \texttt{Prompt} five particle tracing states are defined, as follows: 
\begin{itemize}
    \item \texttt{SURFACE}$:$ A particle reaches the geometry boundary of a volume,
    \item \texttt{ENTRY}$:$ A particle entries a volume,
    \item \texttt{ABSORB}$:$ A particle is absorbed in a volume,
    \item \texttt{PROPAGATE}$:$ A particle propagates in a volume,
    \item \texttt{EXIT}$:$ A particle exits a volume.
\end{itemize}
All of them are available for all scorers, except that \texttt{MultiScat} is immutably set to \texttt{PROPAGATE}. 

\begin{figure}
\centering
\includegraphics[width=\linewidth]{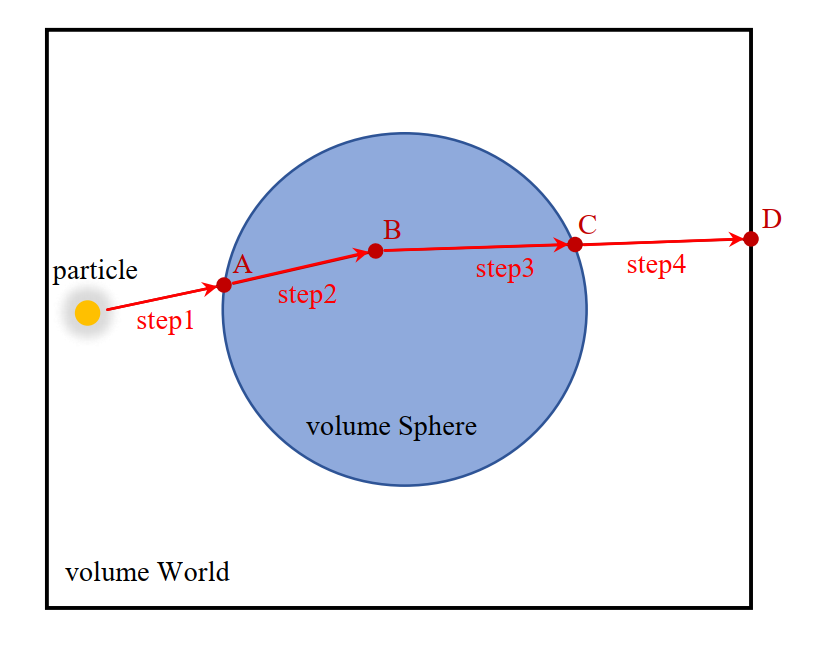}
\caption{Transport process of a particle in a simple geometry}\label{fscorertrigger}
\end{figure}

To illustrate the triggering mechanism of scorers, the transport process of a particle in the geometry of Listing~\ref{LisStructure} is shown in Fig.~\ref{fscorertrigger}. 
The particle transports four steps in this simple geometry. 
In step 1, when the particle reaches point \texttt{A} at the boundary of volume Sphere, it simultaneously exits volume World and entries volume Sphere.
Therefore, it can trigger the scorer that is listening to  the \texttt{EXIT} PTS of the volume World and the \texttt{SURFACE} and \texttt{ENTRY} PTS of the volume Sphere. 
In step 2, when the particle arrives at point \texttt{B}, it is possible to  trigger the \texttt{PROPAGATE} or \texttt{ABSORB} PTS of the volume Sphere. 
In this example, it is chosen to be \texttt{PROPAGATE} so that the simulation continues.  
Similar to step 1, when the particle reaches point C at the end of step 3, it can trigger the scorer on volume Sphere set to \texttt{EXIT} and the scorer on volume World set to \texttt{ENTRY}. 
When the particle reaches point D at the end of step 4, it is at the boundary of volume World and will be out of volume World. 
Hence, it is killed and can trigger the scorer on volume World set to \texttt{EXIT}.

It is worth noting that the \texttt{SURFACE} and \texttt{ENTRY} can be used interchangeably for scorers. 
These two PTS are used in the core of the system to indicate whether the interaction is treated using the surface or bulk material models. 

A scorer creates one or several histograms showing the statistical results. They are encoded in the \texttt{Numpy} array format~\cite{Numpy} which are then stored in the header of output MCPL~\cite{MCPL} files. In addition, a Python script is generated along with each MCPL file to show the basic read and plot functions for post-analysis.

\subsection{Simulation physics}
\label{ssMaterial}
\texttt{Prompt} defines three types of physics, namely, bulk material physics, mirror physics and ray-tracing physics. 

All bulk material cross sections are computed by the backend engine \texttt{NCrystal v3.0.0}. 
The syntax for defining an ideal isotropic Al polycrystal is shown in Listing~\ref{materialAl}. 
The physics configuration string that is passed to \texttt{NCrystal} for the calculation of averaged cross section per atom is enclosed in single quotation marks. 
The configuration string syntax and basic crystal parameters in the \texttt{.ncmat} file are explained in the reference paper~\cite{Cai2019} and the official Wikipedia~\cite{NCWiki} of \texttt{NCrystal} in detail.    
In this example, the cross section biasing factors for both the  scattering and absorption models are doubled. 
Notice that in the special case, where a factor is set to zero, the corresponding model is removed from the simulation.

\begin{lstlisting}[language=XML, 
label=materialAl,
caption=Example of defining an ideal Al polycrystal material with user specified cross section biasing factors,
frame=single
]
<material name="Al_biased">
    <atom value="physics=ncrystal;nccfg='Al_sg225.ncmat';
                 scatter_bias=2.0;abs_bias=2.0"/>
</material>
\end{lstlisting}

The biasing factors are unity by default. Indicated by Eq.~\ref{eWeight}, which effectively disables the cross section biasing mechanism. 
Equivalently, the material definition can be simplified by only passing the \texttt{NCrystal} physics configuration string, as Listing~\ref{materialAl2}. 
\begin{lstlisting}[language=XML, 
label=materialAl2,
caption=Defining the same material as Listing~\ref{materialAl} but with unity cross section biasing factors,
frame=single
]
<material name="Al">
    <atom value="Al_sg225.ncmat"/>
</material>
\end{lstlisting}

For oriented isotropic Gaussian mosaic single crystals, in order to unambiguously define the unit cell orientations, two non-parallel vectors in the Miller indices  and their corresponding laboratory directions should be specified~\cite{Cai2020}. 
For example, the definition of Ge-511 monochromator is followed.

\begin{lstlisting}[language=XML, 
label=materialGe,
caption= Ge-511 single crystal with \SI{0.3}{\degree} isotropic Gaussian mosaicity,
frame=single
]        
<material name="Ge">
  <atom value="Ge_sg227.ncmat;mos=0.3deg;
               dir1=@crys_hkl:5,1,1@lab:0,0,1;
               dir2=@crys_hkl:0,1,-1@lab:1,0,0"/>
</material>
\end{lstlisting}

For layered crystals, i.e. crystals consisting of parallel layers, additional \texttt{lcaxis}, the averaged normal of the layers should be specified~\cite{Kittelmann2021} as showing in the Listing~\ref{materialGraphite}. Notice that, repetition of the \texttt{lcaxis} in the configuration string is not required, if it is already embedded in the \texttt{.ncmat} input file.

\begin{lstlisting}[language=XML, 
label=materialGraphite,
caption= {Pyrolytic graphite-002 layered crystal with \SI{2}{\degree} mosaicity},
frame=single
]        
<material name="Graphite">
  <atom value="C_sg194_pyrolytic_graphite.ncmat;mos=2deg;
               dir1=@crys_hkl:0,0,2@lab:0,0,1;
               dir2=@crys_hkl:1,0,0@lab:1,0,0;
               lcaxis=0,0,1"/>
</material>
\end{lstlisting}

The backend engine also supports the so-called ``unstructured materials''~\cite{NCWiki}, \textit{freegas::} and \textit{solid::}. They are well-suited to describe gaseous or strong absorbing materials in which atomic scale configurations contribute insignificantly to the total cross section. 
The \textit{freegas::} type materials use the classical two-body collision model to treat the scattering, 
while the \textit{solid::}
type material models apply the phonon expansion method using the Debye density of states to prepare the inelastic scattering function. 
For the later material type, additional incoherent elastic scatterings are also included. 

\begin{lstlisting}[language=XML, 
label=materialUnstructured,
caption={Examples of defining dry air with 78\% N and 22\% O and enriched B$_4$C solid with 95\% $^{10}$B and 5\% $^{11}$B},
frame=single
]        
<material name="Air">
    <atom value="freegas::N78O22/1.225kgm3"/>
</material>
<material name="enriched_B4C">
    <atom value="solid::B4C/2.5gcm3/B_is_0.95_B10_0.05_B11"/>
</material>
\end{lstlisting}

The mirror model is implemented as a surface process. It should be attached to a given volume to describe the specular reflection. 
A mirror can be specified by associate a \texttt{SurfaceProcess} to a volume. An example is given in Listing~\ref{materialMirror}. See more discussions on surface type physics at the end of section~\ref{ssScorer}.

\begin{lstlisting}[language=XML, 
label=materialMirror,
caption={Examples of defining a mirror with  \texttt{m}=2, see Eq.~\ref{eMirrowRef}},
frame=single
]        
<volume name="lv_Mirror">
    <solidref ref="solid_mirror"/>
    <materialref ref="Si"/>
    <auxiliary auxtype="BoundaryPhysics" auxvalue="physics=MirrorPhyiscs;m=2.0"/>
</volume>
\end{lstlisting}

\begin{figure}
    \centering
    \includegraphics[width=0.4\textwidth]{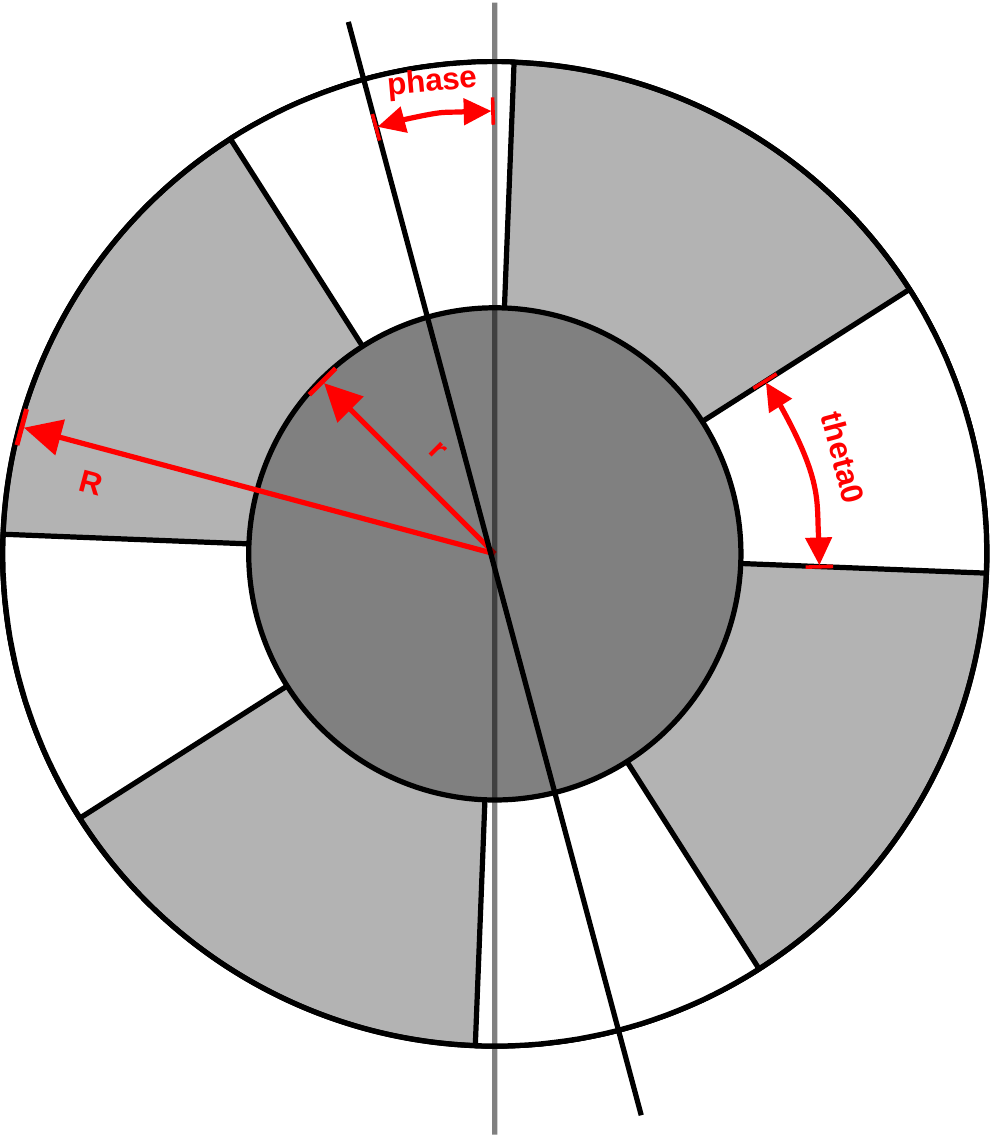}
    \caption{An illustration of the \texttt{DiskChopper} with $n=4$ opening windows in \texttt{Prompt}, with \texttt{phase} the angle deviated from the origin (position at time zero), \texttt{theta0} the opening angle of each window, \texttt{r} the inner radius, \texttt{R} the outer radius (which is the radius of the \texttt{tube} to which the \texttt{DiskChopper} is attached).}
    \label{fig:sChopperSketch}
\end{figure}

A \texttt{DiskChopper} of \texttt{Prompt} is a ray-tracing physics similar to the \texttt{DiskChopper} component of McStas. An example is provided in Listing~\ref{materialChopper}, which adds a \texttt{DiskChopper} to the simulation. The parameters are listed in Table.~\ref{tableDiskChopper}.
The geometry of the disk is illustrated in Fig.~\ref{fig:sChopperSketch}. The central circular area of radius $r$ is completely black to neutrons. There are $n=4$ periodic opening windows that are evenly distributed in 2$\pi$. The angle of a window is defined by $\theta_0$. The positive rotational frequency indicates the rotational direction that follows the right-hand rule.

\begin{lstlisting}[language=XML, 
label=materialChopper,
caption={Examples of defining a disc chopper like Fig.~\ref{fig:sChopperSketch}},
frame=single
]        
<solids>
    <tube lunit="mm" name="solid_DiskChopper" rmin="0.0" rmax="250.0" z="0.001" deltaphi="360.0" startphi="0.0"/>
</solids>
<structure>
    <volume name="lv_DiskChopper">
        <solidref ref="solid_DiskChopper" />
        <materialref ref="Vacuum" />
        <auxiliary auxtype="SurfaceProcess" auxvalue="physics=DiskChopper; rotFreq=100; r=130; theta0=20; n=4; phase=0"/>           
    </volume>
</structure>
\end{lstlisting}

\begin{table}
	\centering                       
	\caption{The parameters of \texttt{DiskChopper}. All of the parameters are required.}
	\label{tableDiskChopper} 
	
    \begin{tabular}{p{0.15\textwidth}p{0.7\textwidth}p{0.15\textwidth}}
    \toprule
    Parameter & Description & Unit
\\ 
    \midrule
    r & Inner radius of the disk chopper & \si{mm}\\
    n & Number of window. If multiple windows are defined, they are evenly arranged in 2$\pi$. & \num{1}\\
    rotFreq & Rotation frequency.  & \SI{}{\hertz}\\
    theta0 & Opening angle of each window. & \SI{}{\degree}\\
    phase & Angle deviated from the position at time zero. & \SI{}{\degree}\\
    \bottomrule
    \end{tabular}
\end{table}

\subsection{A minimal example}
\label{sSample}
An example, minimized but involving most of the introduced elements is presented. 
As a first step, an input is prepared in \texttt{GDML} format, named \texttt{example.gdml}, as shown in Listing~\ref{LisMinmumExample}. In this example, the Debye-Scherrer cones of \SI{2.2}{\angstrom} incident neutrons from an Al powder sample are measured by a position sensitive detector.

\begin{figure}
    \centering
    \begin{subfigure}[b]{0.40\textwidth}
        \centering
        \includegraphics[width=\textwidth]{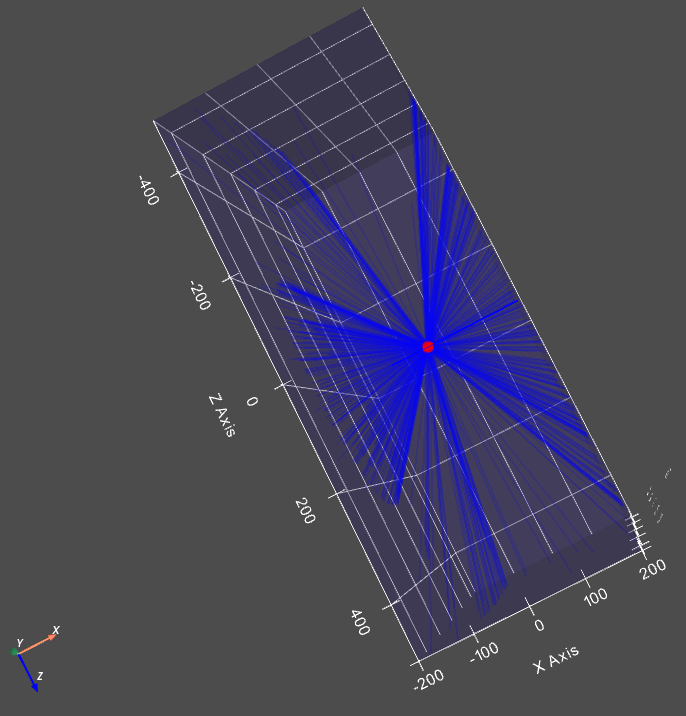}
        \caption{Visualization of debye-scherrer cone reproduced by using \texttt{Prompt}, with Al powder}
        \label{fig:sConeViz}
    \end{subfigure}
    \hfill
    \begin{subfigure}[b]{0.59\textwidth}
        \centering
        \includegraphics[width=\textwidth]{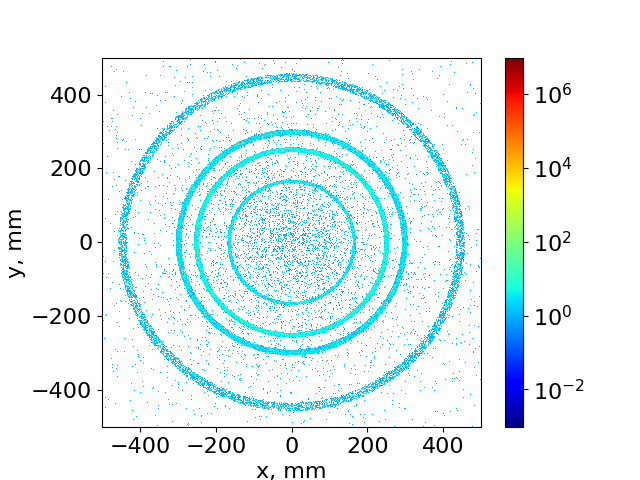}
        \caption{Neutrons captured by \texttt{PSD} scorer, Debye-Scherrer cones manifested with four circles}
        \label{fig:sCone_score}
    \end{subfigure}
    \caption{Visualization and scoring results of the simulation mission}
    \label{fig:sCone}
\end{figure}

Fig.~\ref{fig:sConeViz} is the visualisation of one thousand neutron trajectories. It can be reproduced by the command 

\begin{lstlisting}[language=bash, frame=single
]
#!/bin/bash
prompt -g example.gdml -v -n 1000
\end{lstlisting}

A production run can be executed without the visualiser as
\begin{lstlisting}[language=bash, frame=single]
#!/bin/bash
prompt -g example.gdml -n 5e6
\end{lstlisting}

After the simulation, the heatmap of the detector and the corresponding python script template is generated. The heatmap, as shown in Fig.~\ref{fig:sCone_score}, of the position sensitive score in logarithmic scale can be visualised by running 
\begin{lstlisting}[language=bash, frame=single]
#!/bin/bash
python ScorerPSD_NeutronHistMap_view.py -l
\end{lstlisting}
It should be noted that the file name of the \texttt{Python} script created depends on the configuration of the scorer. 

\begin{lstlisting}[language=XML,
label=LisMinmumExample,
caption={A minimum example of input script in \texttt{Prompt}, \texttt{example.gdml}},
frame=single]
<gdml>
    <userinfo>
        <!-- Defining gun -->
        <auxiliary auxtype="PrimaryGun" auxvalue="gun=SimpleThermalGun; energy=0.0169; position=-80.,0.,0.; direction=1,0,0" />
    </userinfo>
    <materials>
        <!-- Defining material -->
        <material name="Universe">
          <atom value="freegas::H1/1e-26kgm3"/>
        </material>
        <material name="Al">
          <atom value="Al_sg225.ncmat"/>
        </material>
    </materials>
    <solids>
        <!-- Defining solid -->
        <box lunit="mm" name="WorldSolid" x="400.0" y="1000.0" z="1000.0" />
        <box lunit="mm" name="SampleSolid" x="10.0" y="10.0" z="10.0" />
    </solids>
    <structure>
        <!-- Defining geometry tree and hierarchy of volumes -->
        <volume name="VolSample">
            <solidref ref="SampleSolid" />
            <materialref ref="Al" />
        </volume>
        <volume name="VolWorld">
            <solidref ref="WorldSolid" />
            <materialref ref="Universe" />
            <!-- Defining PSD scorer and attaching it to World -->
            <auxiliary auxtype="Scorer" auxvalue="Scorer=PSD; name=NeutronHistMap; xmin=-500; xmax=500; numBins_x=1000; ymin=-500; ymax=500; numBins_y=1000; ptstate=EXIT; type=YZ"/>
            <physvol name="PhysVol_Sample">
                <volumeref ref="VolSample" />
            </physvol>
        </volume>
    </structure>
    <setup name="Default" version="1.0">
        <world ref="VolWorld" />
    </setup>
</gdml>
\end{lstlisting}

\section{Application and Benchmark}
\label{sBenchmark}

In section~\ref{ssTotal}, the total scattering technique is applied to a heavy water sample to investigate the effectiveness of the cross section biasing technique for multiple scatterings. A minimal static source powder diffractometer, including a single crystal monochromator,  pattern is simulated in section~\ref{ssPowder}.
Section~\ref{ssGuide} introduces the simulation of a squared neutron guide as a function of incident wavelength.
A complex geometry is in section~\ref{ssRabbit} to discuss the impact of the geometry on the overall computational speed. A moving optical component, disk chopper, is presented in section~\ref{ssChopper}, showing the capability in ray-tracing process simulation.

\subsection{Total scattering}
\label{ssTotal}
\begin{figure}
\centering
\includegraphics[width=\linewidth]{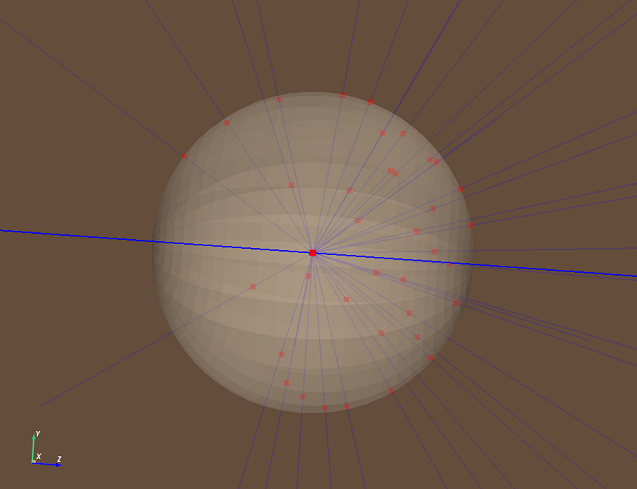}
\caption{Geometry visualisation of the total scattering simulation}\label{ftotal_geo}
\end{figure}

A heavy water sample that is surrounded in a perfect spherical detector is simulated. 
Fig.~\ref{ftotal_geo} shows the visualisation of the geometry and 100 neutron trajectories.
In the simulation, a \texttt{SimpleThermalGun} is used to emit neutrons in the positive z-axis direction at \SI{3.635}{\eV}.
At \SI{10}{\meter} downstream of the gun,
a \SI{12}{\mm} radius spherical sample is located in the center of the detector, of which the inner radius is \SI{2000}{\mm} and the thickness is \SI{1}{\mm}.
A \texttt{MultiScat} scorer is attached to the sample and used in cooperation with the detector to analyse the scattering number of a detected event.  

A few \texttt{DeltaMomentum} scorers triggered by the \texttt{ENTRY} PTS are attached to the detector to obtain momentum transfer distribution of neutrons which are scattered at different times in the sample. 
Due to the special geometry of the detector, the accumulated single scattering $P(Q)$ is proportional to the static structure factor  $S(Q)$. 
On the other hand,
the employed CAB~\cite{CAB2014} cross section for heavy water is generated in the Sk\"old approximation using Eq.~\ref{eSkoldPartial}.  Hence the simulated structure factor is related to  ${S_{j,j'}}(Q)$, the partial static structure factor, used in the cross section generation. 

\begin{figure}
\centering
\includegraphics[width=\linewidth]{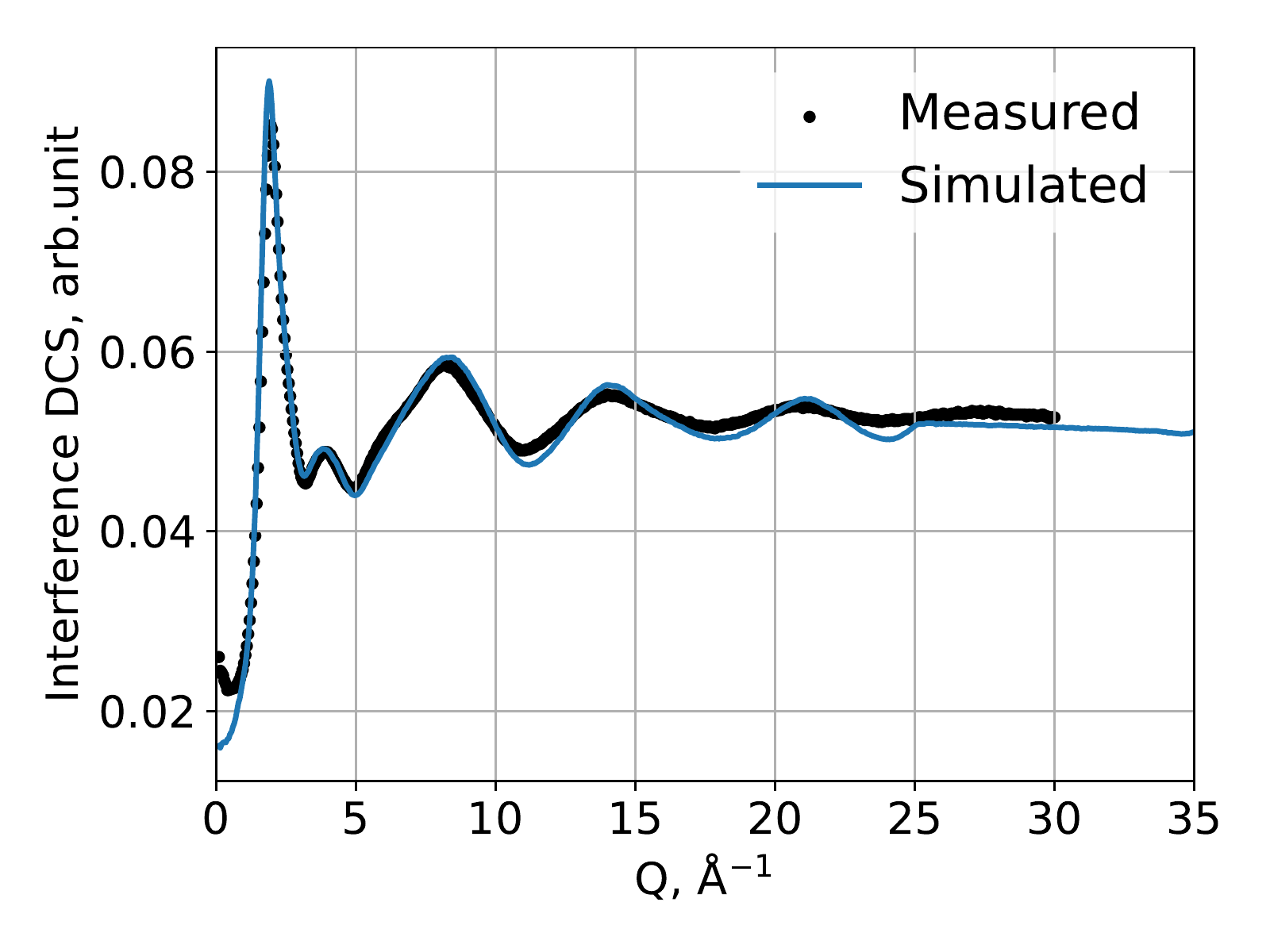}
\caption{Comparisons between normalisation of the simulation result of \texttt{Prompt} and the measurements from Soper~\cite{soper2013radial}
}\label{ftotal_simulation1}
\end{figure}

In Fig.~\ref{ftotal_simulation1}, the single scattering momentum transfer distribution of the non-biased sample is compared with the interference differential scattering cross section (DCS) for heavy water measured by Soper~\cite{soper2013radial}.
The structures of the simulated result are broadly similar to those in measurement. However, the widths of the peaks are slightly different. Also, the raising of the curve towards the lowest $Q$ is not reproduced by the simulation.  The observed discrepancies can be explained by the discrepancies between the static structure factor used in the cross section calculated and that from measurement. 

As indicated by Eq.~\ref{eSkoldPartial}, the partial static structure factors should be used for polyatomic systems. 
To obtain that, the measured data need to be fitted to a microscopic model to determine the atomic structure candidate. 
Therefore, the peak width discrepancies are likely introduced by the fitting procedure. 
Secondly, the raising of the experimental data at low $Q$ is caused by the liquid density fluctuation, which may not be included in the fitting model, hence not included in the cross section.

\begin{figure}
\centering
\includegraphics[width=\linewidth]{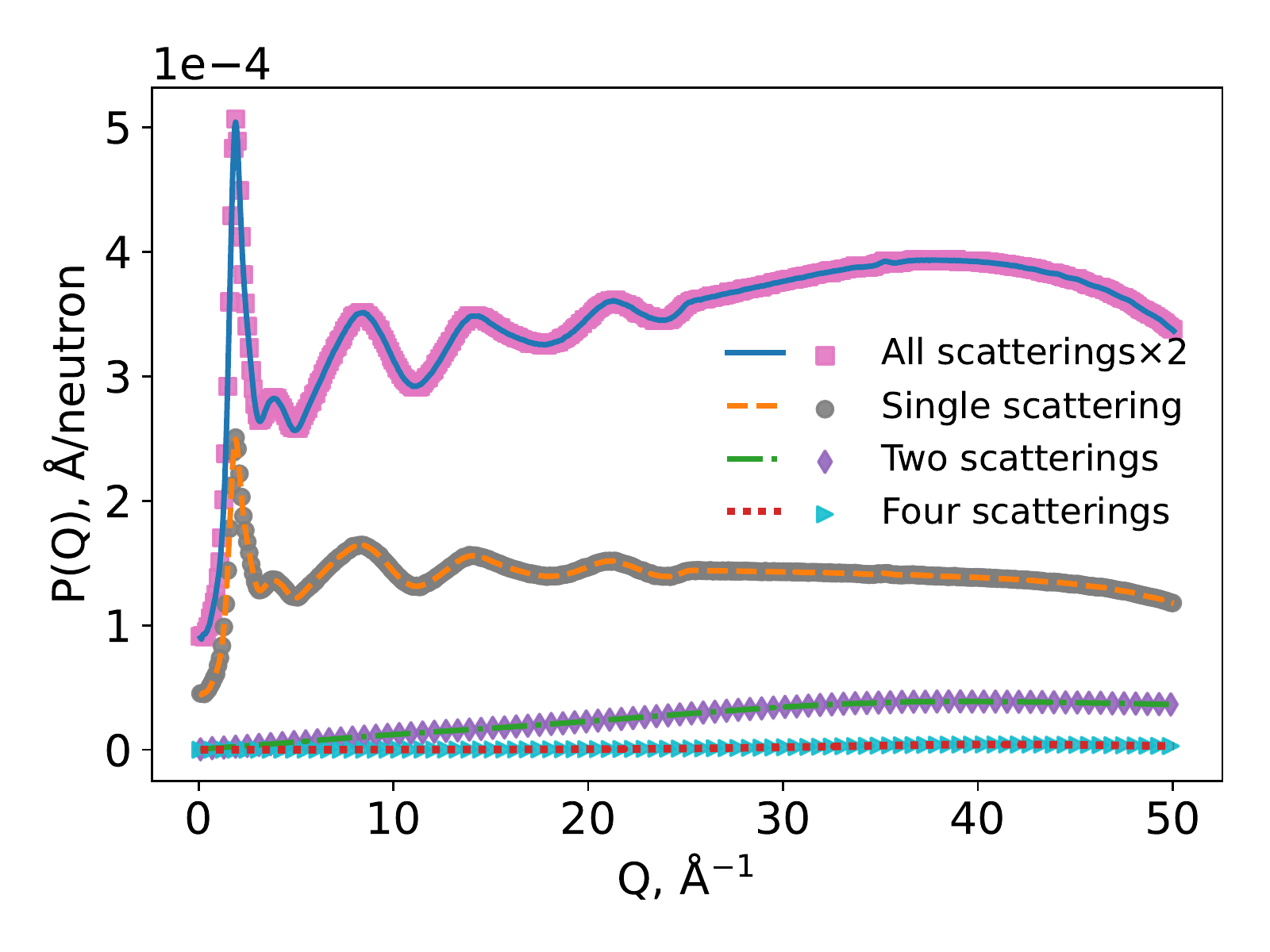}
\caption{Comparisons between $P(Q)$ for different scattering numbers obtained from non-biased and biased runs. The solid, dashed, dash-dotted and dotted lines are respectively the results of all scatterings, single scattering, two scatterings and four scatterings from the non-biased run, while the squares, circles,
diamonds, and triangles right are results of the biased run.}\label{ftotal_simulation2}
\end{figure}

Fig.~\ref{ftotal_simulation2} shows the comparison of momentum transfer distribution $P(Q)$ of neutrons with different scattering numbers obtained from non-biased and biased runs. 
When $Q$ is less than \SI{3}{\per\angstrom}, $P(Q)$ for all scatterings and single scattering are almost not distinguishable. 
However, the discrepancies between them grow with a greater  $Q$ value, suggesting that multiple scattering becomes more and more significant.

The results of two runs for the same scattering number are overlapped, hence the probability is conserved when the cross section biasing technique is applied. 
Although the number of incident neutrons of the biased run is only 10$\%$ of that of the non-biased run, their results are almost statistically equivalent. 
Therefore, the cross section biasing technique is able to improve the computational efficiency significantly.

\subsection{Powder diffraction}
\label{ssPowder}
\begin{figure}
\centering
\includegraphics[width=\linewidth]{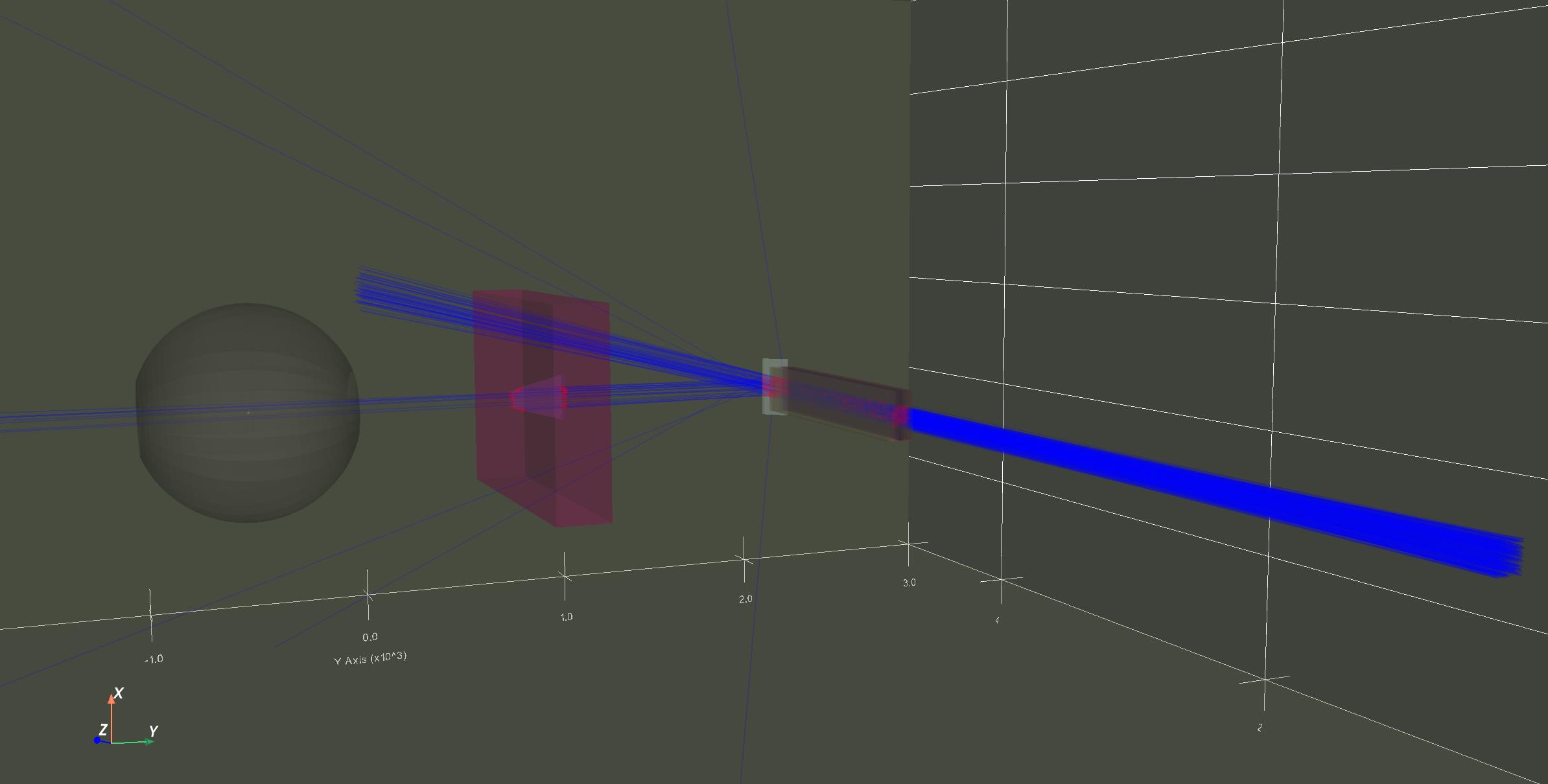}
\caption{Visualisation of the simulated PUS instrument}\label{vis-PUS}
\end{figure}

A powder neutron diffraction example is simulated in \texttt{Prompt}, using the experiment configurations from PUS at the JEEP II Reactor in Norway~\cite{bjorn2000pus}. Fig.~\ref{vis-PUS} shows the visualisation of the geometry and 100 neutron trajectories.

In this example, a \texttt{UniModerator} particle gun is used for the moderator, with the widths and heights of \SI{40}{\mm} for both moderator and slit, the moderator is \SI{1.9}{\m} away from the downstream slit.
Neutrons are uniformly distributed in a \SI{0.05}{\angstrom} interval and the mean wavelength is \SI{1.54}{\angstrom}. A \SI{15}{\arcminute} $\alpha_{1}$ Soller collimator system is located downstream of the moderator, and composed of 22 pieces of B$_4$C blades. 
A single bent monochromator that contains nine Ge-511 single crystals is  modelled as a single slab geometry. 
Such simplification neglects the focusing effects offered by the monochromator. It may have a large impact on the flux on sample and slightly change the instrument resolution function.
In the simulation, a \SI{54}{\mm}$\times$\SI{116}{\mm}$\times$ \SI{8}{\mm} single crystal germanium with \SI{16}{\arcminute} mosaicity is placed at the origin.
Its front surface is  parallel with the 511 reflection planes. The nominal glancing angle of the incident neutron is  \SI{45}{\degree}. In the experiment, the mean wavelength of the monochromatic neutron is
\SI{1.5449}{\angstrom}~\cite{bjorn2000pus}. 
To match that, the lattice length of the original \texttt{NCrystal} \texttt{Ge\_sg227.ncmat} file is adjusted accordingly to be \SI{5.7131}{\angstrom}. 
Downstream of the monochromator, a \SI{1015}{\mm} long trapezoid  beam channel is embedded in the shielding wall and acts as  the $\alpha_{2}$ collimator of the instrument. Its effective divergent is \SI{1.41}{\degree}.
The  sample is placed \SI{2.75}{\m} away from the monochromator along the negative y axis. 
To skip the detector calibration procedures as those in typical experiments, a spherical 4$\pi$ detector is used to cover all solid angles in between \SI{20}{\degree} and  \SI{160}{\degree} take-off angles. 
To eliminate any anisotropic effects, the sample is also made spherical with a \SI{2.5}{\mm} radius.

\begin{figure}
\centering
\includegraphics[width=\linewidth]{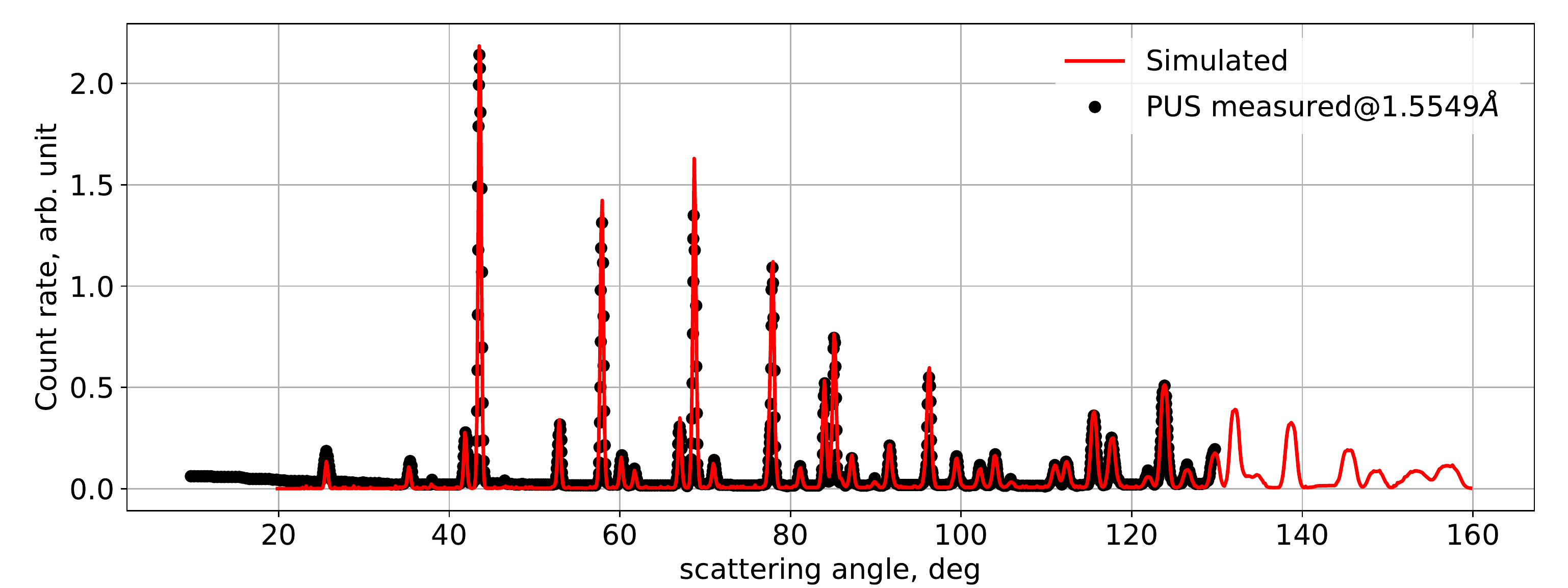}
\caption{Simulated $\rm{Al}_{2}\rm{O}_{3}$ powder diffraction pattern at PUS}\label{pus-result}
\end{figure}

Fig.~\ref{pus-result} shows the simulated powder diffraction pattern for the $\rm{Al}_{2}\rm{O}_{3}$ sample.  
The simulated and measured data are normalised by aligning the 143 peak at \SI{124.1}{\degree}.
The intensity and resolution behaviours of the pattern are well captured by the simulation, and the observed agreements are very good.
However, the experimental data is rising towards zero degrees. Similar behaviour is not observed in the simulation. 
In the experiments, the scattering environment between the sample and the detector is filled with helium; while in the simplified simulation, the environment is effectively vacuum.  Likely, this discrepancy is introduced by the fact that the scattering in the gas is not modelled. 

\subsection{Neutron guide}
\label{ssGuide}
\begin{figure}
\centering
\includegraphics[width=\linewidth]{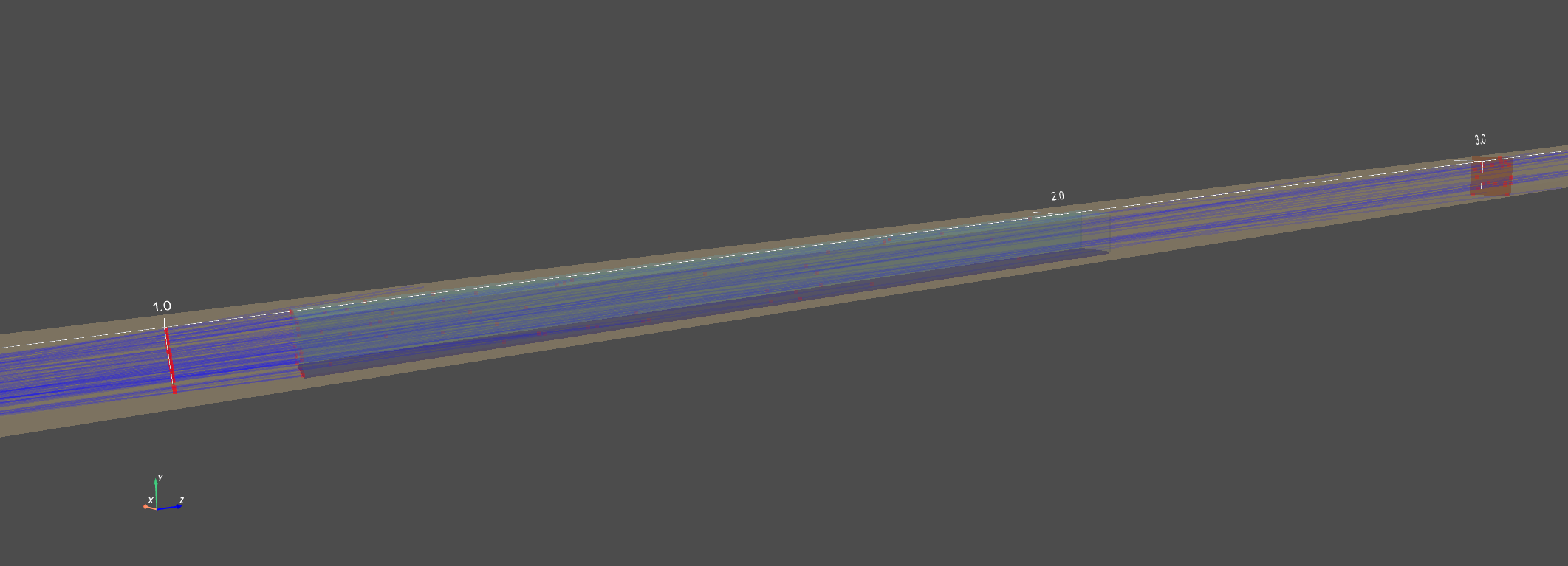}
\caption{Geometry visualisation of the guide simulation }\label{fGuideVis}
\end{figure}

The \texttt{Test\_Guides.instr} setup in the McStas optical examples is reproduced in \texttt{Prompt}. The simulation world contains two position sensitive detectors and a guide. Fig.~\ref{fGuideVis} shows the visualisation of the geometry and 100 neutron trajectories.

A front position sensitive detector, i.e. vacuum box attached by a \texttt{PSD} \texttt{ENTRY} type scorer, is placed \SI{1}{\meter} away from the \SI{50}{\mm}$\times$\SI{50}{\mm} moderator surface  that is emitting \SI{3.39}{\angstrom} neutrons. 
Four boxes are used to form an $m=1$ straight guide of which the inner opening is a \SI{50}{\mm}$\times$\SI{50}{\mm}$\times$\SI{1000}{\mm} box.
The entrance of the guide is \SI{100}{\mm} after the front detector.
In addition, a rear position sensitive detector is placed  \SI{3.2}{\meter} away from the source along the line.  

\begin{figure}
\centering
\includegraphics[width=\linewidth]{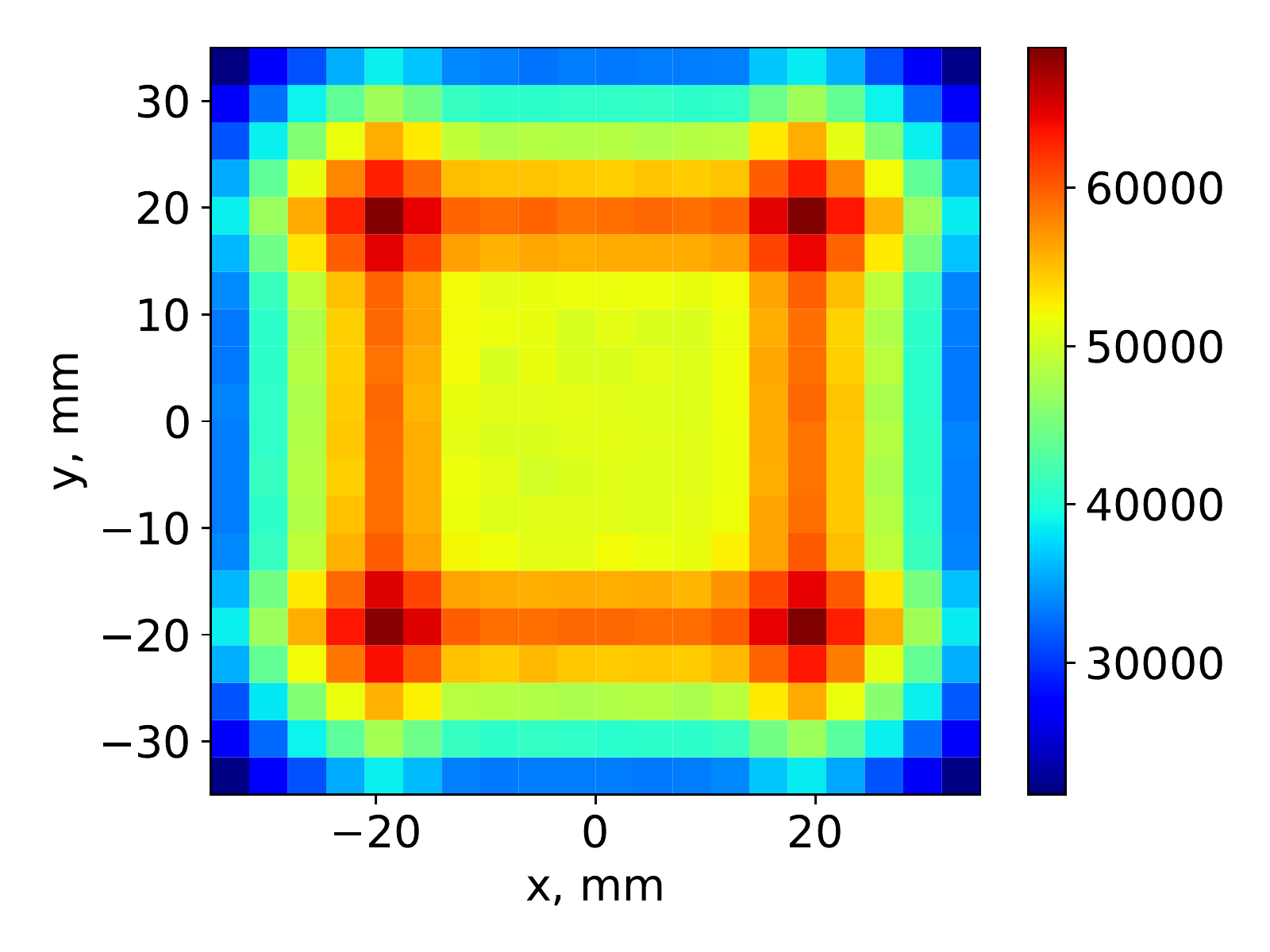}
\caption{Rear position sensitive detector pattern }\label{fGuide2d}
\end{figure}

In the benchmark, \num{e8} neutrons are simulated in about \SI{140}{\second} using a mid-2021 mainstream laptop. As the $m$ value is too low to perfectly reflect such a divergent monochromatic beam, only about 19.34\% of the neutrons reach the rear detector. 
The pattern on the detector is shown in  Fig.~\ref{fGuide2d}. 
The square shape of the guide is  visible and the four corners are highly distinguishable. The pattern from the McStas simulation of the same detector appears to be identical to that from \texttt{Prompt}, hence is not shown. 

\begin{figure}
\centering
\includegraphics[width=\linewidth]{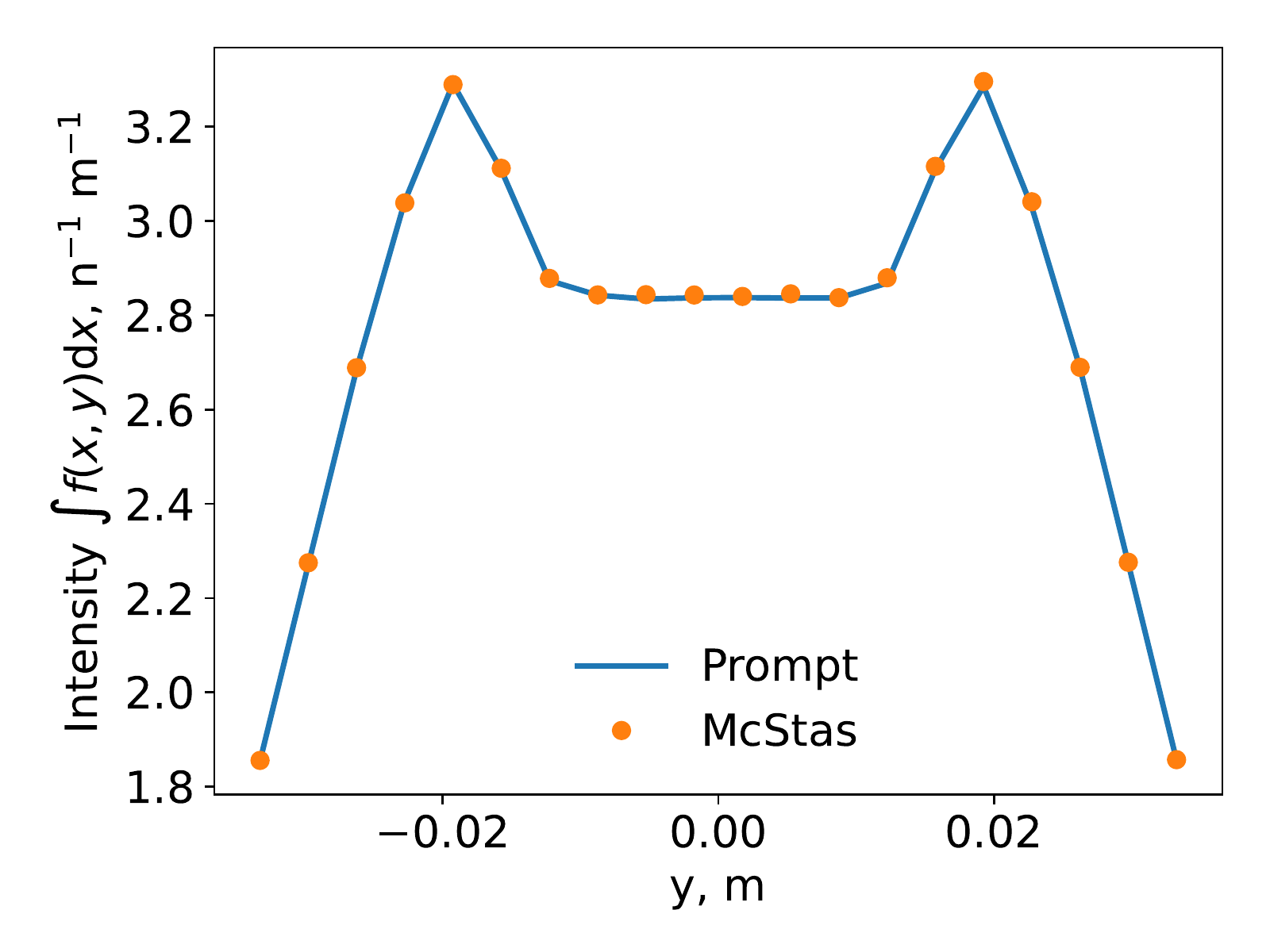}
\caption{Comparison between the integral weight simulated by  \texttt{Prompt} and that from McStas}\label{fGuide1d}
\end{figure}

The integral intensity of the rear detector, normalised by the front detector count, is compared with that from McStas. Not surprisingly, as the reflectivity models are identical, the curves from these two packages are overlapped. However, it is notable that McStas takes about half of the time to complete the simulation. Likely, the geometry module, i.e. VecGeom, in \texttt{Prompt} is not as effective as the ray-tracing algorithm used in McStas.  
On the other hand, \texttt{Prompt} is capable of modeling complex geometries as discussed in section~\ref{ssRabbit}.

\subsection{Stanford bunny}
\label{ssRabbit}
\begin{figure}
    \centering
    \includegraphics[width=\linewidth]{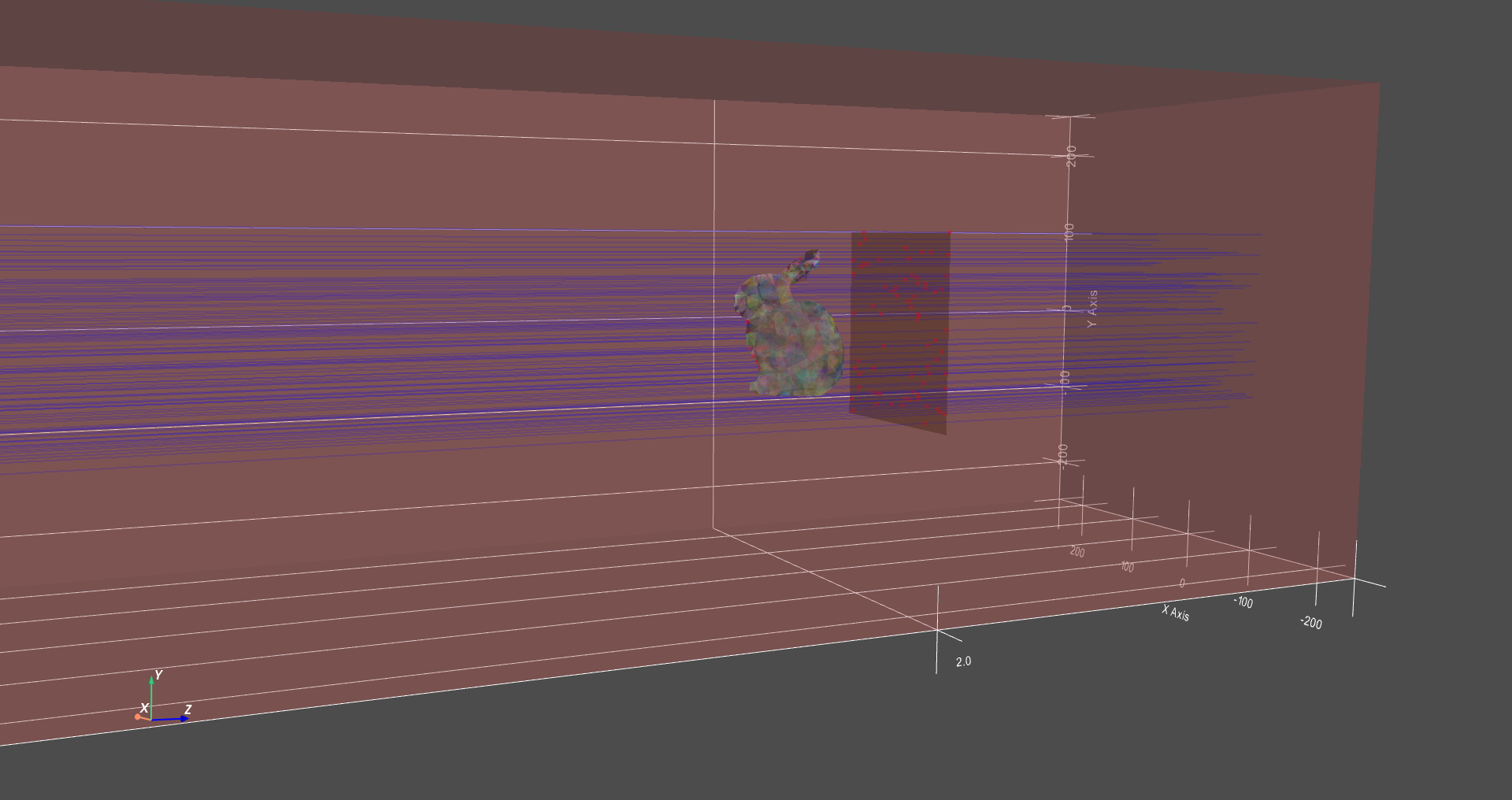}
    \caption{Visualization of simulation mission concerning the neutron imaging of Stanford bunny. $100$ neutrons are shown for clarity}
    \label{fig:BunnySimu}
\end{figure}

A Stanford bunny~\cite{turk1994zippered} is simulated using \texttt{Prompt} in an imaging setup.  
The geometry visualisation is shown in Fig.~\ref{fig:BunnySimu}. 
The bunny is in a surface mesh format originally, and the TetGen~\cite{hang2015tetgen} code is used to generate a tetrahedral mesh of 2441 solids in total.
In the simulation, the bunny is made of vanadium and approximately \SI{15}{\cm} in height. 
The bunny is irradiated in a parallel monochromatic \SI{2}{\angstrom} neutrons beam. The beam size is \SI{0.2}{\meter}$\times$\SI{0.2}{\meter}.
Behind the bunny, an \texttt{ENTRY}  type position detector is set up to measure the transmission pattern. 

\begin{figure}
    \centering
    \begin{subfigure}[b]{0.49\textwidth}
        \centering
        \includegraphics[width=\textwidth]{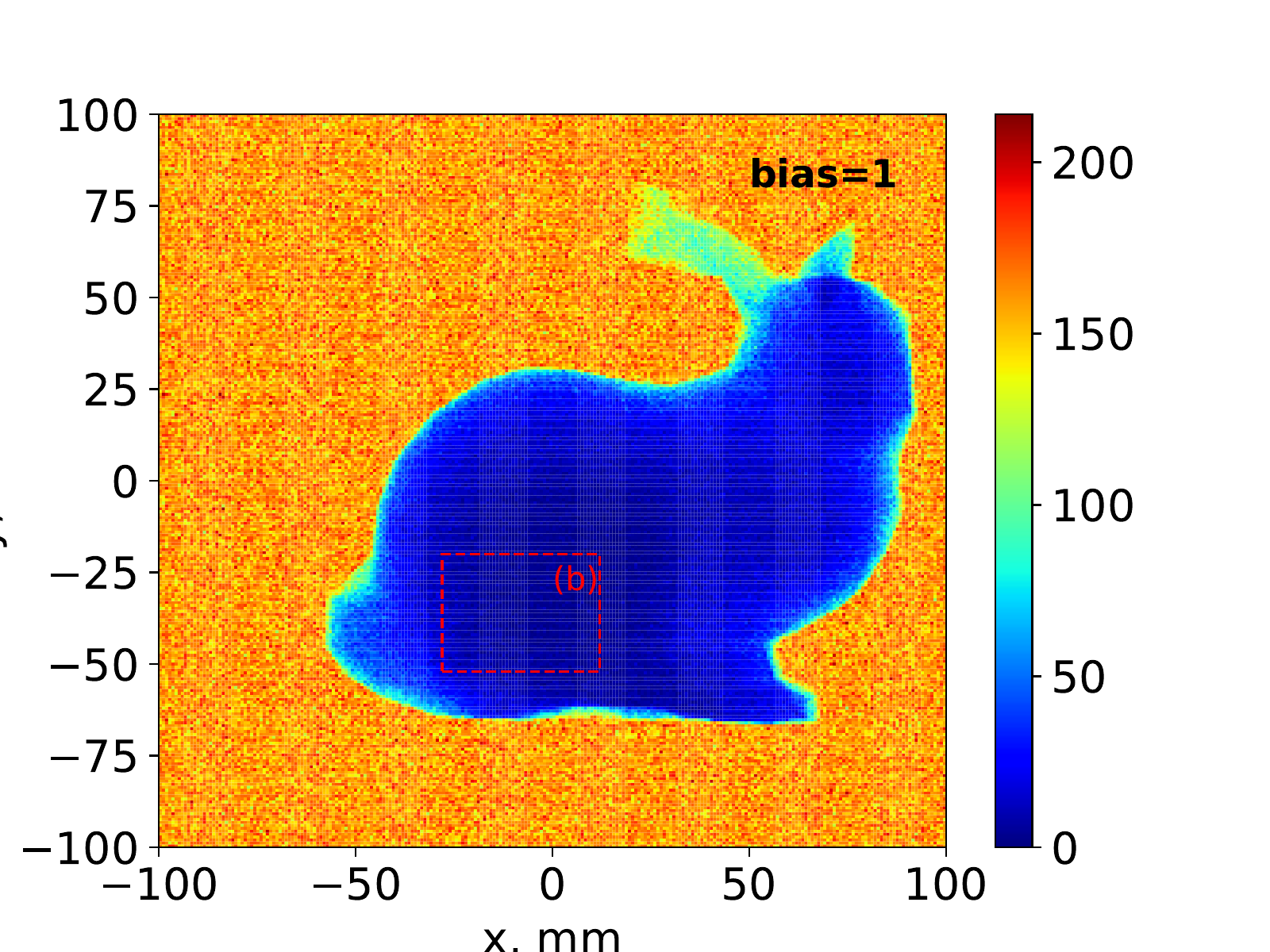}
        \caption{\texttt{PSD} image, absorption bias set to $1.0$, \num{1e7} neutrons}
        \label{fig:sOVAbsBias1}
    \end{subfigure}
    \hfill
    \begin{subfigure}[b]{0.49\textwidth}
        \centering
        \includegraphics[width=\textwidth]{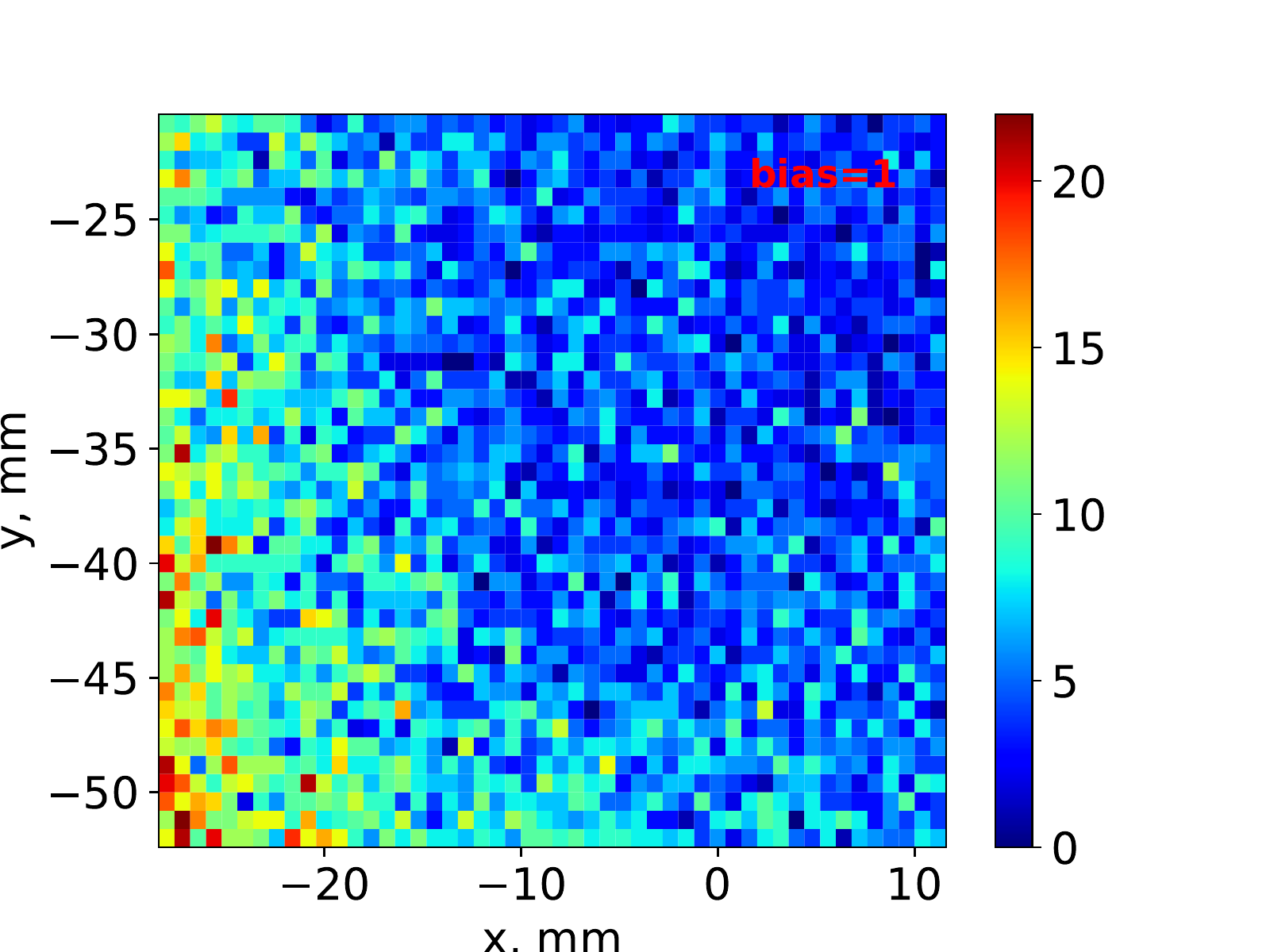}
        \caption{Zoom-in \texttt{PSD} image, absorption bias set to $1.0$, \num{1e7} neutrons}
        \label{fig:sZMAbsBias1}
    \end{subfigure}
    \begin{subfigure}[b]{0.49\textwidth}
        \centering
        \includegraphics[width=\textwidth]{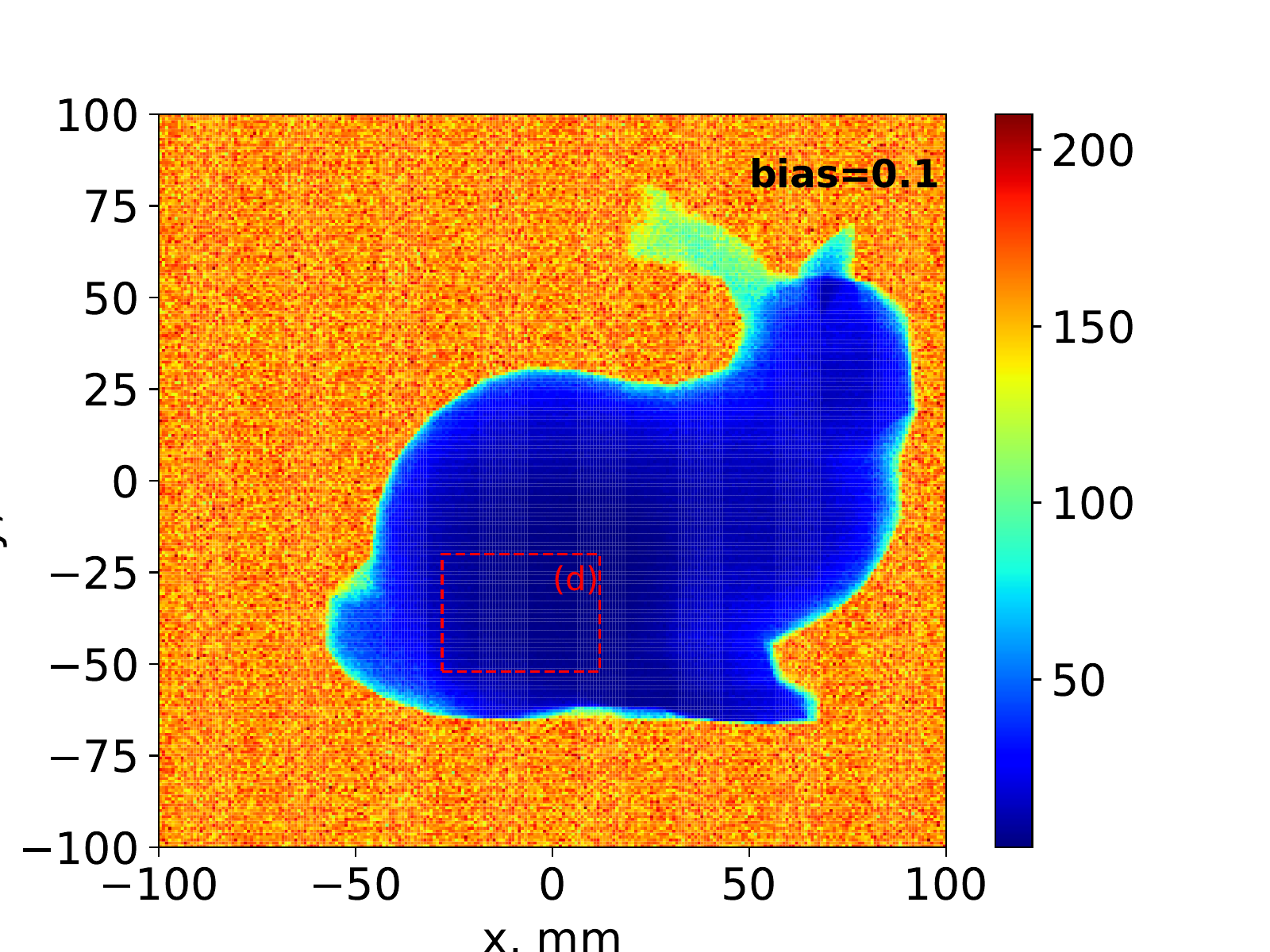}
        \caption{\texttt{PSD} image, absorption bias set to $0.1$, \num{1e7} neutrons}
        \label{fig:sOVAbsBias0p1}
    \end{subfigure}
    \begin{subfigure}[b]{0.49\textwidth}
        \centering
        \includegraphics[width=\textwidth]{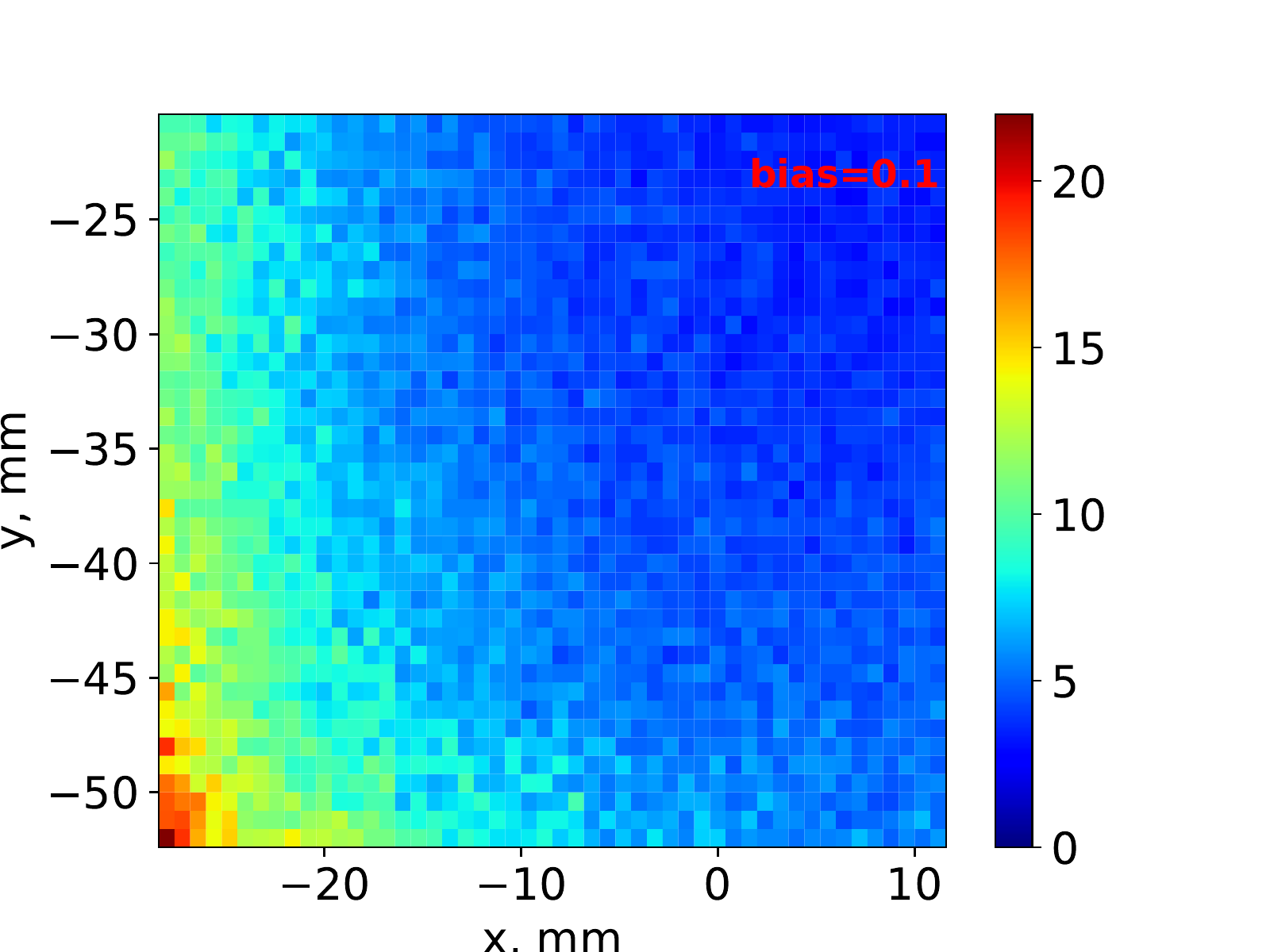}
        \caption{Zoom-in \texttt{PSD} image, absorption bias set to $0.1$, \num{1e7} neutrons}
        \label{fig:sZMAbsBias0p1}
    \end{subfigure}
    \caption{Images recorded from the \texttt{PSD scorer} and their zoom-in images. The absorption biases are set to 1 and 0.1, respectively. The images with 0.1 absorption bias looks 'smoother'}
    \label{fig:sRabbit}
\end{figure}

Two simulations are performed with two different absorption biases, \num{1} and \num{0.1}. The images recorded by the detector are shown in Fig.~\ref{fig:sRabbit}. 
In both Fig.~\ref{fig:sOVAbsBias1} and Fig.~\ref{fig:sOVAbsBias0p1}, the shape of the object can be clearly observed. 
The biased simulation produces visibly identical results as the unbiased case. 

To compare the statistical uncertainties in these two simulations, the areas included in the rectangle of Fig.~\ref{fig:sOVAbsBias1} and Fig.~\ref{fig:sOVAbsBias0p1} are then zoomed-in to Fig.~\ref{fig:sZMAbsBias1} and Fig.~\ref{fig:sZMAbsBias0p1}, respectively. Evidently, there are much less fluctuation for the simulation where the absorption bias is set to \num{0.1} (Fig.~\ref{fig:sOVAbsBias0p1} and Fig.~\ref{fig:sZMAbsBias0p1}) than that with bias \num{1} (Fig.~\ref{fig:sOVAbsBias1} and Fig.~\ref{fig:sZMAbsBias1}), indicating that the cross section biasing mechanism shows its capability in variance reduction in this example.

\subsection{Chopper}
\label{ssChopper}
A simulation example of disk chopper is presented where the results are benchmarked with McStas, to demonstrate the ability of \texttt{Prompt} in neutron optical component simulation.

\begin{figure}
    \centering
    \includegraphics[width=\linewidth]{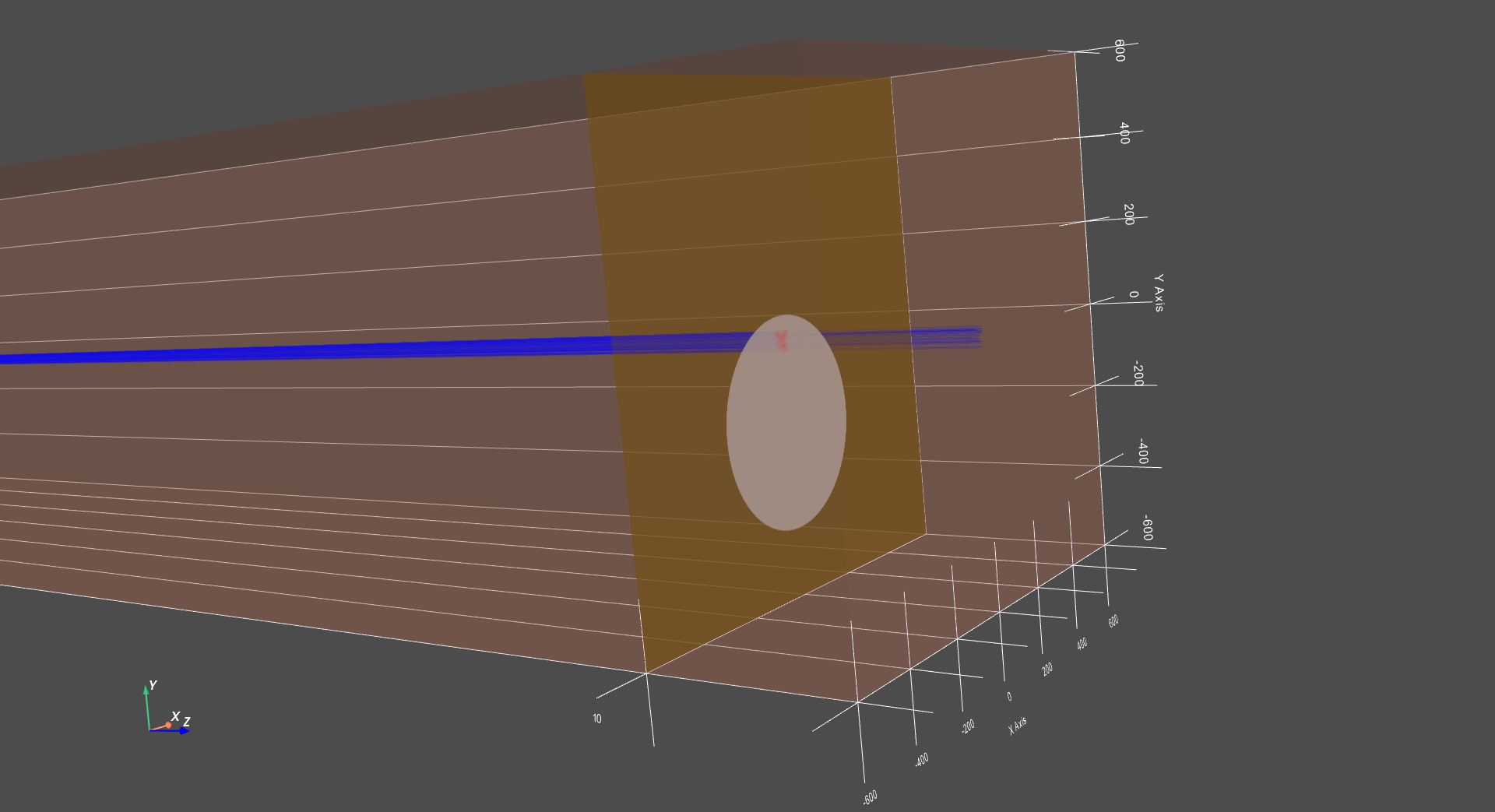}
    \caption{Visualization of \texttt{DiskChopper} simulation in \texttt{Prompt}, with \SI{100}{} neutrons}
    \label{fig:figVizChopper}
\end{figure}

The simulation is visualised in Fig.~\ref{fig:figVizChopper}. A disc-shaped (tube of negligible thickness) chopper with \SI{4}{} axisymetrically arranged \SI{20}{\degree} windows, of which the outer radius and inner radius are respectively \SI{0.25}{\meter} and \SI{0.13}{\meter} and the rotation frequency is \SI{100}{\hertz}, is located at \SI{10}{\meter} away from a \texttt{MaxwellianGun}, which produces \SI{293}{\kelvin} Maxwellian distributed neutrons from a \SI{1}{\mm}$\times$\SI{1}{\mm} opening to another \SI{10}{\meter} away \SI{45}{\mm}$\times$\SI{45}{\mm} opening. 
The identical setup is reproduced in McStas as well to benchmark the implementation in \texttt{Prompt}.
A time-of-flight scorer is positioned just behind and after the chopper to measure the impact of the chopper on the beam. 

\begin{figure}
    \centering
    \includegraphics[width=\linewidth]{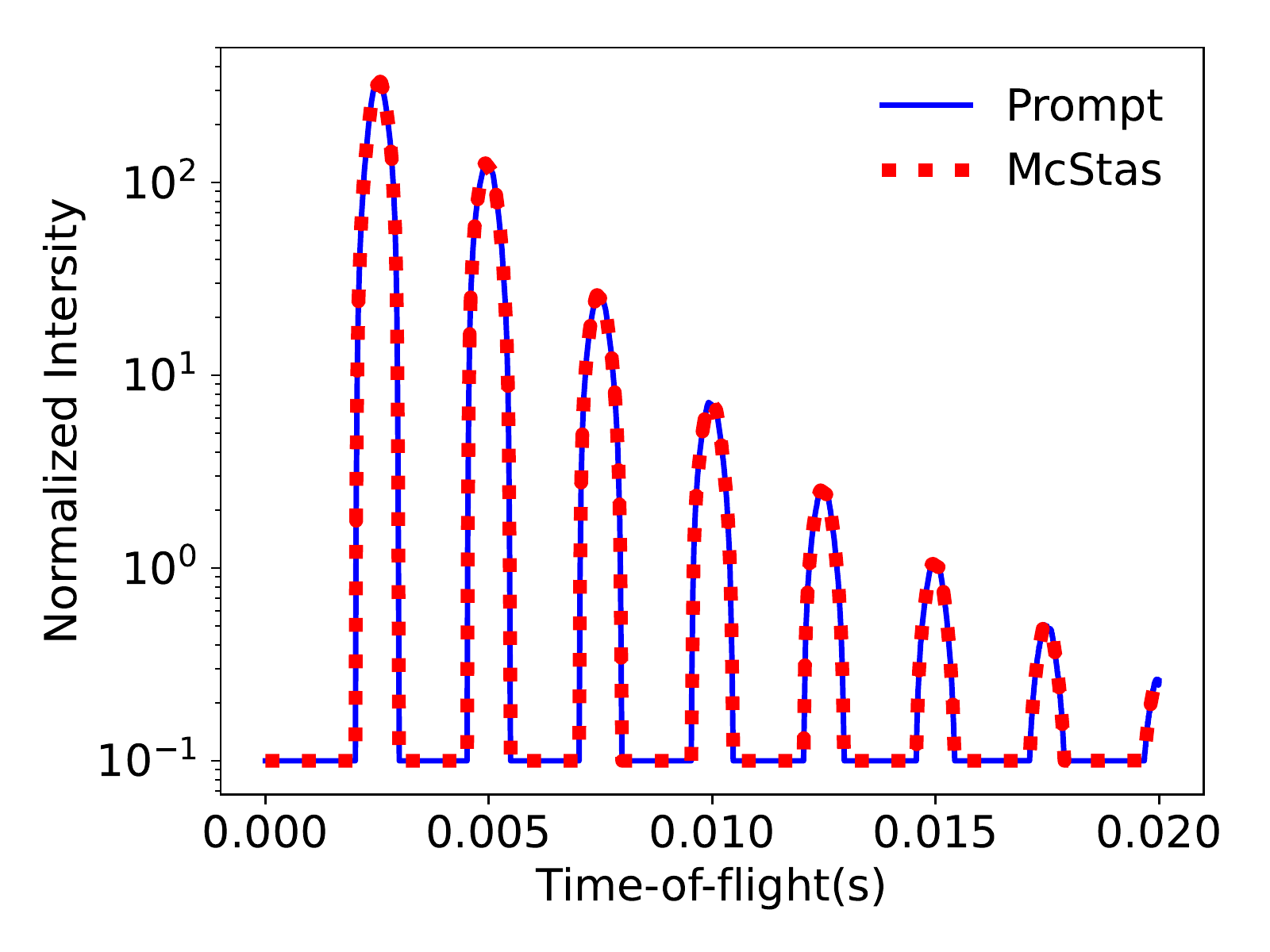}
    \caption{Benchmarking \texttt{DiskChopper} of \texttt{Prompt} with \texttt{Diskchopper} of \texttt{McStas}}
    \label{fig:chopper}
\end{figure}

The time-of-flight spectra are compared in Fig.~\ref{fig:chopper}.
\texttt{Prompt} produced the identical pattern as that from McStas.
Eight peaks are evenly distributed in the time axis with \SI{25}{\ms} spacing. That is consistent with the \SI{400}{\hertz} frequency that the neutron beam sees the opening window.

\section{Outlook}
\label{sConclusion}
A thermal neutron simulation system is presented. As a standalone system, it is expected to be applicable in the areas of neutron instrument characterisation, optimisation as well as data analysis. 
 
This system is the Monte Carlo module of the China Spallation Neutron Source Simulation System (\texttt{Cinema}). Two associated cross section calculation modules, Trajectory Analysis Toolkit(Tak)~\cite{Du2022} and PiXiu, are currently under active development. They are essentially the post-analysis modules for molecular dynamics and density functional theory calculations, respectively.
It is expected that the overall system should provide a full tool-chain to predict data that are directly comparable with instrument detector raw data from sample atomic structures and potentials. 
In parallel with \texttt{Cinema}, a machine learning framework is currently in the conceptual design phase. 
As the simulation tool-chain can describe the instruments and samples as parameters, the initial aim of this framework is for the optimisation of these parameters.
To cope with this framework, \texttt{Prompt} is expected to become purely pythonic at the very top level shortly. It could be implemented as a python \texttt{GDML} manipulator or a direct interface to the \texttt{VecGeom} code.  
In any case, the system will be backward compatibility with the \texttt{GDML} user input introduced in this paper. 

As an open source software project, it is straightforward to expand the system to cover more applications. Therefore, the future development of \texttt{Prompt} depends highly on community interest. 
For instance, a boundary overlap testing logic may be required for simulations with more complex geometries; supporting more particle types and physics models can
enable the designs of moderators and shielding. 
The release of this system opens up many possibilities, feedback and inputs are welcome.

\section*{Acknowledgements}
This research is supported by the National Natural Science Foundation of China (Grant No. 12075266) and the National Key Research and Development Program of China (Grant No. 2022YFA1604100). The authors are grateful for valuable discussions and substantial support.

\bibliographystyle{alpha}
\bibliography{main}

\appendix
\phantomsection\label{\detokenize{./prompt/usersmanual/geometry::doc}}

\setcounter{table}{0}
\setcounter{figure}{0}

\section{Supported solids}
\label{\detokenize{./prompt/usersmanual/geometry:geometry}}

\subsection{Arbitrary\_trapezoid}
\label{\detokenize{./prompt/usersmanual/geometry:arbitrary-trapezoid}}
\sphinxSetupCaptionForVerbatim{An example: definition of an arbitrary\_trapezoid}
\def\sphinxLiteralBlockLabel{\label{\detokenize{./prompt/usersmanual/geometry:id29}}}
\begin{sphinxVerbatim}[commandchars=\\\{\}]
\PYG{n+nt}{\PYGZlt{}arb8} \PYG{n+na}{lunit=}\PYG{l+s}{\PYGZdq{}mm\PYGZdq{}} \PYG{n+na}{name=}\PYG{l+s}{\PYGZdq{}Arb8Solid\PYGZdq{}} \PYG{n+na}{v1x=}\PYG{l+s}{\PYGZdq{}\PYGZhy{}30\PYGZdq{}} \PYG{n+na}{v1y=}\PYG{l+s}{\PYGZdq{}\PYGZhy{}60\PYGZdq{}} \PYG{n+na}{v2x=}\PYG{l+s}{\PYGZdq{}30\PYGZdq{}} \PYG{n+na}{v2y=}\PYG{l+s}{\PYGZdq{}\PYGZhy{}60\PYGZdq{}} \PYG{n+na}{v3x=}\PYG{l+s}{\PYGZdq{}50\PYGZdq{}} \PYG{n+na}{v3y=}\PYG{l+s}{\PYGZdq{}60\PYGZdq{}} \PYG{n+na}{v4x=}\PYG{l+s}{\PYGZdq{}\PYGZhy{}50\PYGZdq{}} \PYG{n+na}{v4y=}\PYG{l+s}{\PYGZdq{}60\PYGZdq{}} \PYG{n+na}{v5x=}\PYG{l+s}{\PYGZdq{}\PYGZhy{}30\PYGZdq{}} \PYG{n+na}{v5y=}\PYG{l+s}{\PYGZdq{}\PYGZhy{}60\PYGZdq{}} \PYG{n+na}{v6x=}\PYG{l+s}{\PYGZdq{}30\PYGZdq{}} \PYG{n+na}{v6y=}\PYG{l+s}{\PYGZdq{}\PYGZhy{}60\PYGZdq{}} \PYG{n+na}{v7x=}\PYG{l+s}{\PYGZdq{}50\PYGZdq{}} \PYG{n+na}{v7y=}\PYG{l+s}{\PYGZdq{}60\PYGZdq{}} \PYG{n+na}{v8x=}\PYG{l+s}{\PYGZdq{}\PYGZhy{}50\PYGZdq{}} \PYG{n+na}{v8y=}\PYG{l+s}{\PYGZdq{}60\PYGZdq{}} \PYG{n+na}{dz=}\PYG{l+s}{\PYGZdq{}70\PYGZdq{}}\PYG{n+nt}{/\PYGZgt{}}
\end{sphinxVerbatim}

\begin{figure}[ht]
    \centering
    \includegraphics[scale=0.15]{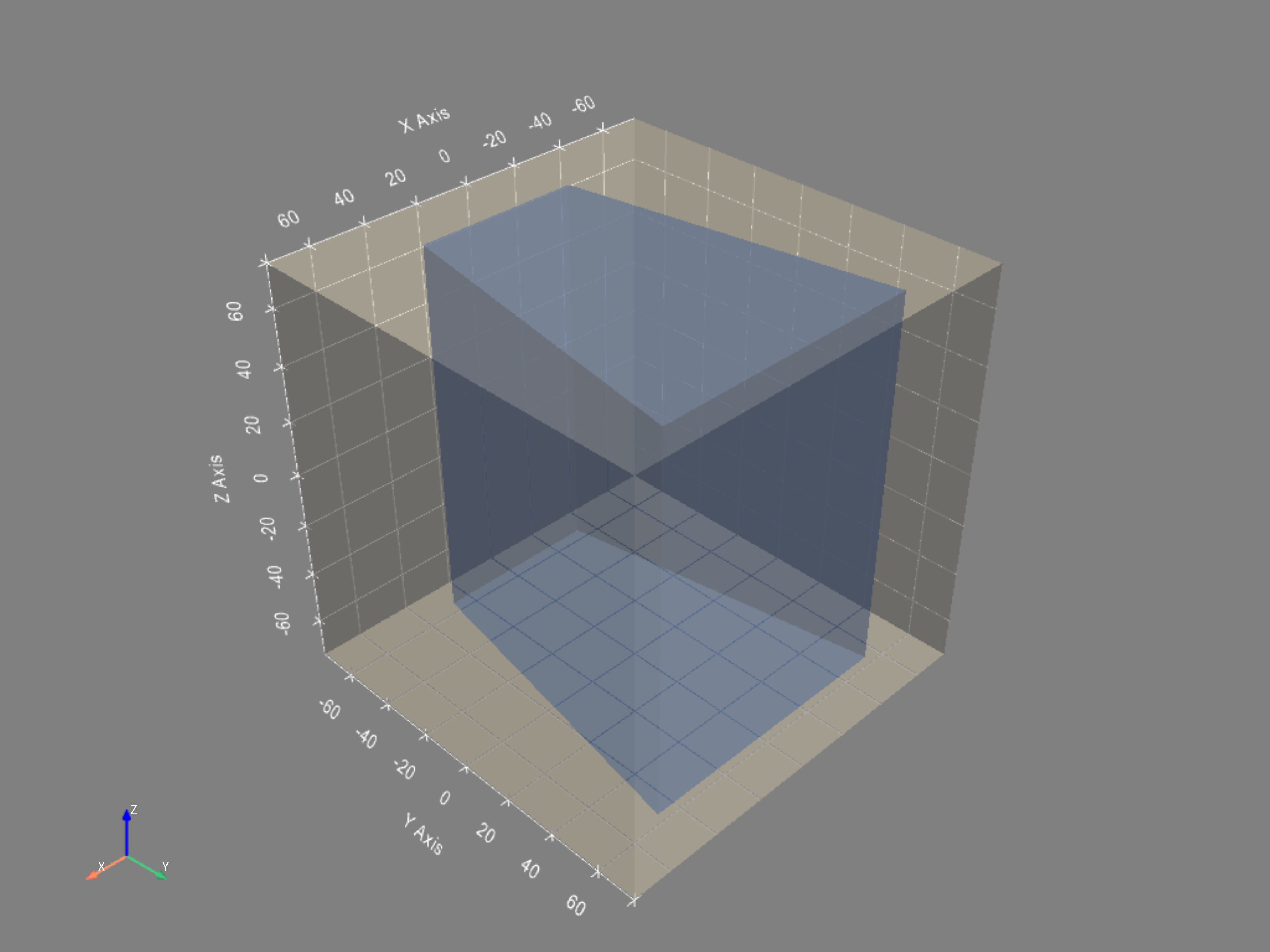}
    \caption{Visualization of an arbitrary\_trapezoid}
    \label{fig:aArbitrary}
\end{figure}

\sphinxattablestart
\sphinxthistablewithglobalstyle
\centering
\sphinxcapstartof{table}
\sphinxthecaptionisattop
\sphinxcaption{Parameters for defining an arbitrary\_trapezoid. If not specified, default length unit is $mm$}\label{\detokenize{./prompt/usersmanual/geometry:id30}}
\sphinxaftertopcaption
\begin{tabulary}{\linewidth}[t]{TT}
\sphinxtoprule
\sphinxstyletheadfamily 
\sphinxAtStartPar
Parameter
&\sphinxstyletheadfamily 
\sphinxAtStartPar
Description
\\
\sphinxmidrule
\sphinxtableatstartofbodyhook
\sphinxAtStartPar
\sphinxcode{\sphinxupquote{v1x}}
&
\sphinxAtStartPar
vertex 1 x position
\\
\sphinxhline
\sphinxAtStartPar
\sphinxcode{\sphinxupquote{v1y}}
&
\sphinxAtStartPar
vertex 1 y position
\\
\sphinxhline
\sphinxAtStartPar
\sphinxcode{\sphinxupquote{v2x}}
&
\sphinxAtStartPar
vertex 2 x position
\\
\sphinxhline
\sphinxAtStartPar
\sphinxcode{\sphinxupquote{v2y}}
&
\sphinxAtStartPar
vertex 2 y position
\\
\sphinxhline
\sphinxAtStartPar
\sphinxcode{\sphinxupquote{v3x}}
&
\sphinxAtStartPar
vertex 3 x position
\\
\sphinxhline
\sphinxAtStartPar
\sphinxcode{\sphinxupquote{v3y}}
&
\sphinxAtStartPar
vertex 3 y position
\\
\sphinxhline
\sphinxAtStartPar
\sphinxcode{\sphinxupquote{v4x}}
&
\sphinxAtStartPar
vertex 4 x position
\\
\sphinxhline
\sphinxAtStartPar
\sphinxcode{\sphinxupquote{v4y}}
&
\sphinxAtStartPar
vertex 4 y position
\\
\sphinxhline
\sphinxAtStartPar
\sphinxcode{\sphinxupquote{v5x}}
&
\sphinxAtStartPar
vertex 5 x position
\\
\sphinxhline
\sphinxAtStartPar
\sphinxcode{\sphinxupquote{v5y}}
&
\sphinxAtStartPar
vertex 5 y position
\\
\sphinxhline
\sphinxAtStartPar
\sphinxcode{\sphinxupquote{v6x}}
&
\sphinxAtStartPar
vertex 6 x position
\\
\sphinxhline
\sphinxAtStartPar
\sphinxcode{\sphinxupquote{v6y}}
&
\sphinxAtStartPar
vertex 6 y position
\\
\sphinxhline
\sphinxAtStartPar
\sphinxcode{\sphinxupquote{v7x}}
&
\sphinxAtStartPar
vertex 7 x position
\\
\sphinxhline
\sphinxAtStartPar
\sphinxcode{\sphinxupquote{v7y}}
&
\sphinxAtStartPar
vertex 7 y position
\\
\sphinxhline
\sphinxAtStartPar
\sphinxcode{\sphinxupquote{v8x}}
&
\sphinxAtStartPar
vertex 8 x position
\\
\sphinxhline
\sphinxAtStartPar
\sphinxcode{\sphinxupquote{v8y}}
&
\sphinxAtStartPar
vertex 8 y position
\\
\sphinxhline
\sphinxAtStartPar
\sphinxcode{\sphinxupquote{dz}}
&
\sphinxAtStartPar
half z length
\\
\sphinxbottomrule
\end{tabulary}
\sphinxtableafterendhook\par
\sphinxattableend

\subsection{Box}
\label{\detokenize{./prompt/usersmanual/geometry:box}}

\sphinxSetupCaptionForVerbatim{An example: definition of a box. If not specified, default length unit is $mm$}
\def\sphinxLiteralBlockLabel{\label{\detokenize{./prompt/usersmanual/geometry:id31}}}
\begin{sphinxVerbatim}[commandchars=\\\{\}]
\PYG{n+nt}{\PYGZlt{}box} \PYG{n+na}{lunit=}\PYG{l+s}{\PYGZdq{}mm\PYGZdq{}} \PYG{n+na}{name=}\PYG{l+s}{\PYGZdq{}BoxSolid\PYGZdq{}} \PYG{n+na}{x=}\PYG{l+s}{\PYGZdq{}100.0\PYGZdq{}} \PYG{n+na}{y=}\PYG{l+s}{\PYGZdq{}100.0\PYGZdq{}} \PYG{n+na}{z=}\PYG{l+s}{\PYGZdq{}100.0\PYGZdq{}}\PYG{n+nt}{/\PYGZgt{}}
\end{sphinxVerbatim}

\begin{figure}[ht]
    \centering
    \includegraphics[scale=0.15]{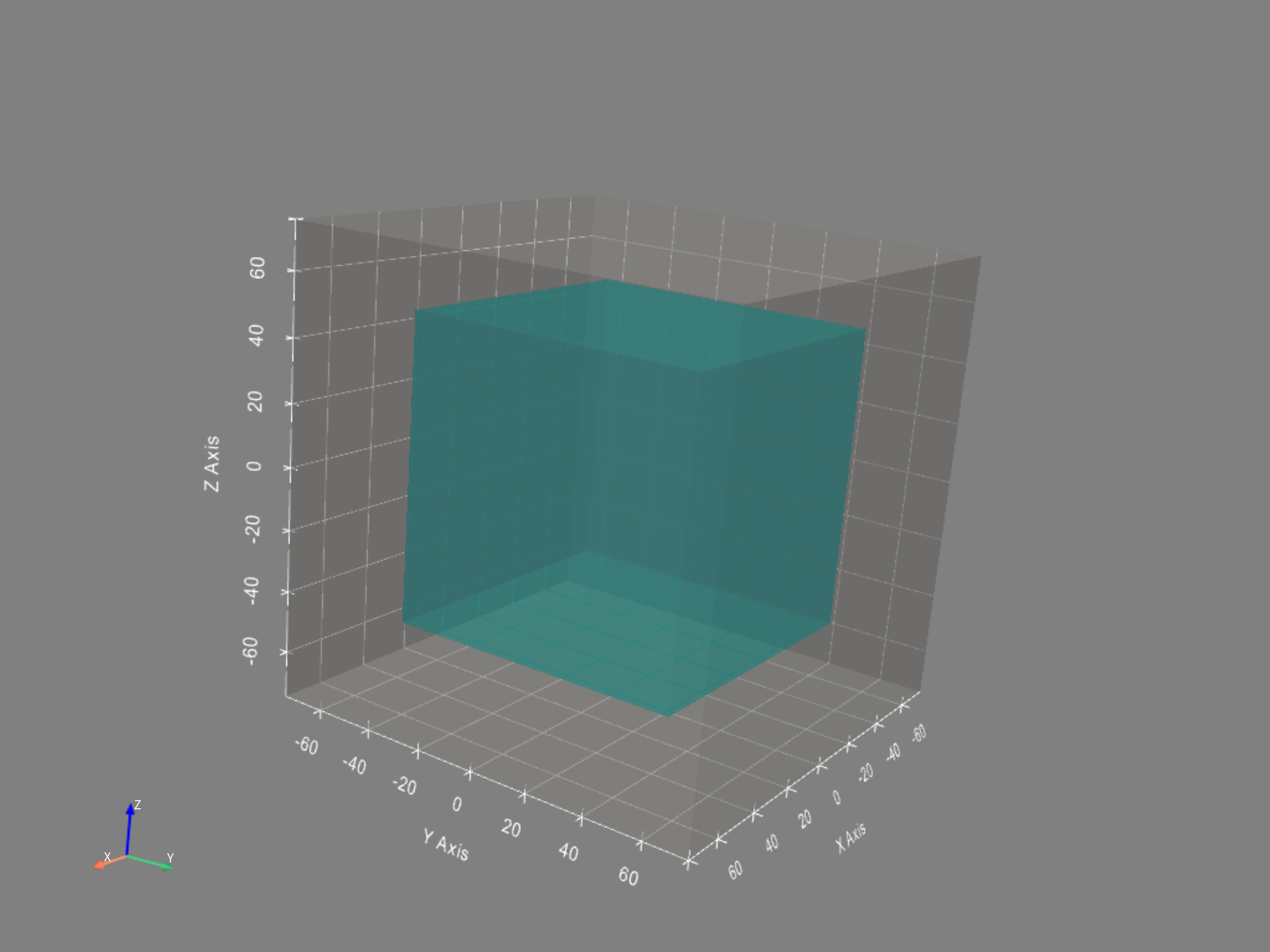}
    \caption{Visualization of a box}
    \label{fig:aBox}
\end{figure}

\sphinxattablestart
\sphinxthistablewithglobalstyle
\centering
\sphinxcapstartof{table}
\sphinxthecaptionisattop
\sphinxcaption{Parameters for defining a box. If not specified, default length unit is $mm$}\label{\detokenize{./prompt/usersmanual/geometry:id32}}
\sphinxaftertopcaption
\begin{tabulary}{\linewidth}[t]{TT}
\sphinxtoprule
\sphinxstyletheadfamily 
\sphinxAtStartPar
Parameter
&\sphinxstyletheadfamily 
\sphinxAtStartPar
Description
\\
\sphinxmidrule
\sphinxtableatstartofbodyhook
\sphinxAtStartPar
\sphinxcode{\sphinxupquote{x}}
&
\sphinxAtStartPar
half length in x direction
\\
\sphinxhline
\sphinxAtStartPar
\sphinxcode{\sphinxupquote{y}}
&
\sphinxAtStartPar
half length in y direction
\\
\sphinxhline
\sphinxAtStartPar
\sphinxcode{\sphinxupquote{z}}
&
\sphinxAtStartPar
half length in z direction
\\
\sphinxbottomrule
\end{tabulary}
\sphinxtableafterendhook\par
\sphinxattableend

\subsection{Cone}
\label{\detokenize{./prompt/usersmanual/geometry:cone}}

\sphinxSetupCaptionForVerbatim{An example: definition of a cone}
\def\sphinxLiteralBlockLabel{\label{\detokenize{./prompt/usersmanual/geometry:id33}}}
\begin{sphinxVerbatim}[commandchars=\\\{\}]
\PYG{n+nt}{\PYGZlt{}cone} \PYG{n+na}{aunit=}\PYG{l+s}{\PYGZdq{}deg\PYGZdq{}} \PYG{n+na}{lunit=}\PYG{l+s}{\PYGZdq{}mm\PYGZdq{}} \PYG{n+na}{name=}\PYG{l+s}{\PYGZdq{}ConeSolid\PYGZdq{}} \PYG{n+na}{rmin1=}\PYG{l+s}{\PYGZdq{}0.0\PYGZdq{}} \PYG{n+na}{rmax1=}\PYG{l+s}{\PYGZdq{}50.0\PYGZdq{}} \PYG{n+na}{rmin2=}\PYG{l+s}{\PYGZdq{}0.0\PYGZdq{}} \PYG{n+na}{rmax2=}\PYG{l+s}{\PYGZdq{}10.0\PYGZdq{}} \PYG{n+na}{z=}\PYG{l+s}{\PYGZdq{}120.0\PYGZdq{}} \PYG{n+na}{deltaphi=}\PYG{l+s}{\PYGZdq{}360.0\PYGZdq{}} \PYG{n+na}{startphi=}\PYG{l+s}{\PYGZdq{}0.0\PYGZdq{}} \PYG{n+nt}{/\PYGZgt{}}
\end{sphinxVerbatim}

\begin{figure}[ht]
    \centering
    \includegraphics[scale=0.15]{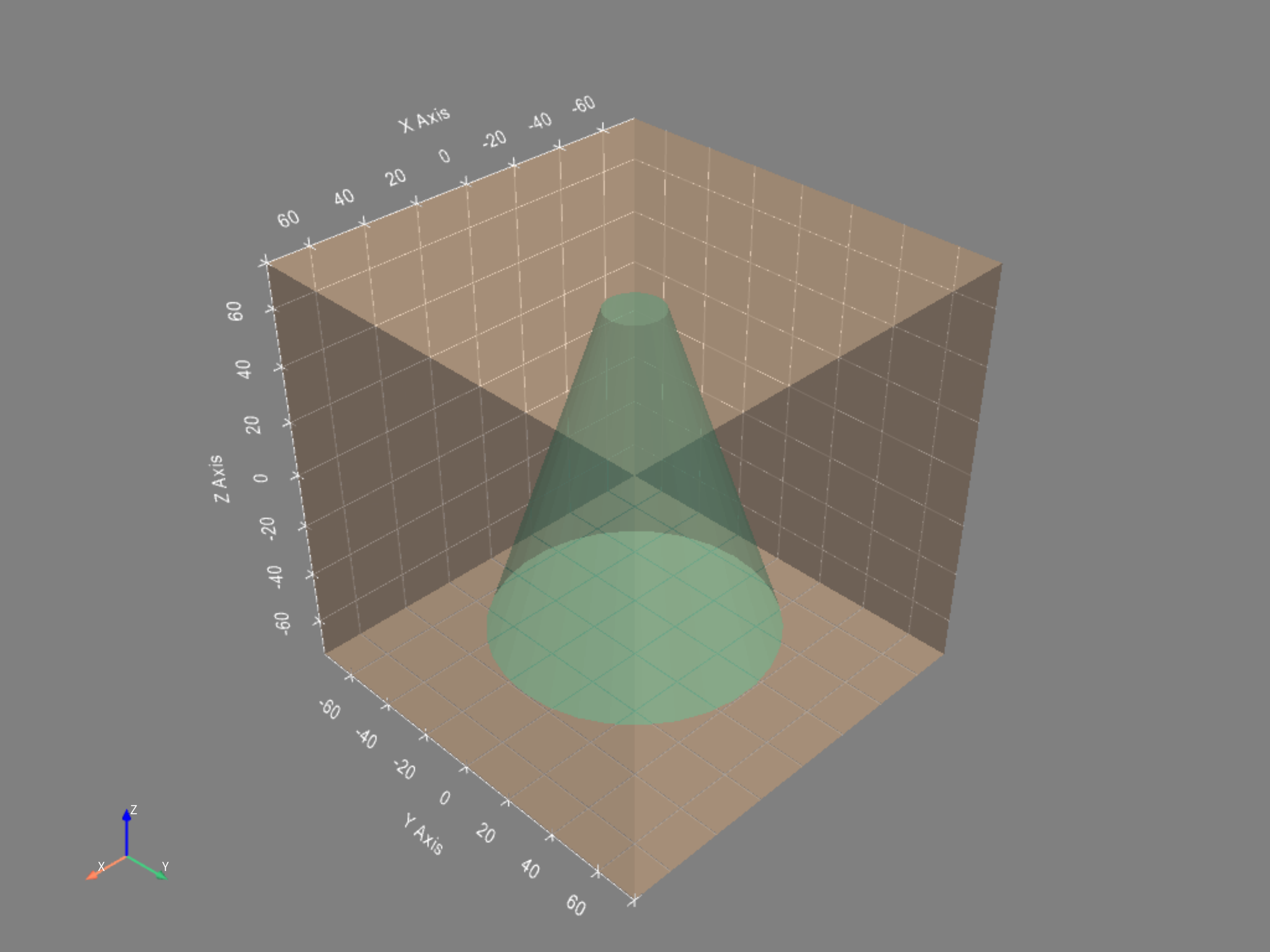}
    \caption{Visualization of a cone}
    \label{fig:aCone}
\end{figure}

\sphinxattablestart
\sphinxthistablewithglobalstyle
\centering
\sphinxcapstartof{table}
\sphinxthecaptionisattop
\sphinxcaption{Parameters for defining a cone. If not specified, default length unit is $mm$, default angle unit is $deg$}\label{\detokenize{./prompt/usersmanual/geometry:id34}}
\sphinxaftertopcaption
\begin{tabulary}{\linewidth}[t]{TT}
\sphinxtoprule
\sphinxstyletheadfamily 
\sphinxAtStartPar
Parameter
&\sphinxstyletheadfamily 
\sphinxAtStartPar
Description
\\
\sphinxmidrule
\sphinxtableatstartofbodyhook
\sphinxAtStartPar
\sphinxcode{\sphinxupquote{rmin1}}
&
\sphinxAtStartPar
inner radius at base of cone
\\
\sphinxhline
\sphinxAtStartPar
\sphinxcode{\sphinxupquote{rmax1}}
&
\sphinxAtStartPar
outer radius at base of cone
\\
\sphinxhline
\sphinxAtStartPar
\sphinxcode{\sphinxupquote{rmin2}}
&
\sphinxAtStartPar
inner radius at top of cone
\\
\sphinxhline
\sphinxAtStartPar
\sphinxcode{\sphinxupquote{rmax2}}
&
\sphinxAtStartPar
outer radius at top of cone
\\
\sphinxhline
\sphinxAtStartPar
\sphinxcode{\sphinxupquote{z}}
&
\sphinxAtStartPar
height of cone segment
\\
\sphinxhline
\sphinxAtStartPar
\sphinxcode{\sphinxupquote{startphi}}
&
\sphinxAtStartPar
start angle of the segment
\\
\sphinxhline
\sphinxAtStartPar
\sphinxcode{\sphinxupquote{deltaphi}}
&
\sphinxAtStartPar
angle of the segment
\\
\sphinxbottomrule
\end{tabulary}
\sphinxtableafterendhook\par
\sphinxattableend

\subsection{CutTube}
\label{\detokenize{./prompt/usersmanual/geometry:cuttube}}

\sphinxSetupCaptionForVerbatim{An example: definition of a cutTube}
\def\sphinxLiteralBlockLabel{\label{\detokenize{./prompt/usersmanual/geometry:id35}}}
\begin{sphinxVerbatim}[commandchars=\\\{\}]
\PYG{n+nt}{\PYGZlt{}cutTube} \PYG{n+na}{aunit=}\PYG{l+s}{\PYGZdq{}deg\PYGZdq{}} \PYG{n+na}{lunit=}\PYG{l+s}{\PYGZdq{}mm\PYGZdq{}} \PYG{n+na}{name=}\PYG{l+s}{\PYGZdq{}CutTubeSolid\PYGZdq{}} \PYG{n+na}{rmin=}\PYG{l+s}{\PYGZdq{}10.0\PYGZdq{}} \PYG{n+na}{rmax=}\PYG{l+s}{\PYGZdq{}25.0\PYGZdq{}} \PYG{n+na}{z=}\PYG{l+s}{\PYGZdq{}50.0\PYGZdq{}} \PYG{n+na}{deltaphi=}\PYG{l+s}{\PYGZdq{}360.0\PYGZdq{}} \PYG{n+na}{startphi=}\PYG{l+s}{\PYGZdq{}0.0\PYGZdq{}} \PYG{n+na}{lowX=}\PYG{l+s}{\PYGZdq{}\PYGZhy{}5\PYGZdq{}} \PYG{n+na}{lowY=}\PYG{l+s}{\PYGZdq{}\PYGZhy{}5\PYGZdq{}} \PYG{n+na}{lowZ=}\PYG{l+s}{\PYGZdq{}\PYGZhy{}5\PYGZdq{}} \PYG{n+na}{highX=}\PYG{l+s}{\PYGZdq{}5\PYGZdq{}} \PYG{n+na}{highY=}\PYG{l+s}{\PYGZdq{}5\PYGZdq{}} \PYG{n+na}{highZ=}\PYG{l+s}{\PYGZdq{}5\PYGZdq{}}\PYG{n+nt}{/\PYGZgt{}}
\end{sphinxVerbatim}

\begin{figure}[ht]
    \centering
    \includegraphics[scale=0.15]{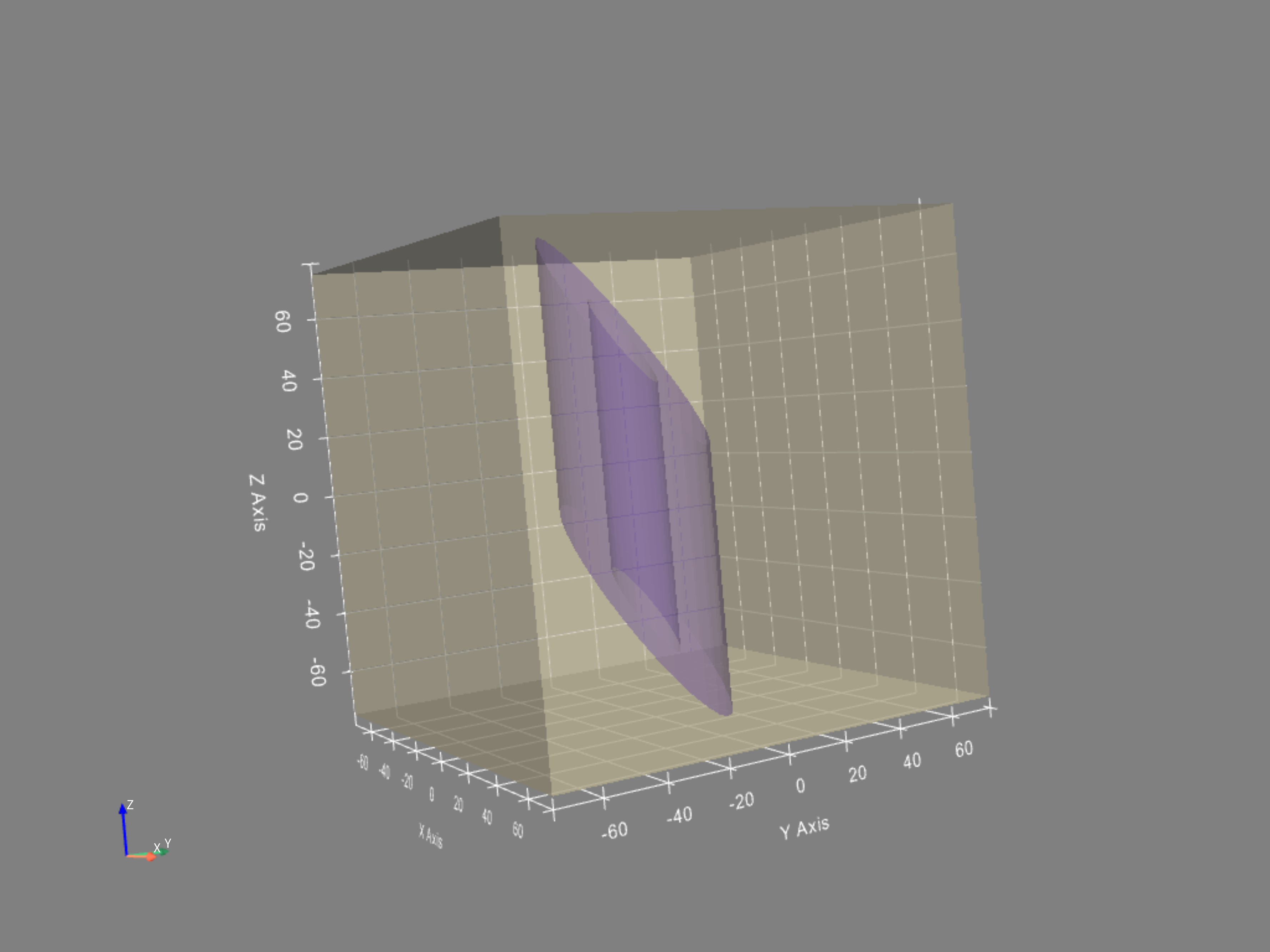}
    \caption{Visualization of a cutTube}
    \label{fig:aCutTube}
\end{figure}

\sphinxattablestart
\sphinxthistablewithglobalstyle
\centering
\sphinxcapstartof{table}
\sphinxthecaptionisattop
\sphinxcaption{Parameters for defining a cutTube. If not specified, default length unit is $mm$, default angle unit is $deg$}\label{\detokenize{./prompt/usersmanual/geometry:id8}}
\sphinxaftertopcaption
\begin{tabulary}{\linewidth}[t]{TTT}
\sphinxtoprule
\sphinxstyletheadfamily 
\sphinxAtStartPar
Parameter
&\sphinxstyletheadfamily 
\sphinxAtStartPar
Default
&\sphinxstyletheadfamily 
\sphinxAtStartPar
Description
\\
\sphinxmidrule
\sphinxtableatstartofbodyhook
\sphinxAtStartPar
\sphinxcode{\sphinxupquote{z}}
&&
\sphinxAtStartPar
length along z axis
\\
\sphinxhline
\sphinxAtStartPar
\sphinxcode{\sphinxupquote{rmin}}
&
\sphinxAtStartPar
0.0
&
\sphinxAtStartPar
inner radius
\\
\sphinxhline
\sphinxAtStartPar
\sphinxcode{\sphinxupquote{rmax}}
&&
\sphinxAtStartPar
outer radius
\\
\sphinxhline
\sphinxAtStartPar
\sphinxcode{\sphinxupquote{startphi}}
&
\sphinxAtStartPar
0.0
&
\sphinxAtStartPar
starting phi angle of segment
\\
\sphinxhline
\sphinxAtStartPar
\sphinxcode{\sphinxupquote{deltaphi}}
&&
\sphinxAtStartPar
delta phi of angle
\\
\sphinxhline
\sphinxAtStartPar
\sphinxcode{\sphinxupquote{lowX}}
&&
\sphinxAtStartPar
normal at lower z plane
\\
\sphinxhline
\sphinxAtStartPar
\sphinxcode{\sphinxupquote{lowY}}
&&
\sphinxAtStartPar
normal at lower z plane
\\
\sphinxhline
\sphinxAtStartPar
\sphinxcode{\sphinxupquote{lowZ}}
&&
\sphinxAtStartPar
normal at lower z plane
\\
\sphinxhline
\sphinxAtStartPar
\sphinxcode{\sphinxupquote{highX}}
&&
\sphinxAtStartPar
normal at upper z plane
\\
\sphinxhline
\sphinxAtStartPar
\sphinxcode{\sphinxupquote{highY}}
&&
\sphinxAtStartPar
normal at upper z plane
\\
\sphinxhline
\sphinxAtStartPar
\sphinxcode{\sphinxupquote{highZ}}
&&
\sphinxAtStartPar
normal at upper z plane
\\
\sphinxbottomrule
\end{tabulary}
\sphinxtableafterendhook\par
\sphinxattableend

\subsection{General\_trapezoid}
\label{\detokenize{./prompt/usersmanual/geometry:general-trapezoid}}

\sphinxSetupCaptionForVerbatim{An example: definition of a general\_trapezoid}
\def\sphinxLiteralBlockLabel{\label{\detokenize{./prompt/usersmanual/geometry:id39}}}
\begin{sphinxVerbatim}[commandchars=\\\{\}]
\PYG{n+nt}{\PYGZlt{}trap} \PYG{n+na}{lunit=}\PYG{l+s}{\PYGZdq{}mm\PYGZdq{}} \PYG{n+na}{name=}\PYG{l+s}{\PYGZdq{}TrapSolid\PYGZdq{}} \PYG{n+na}{z=}\PYG{l+s}{\PYGZdq{}130\PYGZdq{}} \PYG{n+na}{thata=}\PYG{l+s}{\PYGZdq{}45\PYGZdq{}} \PYG{n+na}{phi=}\PYG{l+s}{\PYGZdq{}45\PYGZdq{}} \PYG{n+na}{y1=}\PYG{l+s}{\PYGZdq{}60\PYGZdq{}} \PYG{n+na}{x1=}\PYG{l+s}{\PYGZdq{}40\PYGZdq{}} \PYG{n+na}{x2=}\PYG{l+s}{\PYGZdq{}40\PYGZdq{}} \PYG{n+na}{alpha1=}\PYG{l+s}{\PYGZdq{}45\PYGZdq{}} \PYG{n+na}{y2=}\PYG{l+s}{\PYGZdq{}60\PYGZdq{}} \PYG{n+na}{x3=}\PYG{l+s}{\PYGZdq{}40\PYGZdq{}} \PYG{n+na}{x4=}\PYG{l+s}{\PYGZdq{}40\PYGZdq{}} \PYG{n+na}{alpha2=}\PYG{l+s}{\PYGZdq{}45\PYGZdq{}}\PYG{n+nt}{/\PYGZgt{}}
\end{sphinxVerbatim}

\begin{figure}[ht]
    \centering
    \includegraphics[scale=0.15]{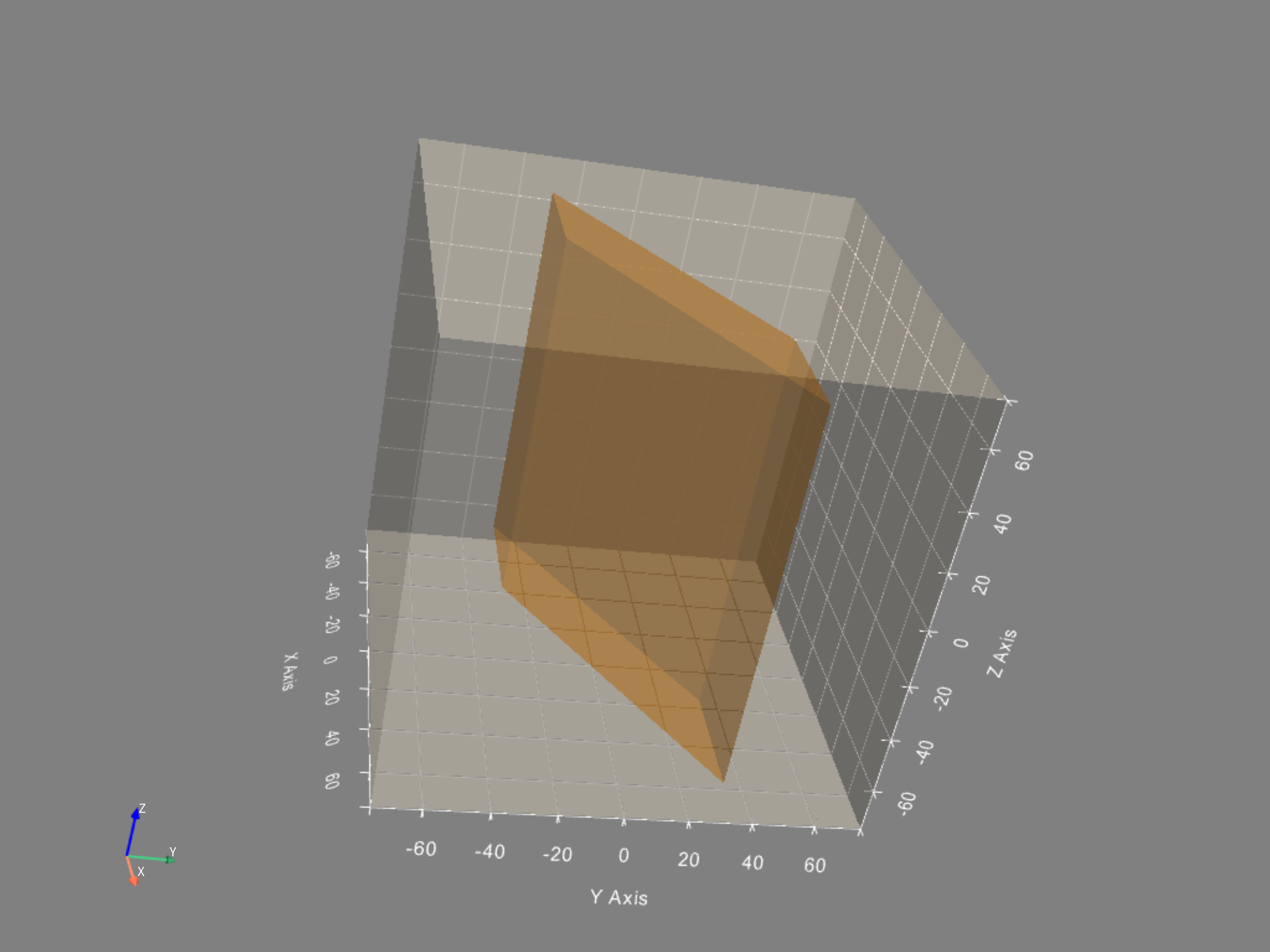}
    \caption{Visualization of a general\_trapezoid}
    \label{fig:aGeneral}
\end{figure}

\sphinxattablestart
\sphinxthistablewithglobalstyle
\centering
\sphinxcapstartof{table}
\sphinxthecaptionisattop
\sphinxcaption{Parameters for defining a general\_trapezoid. If not specified, default length unit is $mm$}\label{\detokenize{./prompt/usersmanual/geometry:id40}}
\sphinxaftertopcaption
\begin{tabular}[t]{\X{20}{100}\X{80}{100}}
\sphinxtoprule
\sphinxstyletheadfamily 
\sphinxAtStartPar
Parameter
&\sphinxstyletheadfamily 
\sphinxAtStartPar
Description
\\
\sphinxmidrule
\sphinxtableatstartofbodyhook
\sphinxAtStartPar
\sphinxcode{\sphinxupquote{z}}
&
\sphinxAtStartPar
length along z axis
\\
\sphinxhline
\sphinxAtStartPar
\sphinxcode{\sphinxupquote{theta}}
&
\sphinxAtStartPar
polar angle to faces joining at \sphinxhyphen{}/+z
\\
\sphinxhline
\sphinxAtStartPar
\sphinxcode{\sphinxupquote{phi}}
&
\sphinxAtStartPar
azimuthal angle of line
joining centre of \textendash{}z face to centre of +z face
\\
\sphinxhline
\sphinxAtStartPar
\sphinxcode{\sphinxupquote{y1}}
&
\sphinxAtStartPar
length along y at the face \sphinxhyphen{}z
\\
\sphinxhline
\sphinxAtStartPar
\sphinxcode{\sphinxupquote{x1}}
&
\sphinxAtStartPar
length along x at side y = \sphinxhyphen{}y1 of the face at \sphinxhyphen{}z
\\
\sphinxhline
\sphinxAtStartPar
\sphinxcode{\sphinxupquote{x2}}
&
\sphinxAtStartPar
length along x at side y = +y1 of the face at \sphinxhyphen{}z
\\
\sphinxhline
\sphinxAtStartPar
\sphinxcode{\sphinxupquote{alpha1}}
&
\sphinxAtStartPar
angle with respect to the y axis
from the centre of side at y = \sphinxhyphen{}y1
to centre of y = +y1 of the face at \sphinxhyphen{}z
\\
\sphinxhline
\sphinxAtStartPar
\sphinxcode{\sphinxupquote{y2}}
&
\sphinxAtStartPar
length along y at the face +z
\\
\sphinxhline
\sphinxAtStartPar
\sphinxcode{\sphinxupquote{x3}}
&
\sphinxAtStartPar
length along x at side y = \sphinxhyphen{}y1 of the face at +z
\\
\sphinxhline
\sphinxAtStartPar
\sphinxcode{\sphinxupquote{x4}}
&
\sphinxAtStartPar
length along x at side y = +y1 of the face at +z
\\
\sphinxhline
\sphinxAtStartPar
\sphinxcode{\sphinxupquote{alpha2}}
&
\sphinxAtStartPar
angle with respect to the y axis
from the centre of side at y = \sphinxhyphen{}y2
to centre of y = +y2 of the face at +z
\\
\sphinxbottomrule
\end{tabular}
\sphinxtableafterendhook\par
\sphinxattableend

\subsection{Hyperbolic\_tube}
\label{\detokenize{./prompt/usersmanual/geometry:hyperbolic-tube}}

\sphinxSetupCaptionForVerbatim{An example: definition of a hyperbolic\_tube}
\def\sphinxLiteralBlockLabel{\label{\detokenize{./prompt/usersmanual/geometry:id41}}}
\begin{sphinxVerbatim}[commandchars=\\\{\}]
\PYG{n+nt}{\PYGZlt{}hype} \PYG{n+na}{lunit=}\PYG{l+s}{\PYGZdq{}mm\PYGZdq{}} \PYG{n+na}{name=}\PYG{l+s}{\PYGZdq{}HypeSolid\PYGZdq{}} \PYG{n+na}{rmin=}\PYG{l+s}{\PYGZdq{}10\PYGZdq{}} \PYG{n+na}{rmax=}\PYG{l+s}{\PYGZdq{}20\PYGZdq{}} \PYG{n+na}{z=}\PYG{l+s}{\PYGZdq{}100\PYGZdq{}} \PYG{n+na}{inst=}\PYG{l+s}{\PYGZdq{}0.785\PYGZdq{}} \PYG{n+na}{outst=}\PYG{l+s}{\PYGZdq{}0.785\PYGZdq{}}\PYG{n+nt}{/\PYGZgt{}}
\end{sphinxVerbatim}

\begin{figure}[ht]
    \centering
    \includegraphics[scale=0.15]{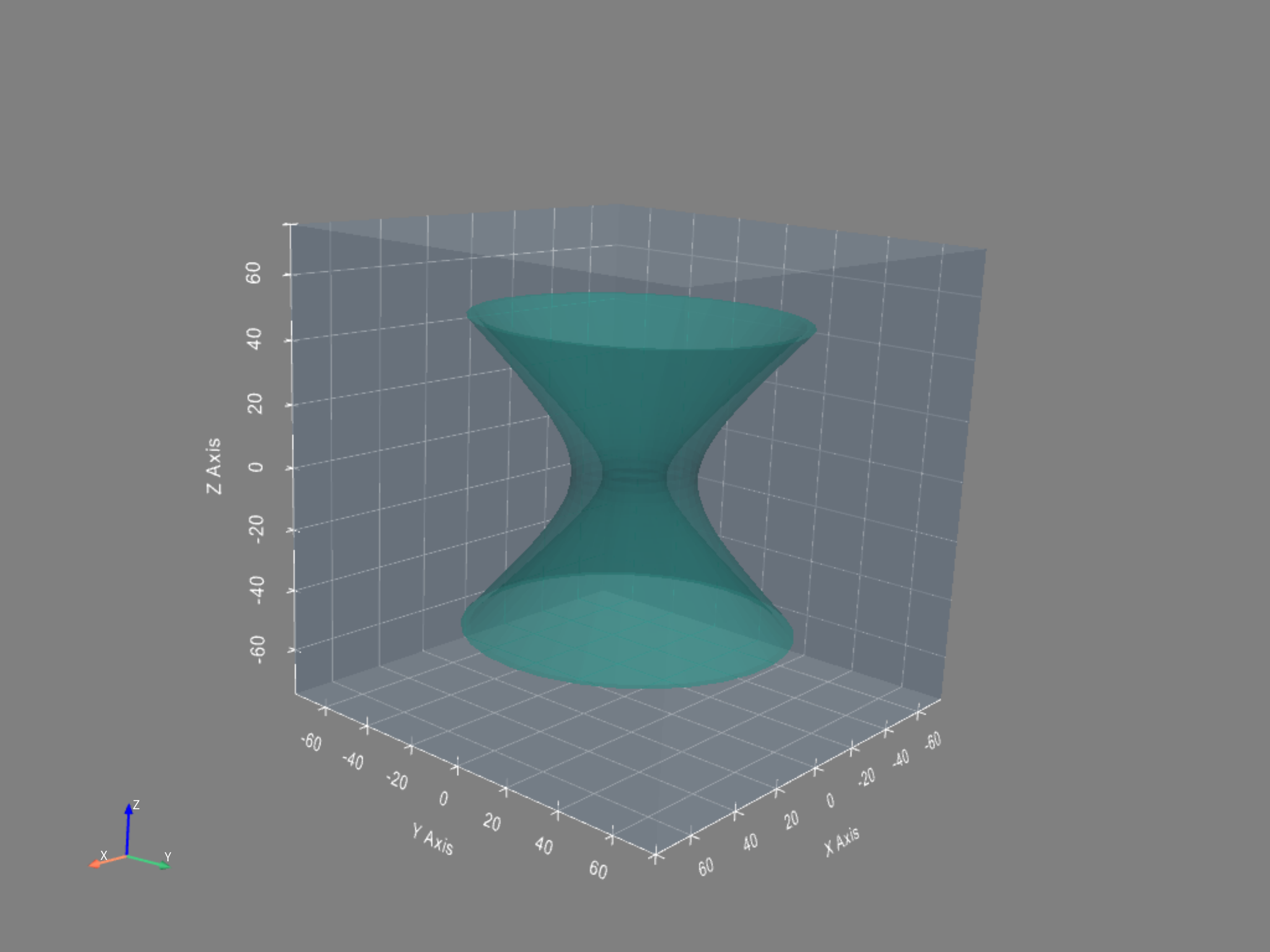}
    \caption{Visualization of a hyperbolic\_tube}
    \label{fig:aHyperbolic}
\end{figure}

\sphinxattablestart
\sphinxthistablewithglobalstyle
\centering
\sphinxcapstartof{table}
\sphinxthecaptionisattop
\sphinxcaption{Parameters for defining a hyperbolic\_tube. If not specified, default length unit is $mm$. Stereo angle unit can only be $rad$}\label{\detokenize{./prompt/usersmanual/geometry:id42}}
\sphinxaftertopcaption
\begin{tabulary}{\linewidth}[t]{TT}
\sphinxtoprule
\sphinxstyletheadfamily 
\sphinxAtStartPar
Parameter
&\sphinxstyletheadfamily 
\sphinxAtStartPar
Description
\\
\sphinxmidrule
\sphinxtableatstartofbodyhook
\sphinxAtStartPar
\sphinxcode{\sphinxupquote{rmin}}
&
\sphinxAtStartPar
inside radius of tube
\\
\sphinxhline
\sphinxAtStartPar
\sphinxcode{\sphinxupquote{rmax}}
&
\sphinxAtStartPar
outside radius of tube
\\
\sphinxhline
\sphinxAtStartPar
\sphinxcode{\sphinxupquote{inst}}
&
\sphinxAtStartPar
inner stereo
\\
\sphinxhline
\sphinxAtStartPar
\sphinxcode{\sphinxupquote{outst}}
&
\sphinxAtStartPar
outer stereo
\\
\sphinxhline
\sphinxAtStartPar
\sphinxcode{\sphinxupquote{z}}
&
\sphinxAtStartPar
z length
\\
\sphinxbottomrule
\end{tabulary}
\sphinxtableafterendhook\par
\sphinxattableend

\subsection{Orb}
\label{\detokenize{./prompt/usersmanual/geometry:orb}}

\sphinxSetupCaptionForVerbatim{An example: definition of an orb}
\def\sphinxLiteralBlockLabel{\label{\detokenize{./prompt/usersmanual/geometry:id43}}}
\begin{sphinxVerbatim}[commandchars=\\\{\}]
\PYG{n+nt}{\PYGZlt{}orb} \PYG{n+na}{lunit=}\PYG{l+s}{\PYGZdq{}mm\PYGZdq{}} \PYG{n+na}{name=}\PYG{l+s}{\PYGZdq{}OrbSolid\PYGZdq{}} \PYG{n+na}{r=}\PYG{l+s}{\PYGZdq{}50\PYGZdq{}}\PYG{n+nt}{/\PYGZgt{}}
\end{sphinxVerbatim}

\begin{figure}[ht]
    \centering
    \includegraphics[scale=0.15]{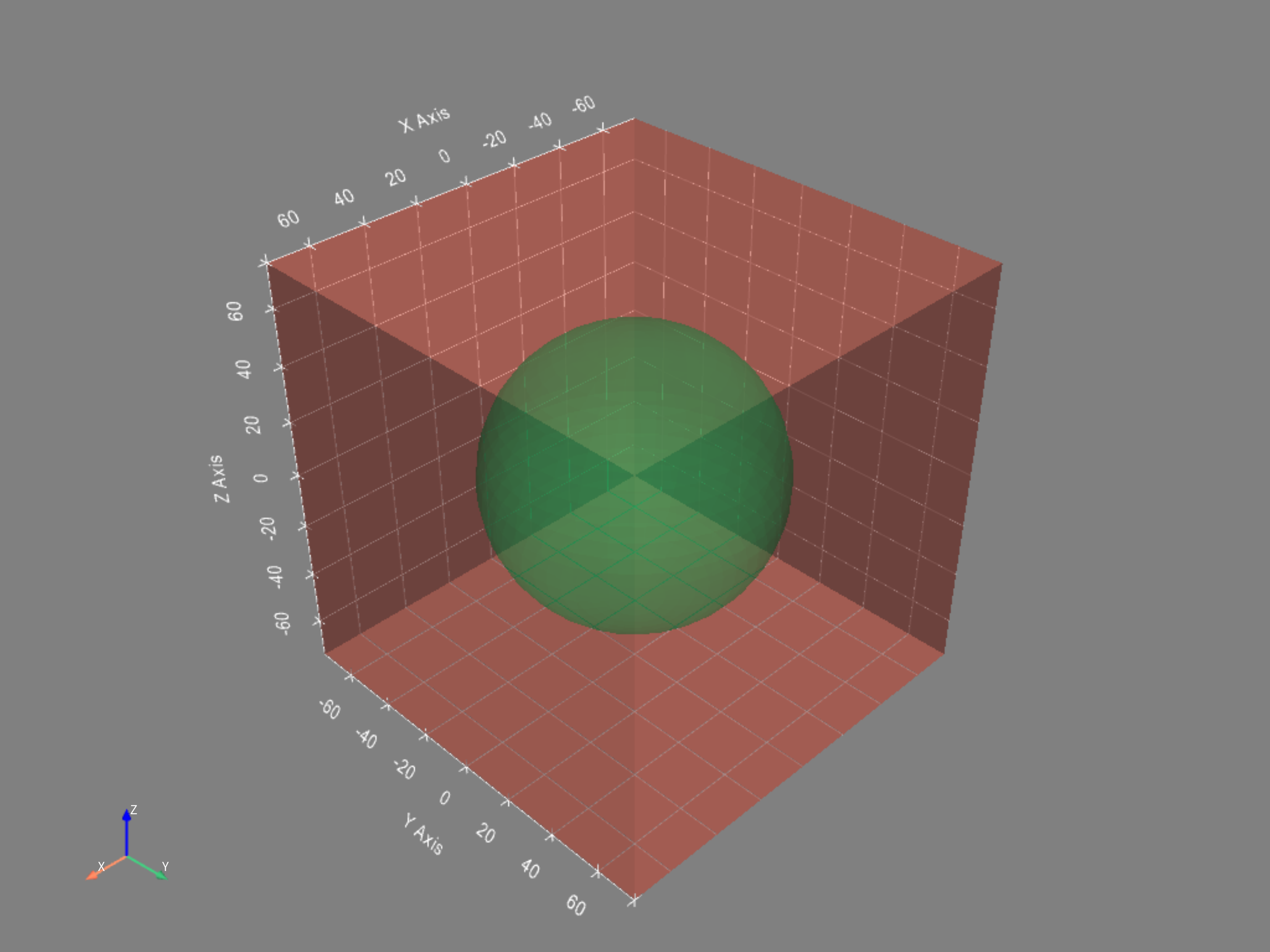}
    \caption{Visualization of an orb}
    \label{fig:aOrb}
\end{figure}

\sphinxattablestart
\sphinxthistablewithglobalstyle
\centering
\sphinxcapstartof{table}
\sphinxthecaptionisattop
\sphinxcaption{Parameters for defining an orb. If not specified, default length unit is $mm$}\label{\detokenize{./prompt/usersmanual/geometry:id44}}
\sphinxaftertopcaption
\begin{tabulary}{\linewidth}[t]{TT}
\sphinxtoprule
\sphinxstyletheadfamily 
\sphinxAtStartPar
Parameter
&\sphinxstyletheadfamily 
\sphinxAtStartPar
Description
\\
\sphinxmidrule
\sphinxtableatstartofbodyhook
\sphinxAtStartPar
\sphinxcode{\sphinxupquote{r}}
&
\sphinxAtStartPar
radius
\\
\sphinxbottomrule
\end{tabulary}
\sphinxtableafterendhook\par
\sphinxattableend

\subsection{Paraboloid}
\label{\detokenize{./prompt/usersmanual/geometry:paraboloid}}

\sphinxSetupCaptionForVerbatim{An example: definition of a paraboloid}
\def\sphinxLiteralBlockLabel{\label{\detokenize{./prompt/usersmanual/geometry:id45}}}
\begin{sphinxVerbatim}[commandchars=\\\{\}]
\PYG{n+nt}{\PYGZlt{}paraboloid} \PYG{n+na}{lunit=}\PYG{l+s}{\PYGZdq{}mm\PYGZdq{}} \PYG{n+na}{name=}\PYG{l+s}{\PYGZdq{}ParaboloidSolid\PYGZdq{}} \PYG{n+na}{rlo=}\PYG{l+s}{\PYGZdq{}60.0\PYGZdq{}} \PYG{n+na}{rhi=}\PYG{l+s}{\PYGZdq{}10.0\PYGZdq{}} \PYG{n+na}{dz=}\PYG{l+s}{\PYGZdq{}60.0\PYGZdq{}}\PYG{n+nt}{/\PYGZgt{}}
\end{sphinxVerbatim}

\begin{figure}[ht]
    \centering
    \includegraphics[scale=0.15]{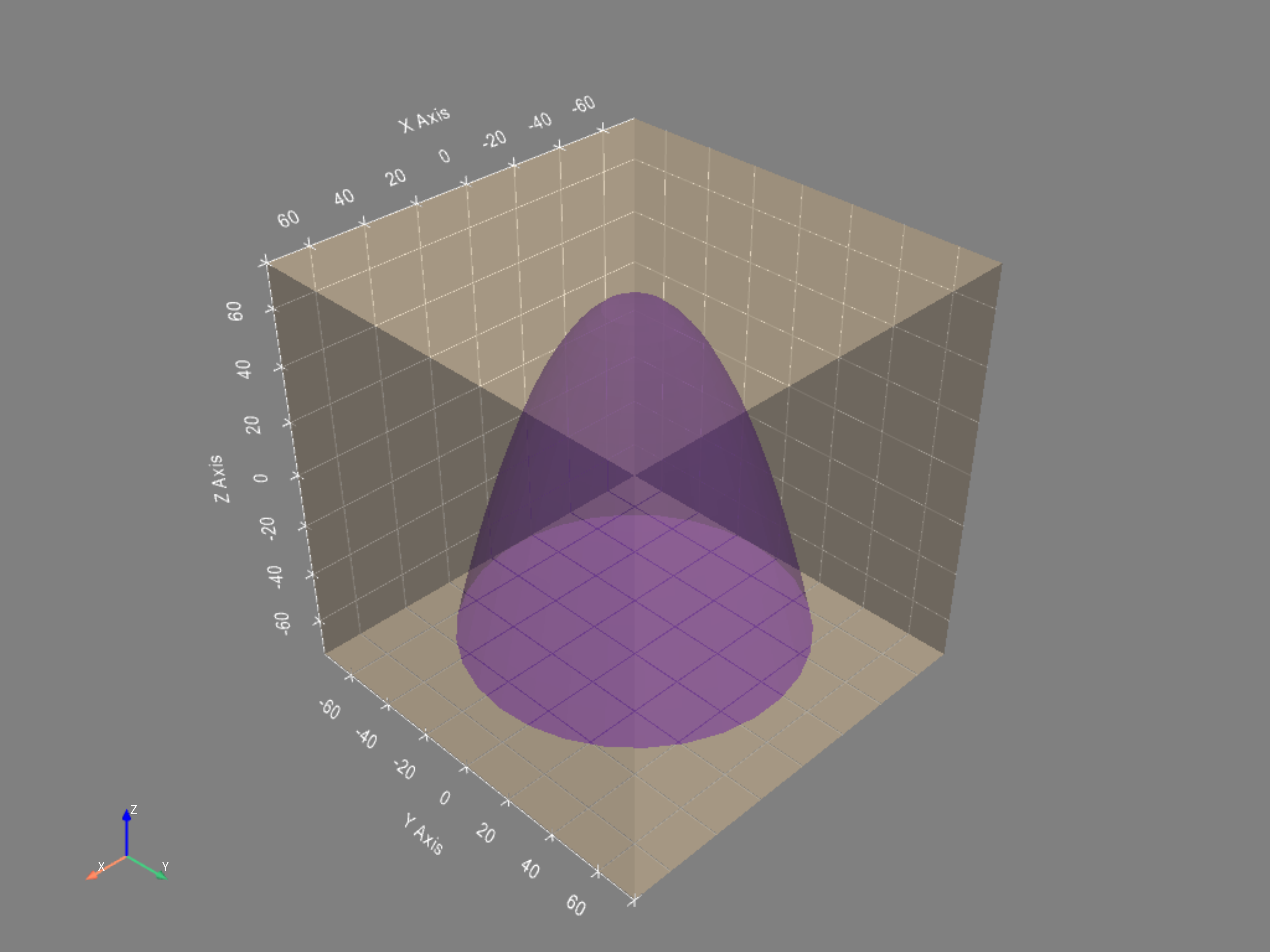}
    \caption{Visualization of a paraboloid}
    \label{fig:aParaboloid}
\end{figure}

\sphinxattablestart
\sphinxthistablewithglobalstyle
\centering
\sphinxcapstartof{table}
\sphinxthecaptionisattop
\sphinxcaption{Parameters for defining a paraboloid. If not specified, default length unit is $mm$}\label{\detokenize{./prompt/usersmanual/geometry:id46}}
\sphinxaftertopcaption
\begin{tabulary}{\linewidth}[t]{TT}
\sphinxtoprule
\sphinxstyletheadfamily 
\sphinxAtStartPar
Parameter
&\sphinxstyletheadfamily 
\sphinxAtStartPar
Description
\\
\sphinxmidrule
\sphinxtableatstartofbodyhook
\sphinxAtStartPar
\sphinxcode{\sphinxupquote{rlo}}
&
\sphinxAtStartPar
radius at \sphinxhyphen{}z
\\
\sphinxhline
\sphinxAtStartPar
\sphinxcode{\sphinxupquote{rhi}}
&
\sphinxAtStartPar
radius at +z
\\
\sphinxhline
\sphinxAtStartPar
\sphinxcode{\sphinxupquote{dz}}
&
\sphinxAtStartPar
z length
\\
\sphinxbottomrule
\end{tabulary}
\sphinxtableafterendhook\par
\sphinxattableend

\subsection{Polycone}
\label{\detokenize{./prompt/usersmanual/geometry:polycone}}

\sphinxSetupCaptionForVerbatim{An example: definition of a polycone}
\def\sphinxLiteralBlockLabel{\label{\detokenize{./prompt/usersmanual/geometry:id47}}}
\begin{sphinxVerbatim}[commandchars=\\\{\}]
\PYG{n+nt}{\PYGZlt{}polycone} \PYG{n+na}{aunit=}\PYG{l+s}{\PYGZdq{}deg\PYGZdq{}} \PYG{n+na}{lunit=}\PYG{l+s}{\PYGZdq{}mm\PYGZdq{}} \PYG{n+na}{name=}\PYG{l+s}{\PYGZdq{}PolyconeSolid\PYGZdq{}} \PYG{n+na}{deltaphi=}\PYG{l+s}{\PYGZdq{}360.0\PYGZdq{}} \PYG{n+na}{startphi=}\PYG{l+s}{\PYGZdq{}0.0\PYGZdq{}}\PYG{n+nt}{\PYGZgt{}}
    \PYG{n+nt}{\PYGZlt{}zplane} \PYG{n+na}{rmin=}\PYG{l+s}{\PYGZdq{}10.0\PYGZdq{}} \PYG{n+na}{rmax=}\PYG{l+s}{\PYGZdq{}20.0\PYGZdq{}} \PYG{n+na}{z=}\PYG{l+s}{\PYGZdq{}60.0\PYGZdq{}}\PYG{n+nt}{/\PYGZgt{}}
    \PYG{n+nt}{\PYGZlt{}zplane} \PYG{n+na}{rmin=}\PYG{l+s}{\PYGZdq{}30.0\PYGZdq{}} \PYG{n+na}{rmax=}\PYG{l+s}{\PYGZdq{}50.0\PYGZdq{}} \PYG{n+na}{z=}\PYG{l+s}{\PYGZdq{}\PYGZhy{}50.0\PYGZdq{}} \PYG{n+nt}{/\PYGZgt{}}
\PYG{n+nt}{\PYGZlt{}/polycone\PYGZgt{}}
\end{sphinxVerbatim}

\begin{figure}[ht]
    \centering
    \includegraphics[scale=0.15]{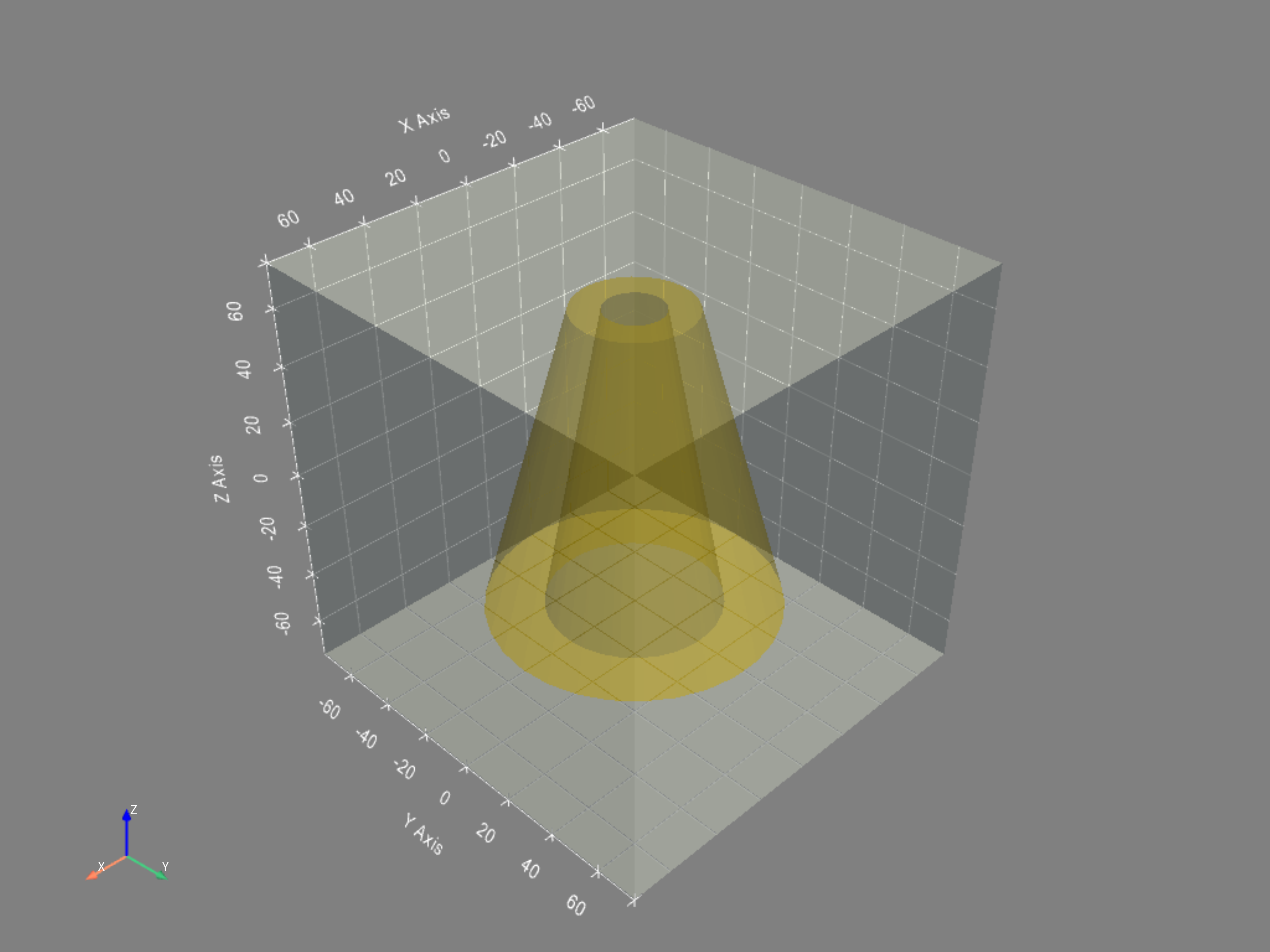}
    \caption{Visualization of a polycone}
    \label{fig:aPolycone}
\end{figure}

\sphinxattablestart
\sphinxthistablewithglobalstyle
\centering
\sphinxcapstartof{table}
\sphinxthecaptionisattop
\sphinxcaption{Parameters for defining a polycone. If not specified, default length unit is $mm$, default angle unit is $deg$}\label{\detokenize{./prompt/usersmanual/geometry:id20}}
\sphinxaftertopcaption
\begin{tabulary}{\linewidth}[t]{TTTT}
\sphinxtoprule
\sphinxstyletheadfamily 
\sphinxAtStartPar
Parameter
&\sphinxstyletheadfamily 
\sphinxAtStartPar
Default
&\sphinxstartmulticolumn{2}%
\begin{varwidth}[t]{\sphinxcolwidth{2}{4}}
\sphinxstyletheadfamily \sphinxAtStartPar
Sub\sphinxhyphen{}parameter and description
\par
\vskip-\baselineskip\vbox{\hbox{\strut}}\end{varwidth}%
\sphinxstopmulticolumn
\\
\sphinxmidrule
\sphinxtableatstartofbodyhook
\sphinxAtStartPar
\sphinxcode{\sphinxupquote{startphi}}
&
\sphinxAtStartPar
0.0
&&
\sphinxAtStartPar
start angle of the segment
\\
\sphinxhline
\sphinxAtStartPar
\sphinxcode{\sphinxupquote{deltaphi}}
&&&
\sphinxAtStartPar
angle of the segment
\\
\sphinxhline\sphinxmultirow{3}{12}{%
\begin{varwidth}[t]{\sphinxcolwidth{1}{4}}
\sphinxAtStartPar
\sphinxcode{\sphinxupquote{zplane}}
\par
\vskip-\baselineskip\vbox{\hbox{\strut}}\end{varwidth}%
}%
&
\sphinxAtStartPar
0.0
&
\sphinxAtStartPar
\sphinxcode{\sphinxupquote{rmin}}
&
\sphinxAtStartPar
inner radius of cone at this point
\\
\sphinxcline{2-4}\sphinxfixclines{4}\sphinxtablestrut{12}&&
\sphinxAtStartPar
\sphinxcode{\sphinxupquote{rmax}}
&
\sphinxAtStartPar
outer radius of cone at this point
\\
\sphinxcline{2-4}\sphinxfixclines{4}\sphinxtablestrut{12}&&
\sphinxAtStartPar
\sphinxcode{\sphinxupquote{z}}
&
\sphinxAtStartPar
z coordinate of the plane
\\
\sphinxbottomrule
\end{tabulary}
\sphinxtableafterendhook\par
\sphinxattableend

\subsection{Polyhedra}
\label{\detokenize{./prompt/usersmanual/geometry:polyhedra}}

\sphinxSetupCaptionForVerbatim{An example: definition of a polyhedra}
\def\sphinxLiteralBlockLabel{\label{\detokenize{./prompt/usersmanual/geometry:id49}}}
\begin{sphinxVerbatim}[commandchars=\\\{\}]
\PYG{n+nt}{\PYGZlt{}polyhedra} \PYG{n+na}{aunit=}\PYG{l+s}{\PYGZdq{}deg\PYGZdq{}} \PYG{n+na}{lunit=}\PYG{l+s}{\PYGZdq{}mm\PYGZdq{}} \PYG{n+na}{name=}\PYG{l+s}{\PYGZdq{}PolyhedraSolid\PYGZdq{}} \PYG{n+na}{deltaphi=}\PYG{l+s}{\PYGZdq{}360.0\PYGZdq{}} \PYG{n+na}{startphi=}\PYG{l+s}{\PYGZdq{}0.0\PYGZdq{}} \PYG{n+na}{numsides=}\PYG{l+s}{\PYGZdq{}6\PYGZdq{}}\PYG{n+nt}{\PYGZgt{}}
    \PYG{n+nt}{\PYGZlt{}zplane} \PYG{n+na}{rmin=}\PYG{l+s}{\PYGZdq{}20.0\PYGZdq{}} \PYG{n+na}{rmax=}\PYG{l+s}{\PYGZdq{}50.0\PYGZdq{}} \PYG{n+na}{z=}\PYG{l+s}{\PYGZdq{}\PYGZhy{}50.0\PYGZdq{}}\PYG{n+nt}{/\PYGZgt{}}
    \PYG{n+nt}{\PYGZlt{}zplane} \PYG{n+na}{rmin=}\PYG{l+s}{\PYGZdq{}5.0\PYGZdq{}} \PYG{n+na}{rmax=}\PYG{l+s}{\PYGZdq{}10.0\PYGZdq{}} \PYG{n+na}{z=}\PYG{l+s}{\PYGZdq{}0.0\PYGZdq{}\PYGZdq{}} \PYG{n+nt}{/\PYGZgt{}}
    \PYG{n+nt}{\PYGZlt{}zplane} \PYG{n+na}{rmin=}\PYG{l+s}{\PYGZdq{}10.0\PYGZdq{}} \PYG{n+na}{rmax=}\PYG{l+s}{\PYGZdq{}30.0\PYGZdq{}} \PYG{n+na}{z=}\PYG{l+s}{\PYGZdq{}60.0\PYGZdq{}} \PYG{n+nt}{/\PYGZgt{}}
\PYG{n+nt}{\PYGZlt{}/polyhedra\PYGZgt{}}
\end{sphinxVerbatim}

\begin{figure}[ht]
    \centering
    \includegraphics[scale=0.15]{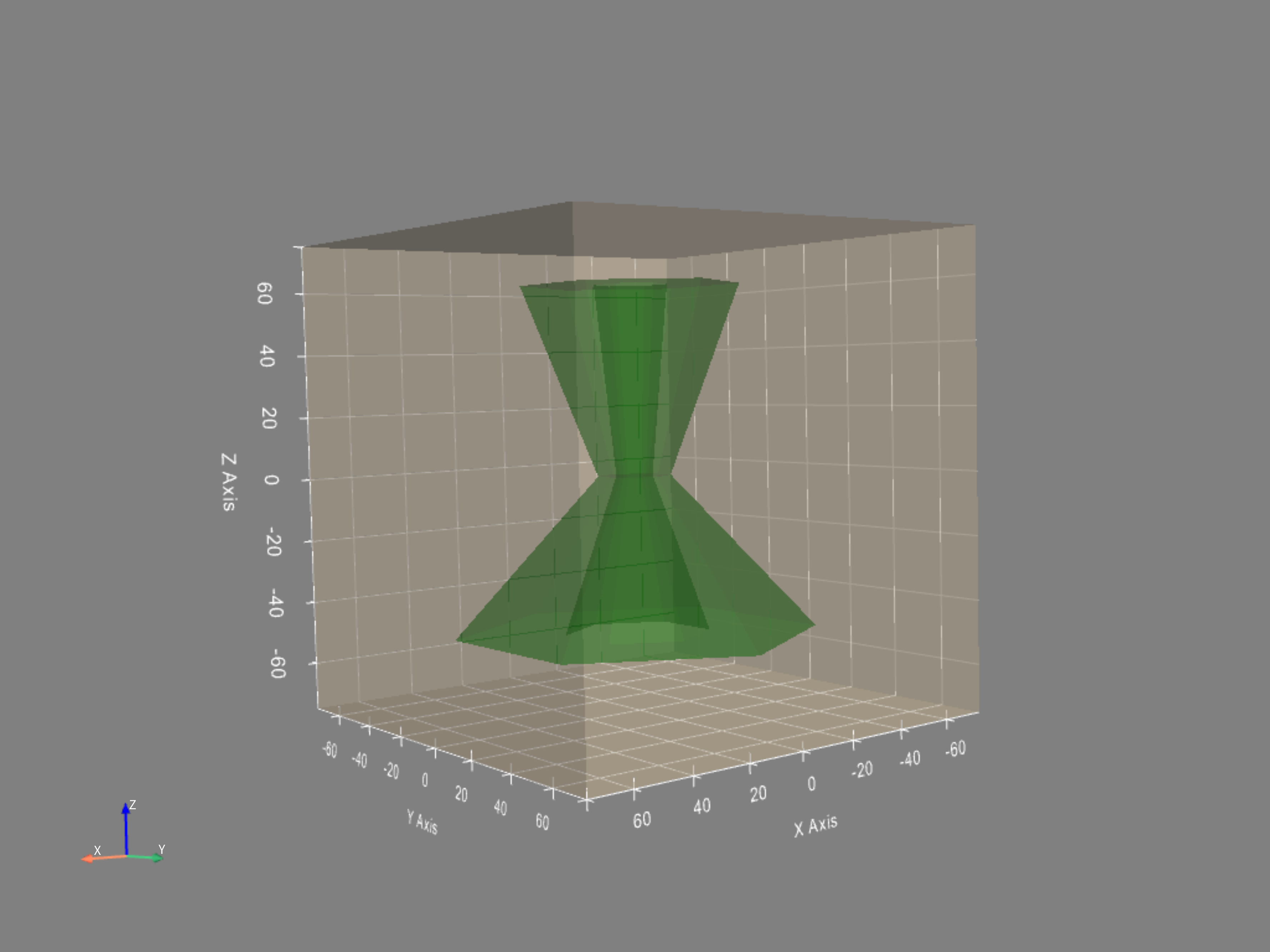}
    \caption{Visualization of a polyhedra}
    \label{fig:aPolyhedra}
\end{figure}

\sphinxattablestart
\sphinxthistablewithglobalstyle
\centering
\sphinxcapstartof{table}
\sphinxthecaptionisattop
\sphinxcaption{Parameters for defining a polyhedra. If not specified, default length unit is $mm$, default angle unit is $deg$}\label{\detokenize{./prompt/usersmanual/geometry:id22}}
\sphinxaftertopcaption
\begin{tabulary}{\linewidth}[t]{TTTT}
\sphinxtoprule
\sphinxstyletheadfamily 
\sphinxAtStartPar
Parameter
&\sphinxstyletheadfamily 
\sphinxAtStartPar
Default
&\sphinxstartmulticolumn{2}%
\begin{varwidth}[t]{\sphinxcolwidth{2}{4}}
\sphinxstyletheadfamily \sphinxAtStartPar
Sub\sphinxhyphen{}parameter and description
\par
\vskip-\baselineskip\vbox{\hbox{\strut}}\end{varwidth}%
\sphinxstopmulticolumn
\\
\sphinxmidrule
\sphinxtableatstartofbodyhook
\sphinxAtStartPar
\sphinxcode{\sphinxupquote{startphi}}
&&&
\sphinxAtStartPar
start angle of the segment
\\
\sphinxhline
\sphinxAtStartPar
\sphinxcode{\sphinxupquote{deltaphi}}
&&&
\sphinxAtStartPar
angle of the segment
\\
\sphinxhline
\sphinxAtStartPar
\sphinxcode{\sphinxupquote{numsides}}
&&&
\sphinxAtStartPar
number of sides
\\
\sphinxhline\sphinxmultirow{3}{16}{%
\begin{varwidth}[t]{\sphinxcolwidth{1}{4}}
\sphinxAtStartPar
\sphinxcode{\sphinxupquote{zplane}}
\par
\vskip-\baselineskip\vbox{\hbox{\strut}}\end{varwidth}%
}%
&
\sphinxAtStartPar
0.0
&
\sphinxAtStartPar
\sphinxcode{\sphinxupquote{rmin}}
&
\sphinxAtStartPar
inner radius of cone at this point
\\
\sphinxcline{2-4}\sphinxfixclines{4}\sphinxtablestrut{16}&&
\sphinxAtStartPar
\sphinxcode{\sphinxupquote{rmax}}
&
\sphinxAtStartPar
outer radius of cone at this point
\\
\sphinxcline{2-4}\sphinxfixclines{4}\sphinxtablestrut{16}&&
\sphinxAtStartPar
\sphinxcode{\sphinxupquote{z}}
&
\sphinxAtStartPar
z coordinate of the plane
\\
\sphinxbottomrule
\end{tabulary}
\sphinxtableafterendhook\par
\sphinxattableend

\subsection{Sphere}
\label{\detokenize{./prompt/usersmanual/geometry:sphere}}

\sphinxSetupCaptionForVerbatim{An example: definition of a sphere}
\def\sphinxLiteralBlockLabel{\label{\detokenize{./prompt/usersmanual/geometry:id51}}}
\begin{sphinxVerbatim}[commandchars=\\\{\}]
\PYG{n+nt}{\PYGZlt{}sphere} \PYG{n+na}{aunit=}\PYG{l+s}{\PYGZdq{}deg\PYGZdq{}} \PYG{n+na}{lunit=}\PYG{l+s}{\PYGZdq{}mm\PYGZdq{}} \PYG{n+na}{name=}\PYG{l+s}{\PYGZdq{}SphereSolid\PYGZdq{}} \PYG{n+na}{rmin=}\PYG{l+s}{\PYGZdq{}0.0\PYGZdq{}} \PYG{n+na}{rmax=}\PYG{l+s}{\PYGZdq{}60.0\PYGZdq{}} \PYG{n+na}{deltaphi=}\PYG{l+s}{\PYGZdq{}180.0\PYGZdq{}} \PYG{n+na}{startphi=}\PYG{l+s}{\PYGZdq{}0.0\PYGZdq{}} \PYG{n+na}{deltatheta=}\PYG{l+s}{\PYGZdq{}90.0\PYGZdq{}} \PYG{n+na}{starttheta=}\PYG{l+s}{\PYGZdq{}0.0\PYGZdq{}}\PYG{n+nt}{/\PYGZgt{}}
\end{sphinxVerbatim}

\begin{figure}[ht]
    \centering
    \includegraphics[scale=0.15]{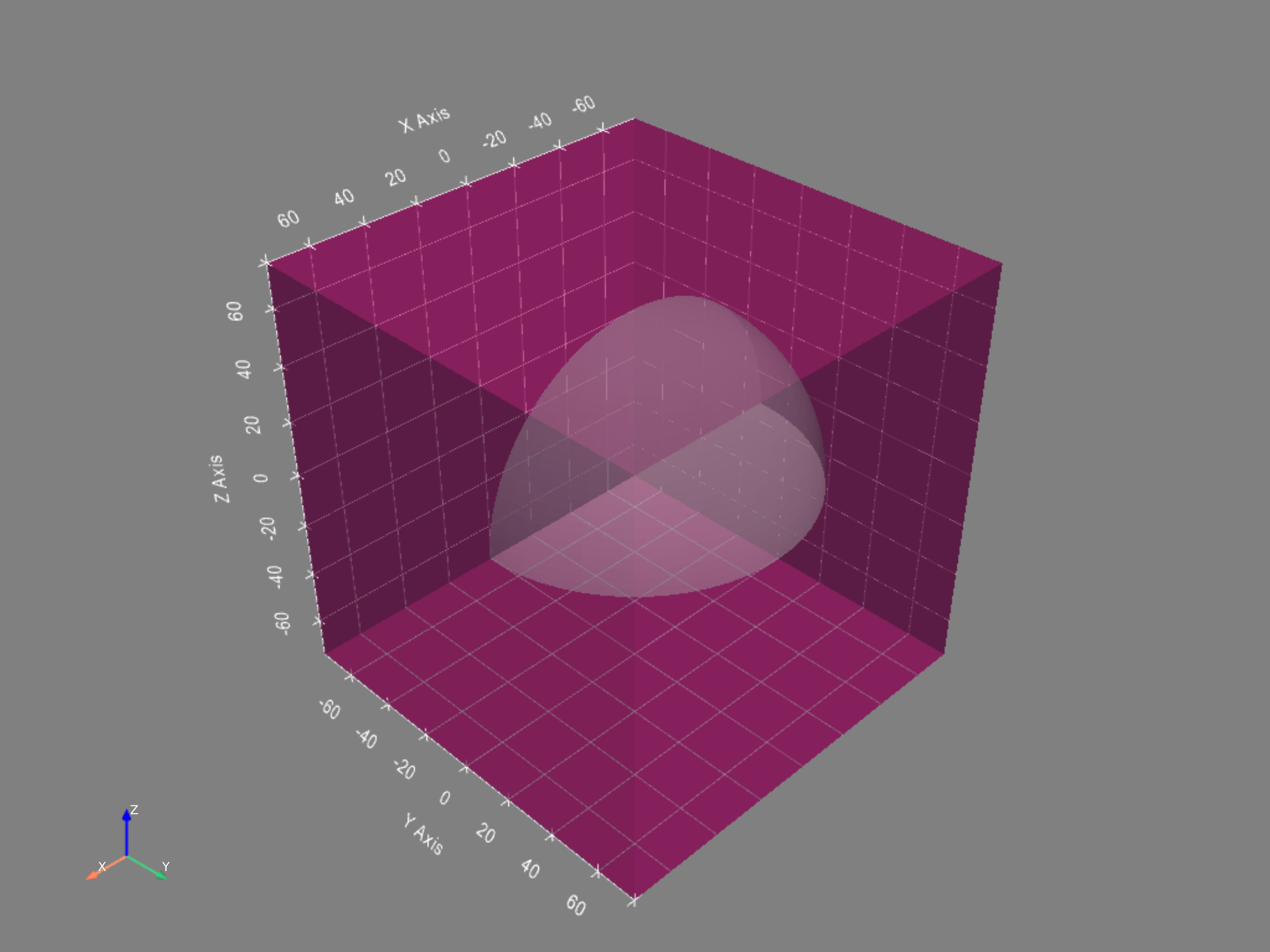}
    \caption{Visualization of a sphere}
    \label{fig:aphere}
\end{figure}

\sphinxattablestart
\sphinxthistablewithglobalstyle
\centering
\sphinxcapstartof{table}
\sphinxthecaptionisattop
\sphinxcaption{Parameters for defining a sphere. If not specified, default length unit is $mm$, default angle unit is $deg$}\label{\detokenize{./prompt/usersmanual/geometry:id24}}
\sphinxaftertopcaption
\begin{tabulary}{\linewidth}[t]{TTT}
\sphinxtoprule
\sphinxstyletheadfamily 
\sphinxAtStartPar
Parameter
&\sphinxstyletheadfamily 
\sphinxAtStartPar
Default
&\sphinxstyletheadfamily 
\sphinxAtStartPar
Description
\\
\sphinxmidrule
\sphinxtableatstartofbodyhook
\sphinxAtStartPar
\sphinxcode{\sphinxupquote{rmin}}
&
\sphinxAtStartPar
0.0
&
\sphinxAtStartPar
inner radius
\\
\sphinxhline
\sphinxAtStartPar
\sphinxcode{\sphinxupquote{rmax}}
&&
\sphinxAtStartPar
outer radius
\\
\sphinxhline
\sphinxAtStartPar
\sphinxcode{\sphinxupquote{startphi}}
&
\sphinxAtStartPar
0.0
&
\sphinxAtStartPar
starting angle of the segment
\\
\sphinxhline
\sphinxAtStartPar
\sphinxcode{\sphinxupquote{deltaphi}}
&&
\sphinxAtStartPar
delta angle of the segment
\\
\sphinxhline
\sphinxAtStartPar
\sphinxcode{\sphinxupquote{starttheta}}
&
\sphinxAtStartPar
0.0
&
\sphinxAtStartPar
starting angle of the segment
\\
\sphinxhline
\sphinxAtStartPar
\sphinxcode{\sphinxupquote{deltatheta}}
&&
\sphinxAtStartPar
delta angle of the segment
\\
\sphinxbottomrule
\end{tabulary}
\sphinxtableafterendhook\par
\sphinxattableend

\subsection{Tetrahedron}
\label{\detokenize{./prompt/usersmanual/geometry:tetrahedron}}

\sphinxSetupCaptionForVerbatim{An example: definition of a tetrahedron}
\def\sphinxLiteralBlockLabel{\label{\detokenize{./prompt/usersmanual/geometry:id53}}}
\begin{sphinxVerbatim}[commandchars=\\\{\}]
\PYG{n+nt}{\PYGZlt{}define}\PYG{n+nt}{\PYGZgt{}}
    \PYG{n+nt}{\PYGZlt{}position} \PYG{n+na}{name=}\PYG{l+s}{\PYGZdq{}v1\PYGZdq{}} \PYG{n+na}{x=}\PYG{l+s}{\PYGZdq{}\PYGZhy{}70\PYGZdq{}} \PYG{n+na}{y=}\PYG{l+s}{\PYGZdq{}\PYGZhy{}70\PYGZdq{}} \PYG{n+na}{z=}\PYG{l+s}{\PYGZdq{}\PYGZhy{}60\PYGZdq{}}\PYG{n+nt}{/\PYGZgt{}}
    \PYG{n+nt}{\PYGZlt{}position} \PYG{n+na}{name=}\PYG{l+s}{\PYGZdq{}v2\PYGZdq{}} \PYG{n+na}{x=}\PYG{l+s}{\PYGZdq{}70\PYGZdq{}} \PYG{n+na}{y=}\PYG{l+s}{\PYGZdq{}\PYGZhy{}40\PYGZdq{}} \PYG{n+na}{z=}\PYG{l+s}{\PYGZdq{}\PYGZhy{}60\PYGZdq{}}\PYG{n+nt}{/\PYGZgt{}}
    \PYG{n+nt}{\PYGZlt{}position} \PYG{n+na}{name=}\PYG{l+s}{\PYGZdq{}v3\PYGZdq{}} \PYG{n+na}{x=}\PYG{l+s}{\PYGZdq{}0\PYGZdq{}} \PYG{n+na}{y=}\PYG{l+s}{\PYGZdq{}60\PYGZdq{}} \PYG{n+na}{z=}\PYG{l+s}{\PYGZdq{}\PYGZhy{}60\PYGZdq{}}\PYG{n+nt}{/\PYGZgt{}}
    \PYG{n+nt}{\PYGZlt{}position} \PYG{n+na}{name=}\PYG{l+s}{\PYGZdq{}v4\PYGZdq{}} \PYG{n+na}{x=}\PYG{l+s}{\PYGZdq{}0\PYGZdq{}} \PYG{n+na}{y=}\PYG{l+s}{\PYGZdq{}0\PYGZdq{}} \PYG{n+na}{z=}\PYG{l+s}{\PYGZdq{}60\PYGZdq{}}\PYG{n+nt}{/\PYGZgt{}}
\PYG{n+nt}{\PYGZlt{}/define\PYGZgt{}}
\PYG{n+nt}{\PYGZlt{}tet} \PYG{n+na}{name=}\PYG{l+s}{\PYGZdq{}TetrahedronSolid\PYGZdq{}} \PYG{n+na}{vertex1=}\PYG{l+s}{\PYGZdq{}v1\PYGZdq{}} \PYG{n+na}{vertex2=}\PYG{l+s}{\PYGZdq{}v2\PYGZdq{}} \PYG{n+na}{vertex3=}\PYG{l+s}{\PYGZdq{}v3\PYGZdq{}} \PYG{n+na}{vertex4=}\PYG{l+s}{\PYGZdq{}v4\PYGZdq{}}\PYG{n+nt}{/\PYGZgt{}}
\end{sphinxVerbatim}

\begin{figure}[ht]
    \centering
    \includegraphics[scale=0.15]{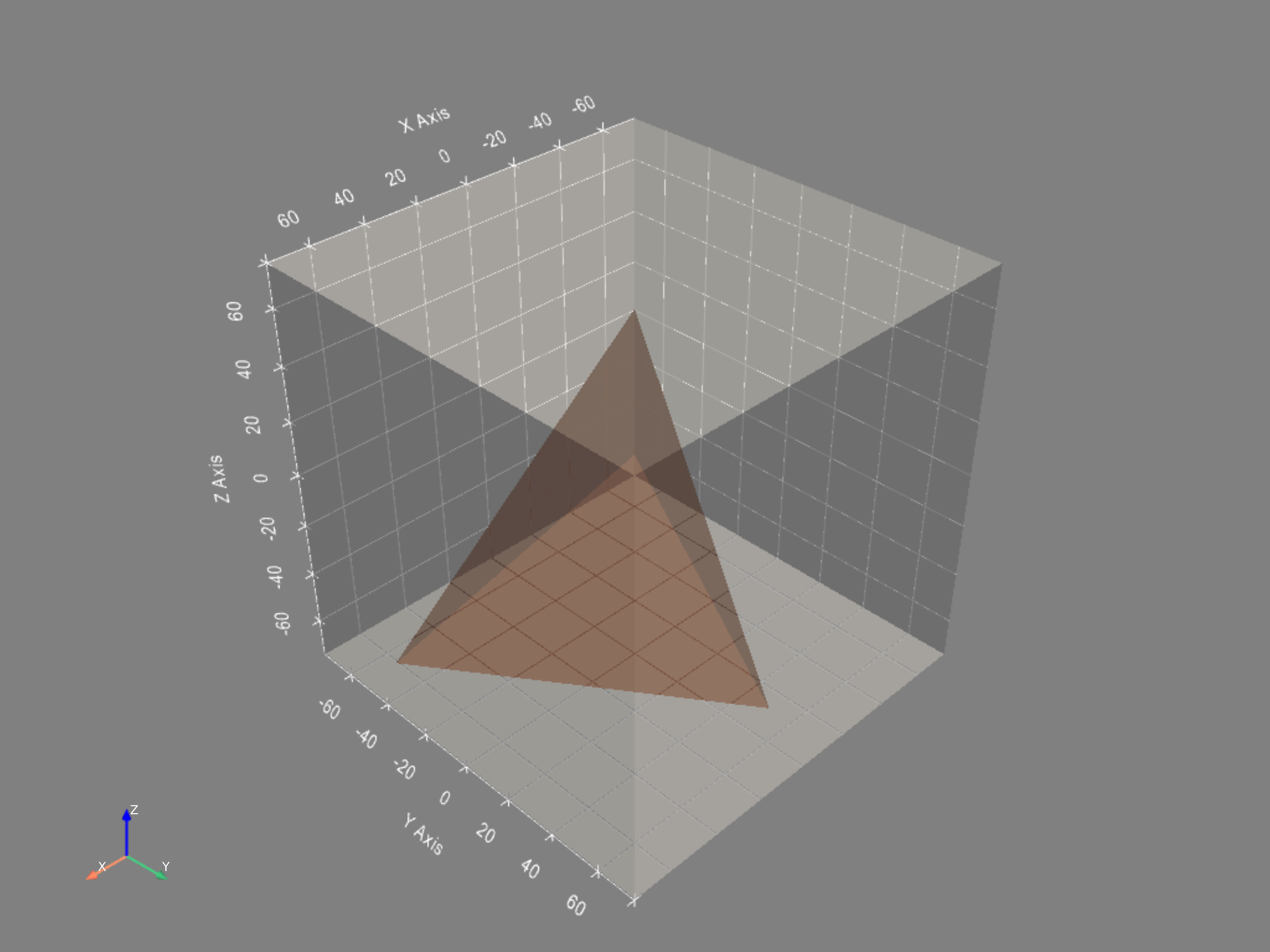}
    \caption{Visualization of a tetrahedron}
    \label{fig:aTetrahedron}
\end{figure}

\sphinxattablestart
\sphinxthistablewithglobalstyle
\centering
\sphinxcapstartof{table}
\sphinxthecaptionisattop
\sphinxcaption{Parameters for defining a tetrahedron. If not specified, default length unit is $mm$}\label{\detokenize{./prompt/usersmanual/geometry:id54}}
\sphinxaftertopcaption
\begin{tabulary}{\linewidth}[t]{TT}
\sphinxtoprule
\sphinxstyletheadfamily 
\sphinxAtStartPar
Parameter
&\sphinxstyletheadfamily 
\sphinxAtStartPar
Description
\\
\sphinxmidrule
\sphinxtableatstartofbodyhook
\sphinxAtStartPar
\sphinxcode{\sphinxupquote{vertex1}}
&
\sphinxAtStartPar
vertex 1 position
\\
\sphinxhline
\sphinxAtStartPar
\sphinxcode{\sphinxupquote{vertex2}}
&
\sphinxAtStartPar
vertex 2 position
\\
\sphinxhline
\sphinxAtStartPar
\sphinxcode{\sphinxupquote{vertex3}}
&
\sphinxAtStartPar
vertex 3 position
\\
\sphinxhline
\sphinxAtStartPar
\sphinxcode{\sphinxupquote{vertex4}}
&
\sphinxAtStartPar
vertex 4 position
\\
\sphinxbottomrule
\end{tabulary}
\sphinxtableafterendhook\par
\sphinxattableend

\subsection{Trapezoid}
\label{\detokenize{./prompt/usersmanual/geometry:trapezoid}}

\sphinxSetupCaptionForVerbatim{An example: definition of a trapezoid}
\def\sphinxLiteralBlockLabel{\label{\detokenize{./prompt/usersmanual/geometry:id55}}}
\begin{sphinxVerbatim}[commandchars=\\\{\}]
\PYG{n+nt}{\PYGZlt{}trd} \PYG{n+na}{lunit=}\PYG{l+s}{\PYGZdq{}mm\PYGZdq{}} \PYG{n+na}{name=}\PYG{l+s}{\PYGZdq{}TrdSolid\PYGZdq{}}  \PYG{n+na}{x1=}\PYG{l+s}{\PYGZdq{}50\PYGZdq{}} \PYG{n+na}{x2=}\PYG{l+s}{\PYGZdq{}100\PYGZdq{}} \PYG{n+na}{y1=}\PYG{l+s}{\PYGZdq{}60\PYGZdq{}} \PYG{n+na}{y2=}\PYG{l+s}{\PYGZdq{}80\PYGZdq{}} \PYG{n+na}{z=}\PYG{l+s}{\PYGZdq{}130\PYGZdq{}}\PYG{n+nt}{/\PYGZgt{}}
\end{sphinxVerbatim}

\begin{figure}[ht]
    \centering
    \includegraphics[scale=0.15]{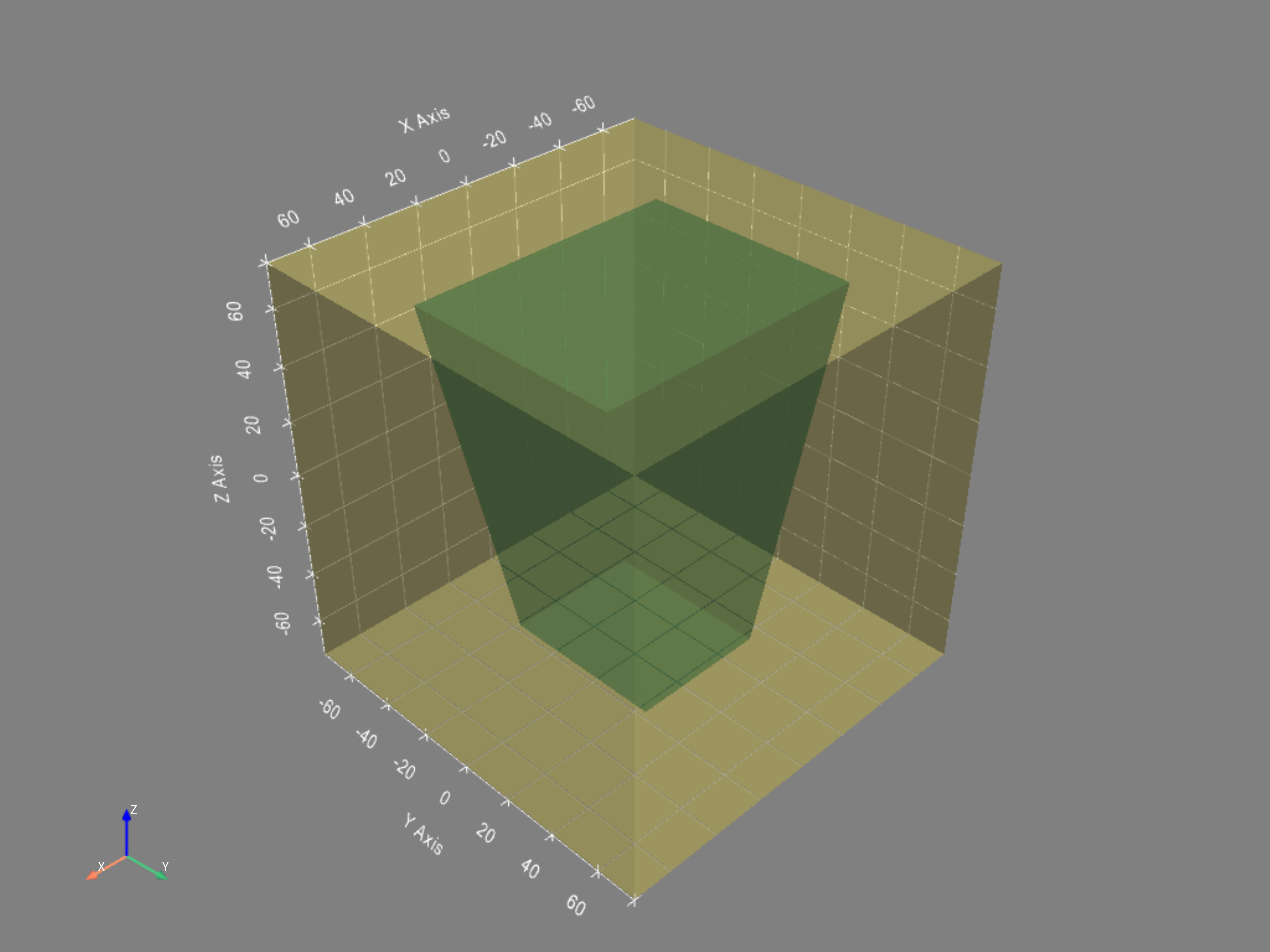}
    \caption{Visualization of a trapezoid}
    \label{fig:aTrapezoid}
\end{figure}

\sphinxattablestart
\sphinxthistablewithglobalstyle
\centering
\sphinxcapstartof{table}
\sphinxthecaptionisattop
\sphinxcaption{Parameters for defining a trapezoid. If not specified, default length unit is $mm$}\label{\detokenize{./prompt/usersmanual/geometry:id56}}
\sphinxaftertopcaption
\begin{tabulary}{\linewidth}[t]{TT}
\sphinxtoprule
\sphinxstyletheadfamily 
\sphinxAtStartPar
Parameter
&\sphinxstyletheadfamily 
\sphinxAtStartPar
Description
\\
\sphinxmidrule
\sphinxtableatstartofbodyhook
\sphinxAtStartPar
\sphinxcode{\sphinxupquote{x1}}
&
\sphinxAtStartPar
x length at \sphinxhyphen{}z
\\
\sphinxhline
\sphinxAtStartPar
\sphinxcode{\sphinxupquote{x2}}
&
\sphinxAtStartPar
x length at +z
\\
\sphinxhline
\sphinxAtStartPar
\sphinxcode{\sphinxupquote{y1}}
&
\sphinxAtStartPar
y length at \sphinxhyphen{}z
\\
\sphinxhline
\sphinxAtStartPar
\sphinxcode{\sphinxupquote{y2}}
&
\sphinxAtStartPar
y length at +z
\\
\sphinxhline
\sphinxAtStartPar
\sphinxcode{\sphinxupquote{z}}
&
\sphinxAtStartPar
z length
\\
\sphinxbottomrule
\end{tabulary}
\sphinxtableafterendhook\par
\sphinxattableend

\subsection{Tube}
\label{\detokenize{./prompt/usersmanual/geometry:tube}}

\sphinxSetupCaptionForVerbatim{An example: definition of a tube}
\def\sphinxLiteralBlockLabel{\label{\detokenize{./prompt/usersmanual/geometry:id57}}}
\begin{sphinxVerbatim}[commandchars=\\\{\}]
\PYG{n+nt}{\PYGZlt{}tube} \PYG{n+na}{aunit=}\PYG{l+s}{\PYGZdq{}deg\PYGZdq{}} \PYG{n+na}{lunit=}\PYG{l+s}{\PYGZdq{}mm\PYGZdq{}} \PYG{n+na}{name=}\PYG{l+s}{\PYGZdq{}TubeSolid\PYGZdq{}} \PYG{n+na}{rmin=}\PYG{l+s}{\PYGZdq{}10.0\PYGZdq{}} \PYG{n+na}{rmax=}\PYG{l+s}{\PYGZdq{}50.0\PYGZdq{}} \PYG{n+na}{z=}\PYG{l+s}{\PYGZdq{}120.0\PYGZdq{}} \PYG{n+na}{deltaphi=}\PYG{l+s}{\PYGZdq{}360.0\PYGZdq{}} \PYG{n+na}{startphi=}\PYG{l+s}{\PYGZdq{}0.0\PYGZdq{}}\PYG{n+nt}{/\PYGZgt{}}
\end{sphinxVerbatim}

\begin{figure}[ht]
    \centering
    \includegraphics[scale=0.15]{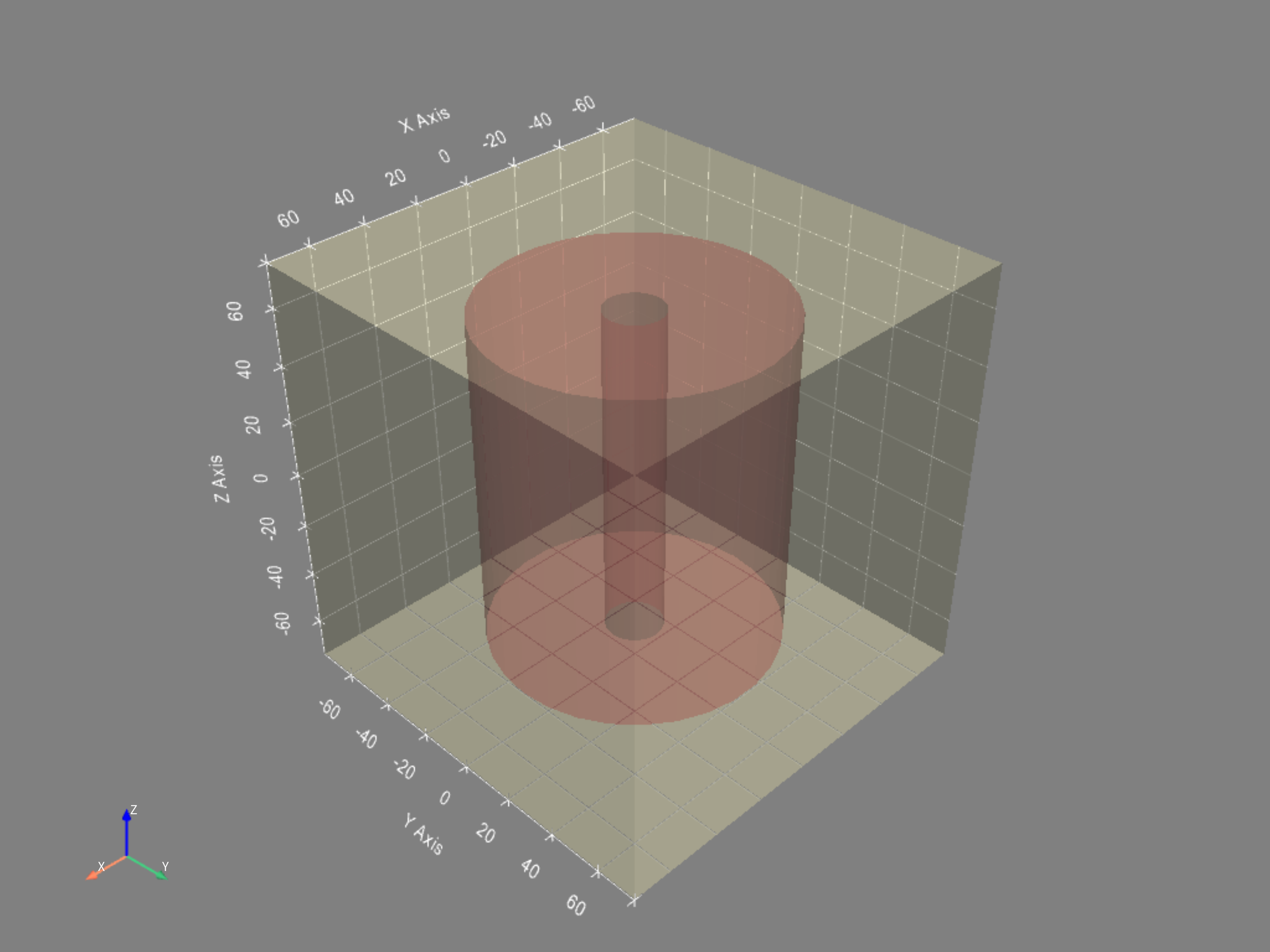}
    \caption{Visualization of a tube}
    \label{fig:aTube}
\end{figure}

\sphinxattablestart
\sphinxthistablewithglobalstyle
\centering
\sphinxcapstartof{table}
\sphinxthecaptionisattop
\sphinxcaption{Parameters for defining a tube. If not specified, default length unit is $mm$, default angle unit is $deg$}\label{\detokenize{./prompt/usersmanual/geometry:id58}}
\sphinxaftertopcaption
\begin{tabulary}{\linewidth}[t]{TTT}
\sphinxtoprule
\sphinxstyletheadfamily 
\sphinxAtStartPar
Parameter
&\sphinxstyletheadfamily 
\sphinxAtStartPar
Default
&\sphinxstyletheadfamily 
\sphinxAtStartPar
Description
\\
\sphinxmidrule
\sphinxtableatstartofbodyhook
\sphinxAtStartPar
\sphinxcode{\sphinxupquote{rmin}}
&
\sphinxAtStartPar
0.0
&
\sphinxAtStartPar
inside radius of segment
\\
\sphinxhline
\sphinxAtStartPar
\sphinxcode{\sphinxupquote{rmax}}
&&
\sphinxAtStartPar
outside radius of segment
\\
\sphinxhline
\sphinxAtStartPar
\sphinxcode{\sphinxupquote{z}}
&&
\sphinxAtStartPar
z length of tube segment
\\
\sphinxhline
\sphinxAtStartPar
\sphinxcode{\sphinxupquote{startphi}}
&
\sphinxAtStartPar
0.0
&
\sphinxAtStartPar
starting phi position angle of segment
\\
\sphinxhline
\sphinxAtStartPar
\sphinxcode{\sphinxupquote{deltaphi}}
&&
\sphinxAtStartPar
delta angle of segment
\\
\sphinxbottomrule
\end{tabulary}
\sphinxtableafterendhook\par
\sphinxattableend

\section{Particle guns}

\setcounter{table}{0}
\label{sParticlegun}

\subsection{\rm\texttt{SimpleThermalGun}}
\label{ssSimpleThermalGun }
\begin{table}[htb]
    \begin{threeparttable}[c]
    	\centering                       
    	\caption{Parameters available in \texttt{SimpleThermalGun} configuration strings. Parameters without default values are required, the others are optional. }     
    	\label{tableSimpleThermalGun} 
    	
        \begin{tabular}{p{5em}p{5em}p{17em}p{3em}} 
            \toprule
            Parameter & Default & Description & Unit   \\ 
        
            \midrule
            \texttt{energy} & 0  & Energy of particles. The value can be
            positive\tnote{1} or zero\tnote{2}. & \si{\eV}  \\
            \texttt{position} &  & Position of primary particles. & \si{\milli\meter}  \\ 
            \texttt{direction} & 0,0,1  & Direction vector of primary particles. &  \si{\milli\meter}\\ 
            \bottomrule
        \end{tabular}
        \begin{tablenotes}
           \item [1] Single energy of input value.
           \item [2] Maxwell’s distribution with temperature of \SI{293}{\kelvin}.
        \end{tablenotes}
    \end{threeparttable}
\end{table}

\subsection{\rm\texttt{IsotropicGun}}
\label{ssIsotropicGun }

\begin{table}[htb]
    \begin{threeparttable}[c]
	\centering                       
	\caption{Parameters available in \texttt{IsotropicGun} configuration strings. Parameters without default values are required, the others are optional. }     
	\label{tableIsotropicGun} 
	
    \begin{tabular}{p{5em}p{5em}p{17em}p{3em}} 
    \toprule
    Parameter & Default & Description & Unit   \\ 

    \midrule
    \texttt{energy} & 0  &Energy of particles. The value can be
            positive\tnote{1} or zero\tnote{2}. & \si{\eV} \\
    \texttt{position} &  & Position of primary particles. & \si{\milli\meter}\\ 
    \bottomrule
    \end{tabular}
    \begin{tablenotes}
        \item [1] Single energy of input value.
        \item [2] Maxwell’s distribution with temperature of \SI{293}{\kelvin}.
    \end{tablenotes}
    \end{threeparttable}
\end{table}

\subsection{\rm\texttt{UniModeratorGun}}
\label{ssUniModeratorGun}

\begin{table}[htb]
	\centering                       
	\caption{Parameters available in \texttt{UniModeratorGun} configuration strings. Parameters without default values are required, the others are optional. }     
	\label{tableUniModeratorGun} 
	
    \begin{tabular}{p{5em}p{5em}p{17em}p{3em}} 
    \toprule
    Parameter & Default & Description & Unit   \\ 

    \midrule
    \texttt{src\_w} &  & X-dimension (width) of particle source surface.  & \si{\milli\meter} \\
    \texttt{src\_h} &  & Y-dimension (height) of particle source surface. & \si{\milli\meter} \\
    \texttt{src\_z} &  & Z-coordinate of particle source center in global reference system. & \si{\milli\meter} \\
    \texttt{slit\_w} &  & X-dimension (width) of slit surface. & \si{\milli\meter} \\
    \texttt{slit\_h} &  & Y-dimension (height) of slit surface. & \si{\milli\meter}\\
    \texttt{slit\_z} &  & Z-coordinate of slit center in global reference system. & \si{\milli\meter} \\
    \texttt{mean\_wl} & 1 & Mean wavelength of primary particles. &\si{\angstrom}\\
    \texttt{range\_wl} & 0.0001  & The width of the uniform wavelength distribution. & \si{\angstrom}  \\ 
    
    \bottomrule
    \end{tabular}
\end{table}

\subsection{\rm\texttt{MaxwellianGun}}
\label{ssMaxwellianGun}

\caption{Parameters available in \texttt{MaxwellianGun} configuration strings. Parameters without default values are required, the others are optional. }     
\label{tableMaxwellianGun}   	
\begin{tabular}{p{5em}p{5em}p{17em}p{3em}} 
    \toprule
    Parameter & Default & Description & Unit   \\ 

    \midrule
    \texttt{src\_w} &  & X-dimension (width) of particle source surface. & \si{\milli\meter} \\
    \texttt{src\_h} &  & Y-dimension (height) of particle source surface. & \si{\milli\meter} \\
    \texttt{src\_z} &  & Z-coordinate of particle source center in global reference system. & \si{\milli\meter} \\
    \texttt{slit\_w} &  & X-dimension (width) of slit surface. & \si{\milli\meter} \\
    \texttt{slit\_h} &  & Y-dimension (height) of slit surface. & \si{\milli\meter} \\
    \texttt{slit\_z} &  & Z-coordinate of slit center in global reference system. & \si{\milli\meter} \\
    \texttt{temperature} & 293.15  & Temperature of Maxwellian energy distribution.  & \si{\kelvin}   \\ 
    
    \bottomrule
\end{tabular}

\subsection{\rm\texttt{MCPLGun}}
\label{ssMCPLGun}

\begin{table}[htb]
	\centering                       
	\caption{Parameters available in \texttt{MCPLGun} configuration strings. Parameters without default values are required, the others are optional. }     
	\label{tableMCPLGun} 
	
    \begin{tabular}{p{5em}p{5em}p{17em}p{3em}} 
    \toprule
    Parameter & Default & Description & Unit   \\ 

    \midrule
    \texttt{mcplfile} &  & Input \texttt{MCPL} file. &  str\\
    \bottomrule
    \end{tabular}
\end{table}

\section{Scorers}
\setcounter{table}{0}
\label{sScorers}

\subsection{\rm\texttt{ESpectrum}}
\label{ssESpectrum}

\begin{table}[htb]
	\centering                       
	\caption{Parameters available in \texttt{ESpectrum} configuration strings. Parameters without default values are required, the others are optional. }     
	\label{tableESpectrum} 
	
    \begin{tabular}{p{5em}p{5em}p{17em}p{3em}} 
    \toprule
    Parameter & Default & Description & Unit   \\ 

    \midrule
    \texttt{name} &  & Name of the histogram. &  str\\
    \texttt{min} &  & Minimum energy of the histogram. & \si{\eV}  \\ 
    \texttt{max} &  & Maximum energy of the histogram. & \si{\eV} \\ 
    \texttt{numbin} & 100 & Bin number of the histogram. & 1 \\ 
    \texttt{ptstate} & \texttt{ENTRY} & Particle tracing state, available options are: \texttt{SURFACE}, \texttt{ENTRY}, \texttt{ABSORB}, \texttt{PROPAGATE}, \texttt{EXIT}. & str\\
    
    \bottomrule
    \end{tabular}
\end{table}

\subsection{\rm\texttt{WlSpectrum}}
\label{ssWlSpectrum}

\begin{table}[htb]
	\centering                       
	\caption{Parameters available in \texttt{WlSpectrum} configuration strings. Parameters without default values are required, the others are optional. }     
	\label{tableWlSpectrum} 
	
    \begin{tabular}{p{5em}p{5em}p{17em}p{3em}} 
    \toprule
    Parameter & Default & Description & Unit   \\ 

    \midrule
    \texttt{name} &  & Name of the histogram. &  str\\
    \texttt{min} &  & Minimum  wavelength of the histogram. & \si{\angstrom}  \\ 
    \texttt{max} &  & Maximum  wavelength of the histogram. & \si{\angstrom} \\ 
    \texttt{numbin} & 100 & Bin number of the histogram. & 1 \\ 
    \texttt{ptstate} & \texttt{ENTRY} &Particle tracing state, available options are: \texttt{SURFACE}, \texttt{ENTRY}, \texttt{ABSORB}, \texttt{PROPAGATE}, \texttt{EXIT}. & str\\
    
    \bottomrule
    \end{tabular}
\end{table}

\subsection{\rm\texttt{TOF}}
\label{ssTOF}

\begin{table}[htb]
	\centering                       
	\caption{Parameters available in \texttt{TOF} configuration strings. Parameters without default values are required, the others are optional. }     
	\label{tableTOF} 
	
    \begin{tabular}{p{5em}p{5em}p{17em}p{3em}} 
    \toprule
    Parameter & Default & Description & Unit   \\ 

    \midrule
    \texttt{name} &  & Name of the histogram. &  str\\
    \texttt{min} &  & Minimum time of the histogram. & \si{\second} \\ 
    \texttt{max} &  & Maximum time of the histogram. & \si{\second} \\ 
    \texttt{numbin} & 100 & Bin number of the histogram. & 1 \\ 
    \texttt{ptstate} & \texttt{ENTRY} & Particle tracing state, available options are: \texttt{SURFACE}, \texttt{ENTRY}, \texttt{ABSORB}, \texttt{PROPAGATE}, \texttt{EXIT}. & str\\
    
    \bottomrule
    \end{tabular}
\end{table}

\subsection{\rm\texttt{PSD}}
\label{ssPSD}

\begin{table}[htb]
    \begin{threeparttable}[c]
	\centering                       
	\caption{Parameters available in \texttt{PSD} configuration strings. Parameters without default values are required, the others are optional. }     
	\label{tablePSD} 
	
    \begin{tabular}{p{5em}p{5em}p{17em}p{3em}} 
    \toprule
    Parameter & Default & Description & Unit   \\ 

    \midrule
    \texttt{name} &  & Name of the histogram. & str \\
    \texttt{xmin} &  & Minimum x of the histogram. & \si{\milli\meter}    \\ 
    \texttt{xmax} &  & Maximum x of the histogram. & \si{\milli\meter}  \\ 
    \texttt{numbin\_x} & 100 & Bin number of X-axis.  & 1 \\ 
    \texttt{ymin} &  & Minimum y of the histogram. & \si{\milli\meter}    \\ 
    \texttt{ymax} &  & Maximum y of the histogram. & \si{\milli\meter}  \\ 
    \texttt{numbin\_y} & 100 & Bin number of Y-axis.  & 1 \\ 
    \texttt{type} & \texttt{XY} & Projection plane of the volume, can be set to: \texttt{XY}\tnote{1}, \texttt{XZ} or \texttt{YZ}. 
    & str\\ 
    \texttt{ptstate} & \texttt{ENTRY} & Particle tracing state, available options are: \texttt{SURFACE}, \texttt{ENTRY}, \texttt{ABSORB}, \texttt{PROPAGATE}, \texttt{EXIT}.& str\\
    \bottomrule
    \end{tabular}
    \begin{tablenotes}
        \item [1] For instance, \texttt{XY} implies the  projection of the volume on XY-plane.
    \end{tablenotes}
    \end{threeparttable}
\end{table}

\subsection{\rm\texttt{MultiScat}}
\label{ssMultiScat}

\begin{threeparttable}

	\centering                       
	\caption{Parameters available in \texttt{MultiScat} configuration strings. Parameters without default values are required, the others are optional. }     
	\label{tableMultiScat} 
	
    \begin{tabular}{p{5em}p{5em}p{17em}p{3em}} 
    \toprule
    Parameter & Default & Description & Unit   \\ 

    \midrule
    \texttt{name} &  & Name of the histogram. &  str\\
    \texttt{min} & 0 & Minimum scattering number of the histogram. & 1   \\ 
    \texttt{max} & 5 & Maximum scattering number of the histogram. & 1 \\ 
    \texttt{linear} & \texttt{yes} & Linear option for X-axis of the histogram\tnote{1}.  & str\\
    \bottomrule
    \end{tabular}
    \begin{tablenotes}
        \item [1] The default, \texttt{yes}, implies the linear X-axis, while \texttt{no} changes the X-axis to log scale.
    \end{tablenotes}
\end{threeparttable}

\subsection{\rm\texttt{DeltaMomentum}}
\label{ssDeltaMomentum}

    \begin{threeparttable}[c]
	\centering                       
	\caption{Parameters available in \texttt{DeltaMomentum} configuration strings. Parameters without default values are required, the others are optional. }     
	\label{tableDeltaMomentum} 
	
    \begin{tabular}{L{5em}L{5em}L{17em}L{3em}} 
    \toprule
    Parameter & Default & Description & Unit   \\ 

    \midrule
    \texttt{name} &  & Name of the histogram. &  str\\
    \texttt{sample\_pos} &  & Position vector of sample in global reference.
 & \si{\milli\meter}  \\
    \texttt{beam\_dir} &  & Direction vector of beam. & \si{\milli\meter}\\
    \texttt{dist} &  & Distance between neutron source and sample. & \si{\milli\meter}  \\
    \texttt{min} &  & Minimum $Q$ of the histogram. & \si{\per\angstrom} \\ 
    \texttt{max} &  & Maximum $Q$ of the histogram. & \si{\per\angstrom} \\
    \texttt{numbin} & 100 & Bin number of the histogram. & 1 \\
    \texttt{method} & 0  & Method for calculating momentum transfer $Q$\tnote{1}.  &  str\\
    \texttt{ptstate} & \texttt{ENTRY} & Particle tracing state, available options are: \texttt{SURFACE}, \texttt{ENTRY}, \texttt{ABSORB}, \texttt{PROPAGATE}, \texttt{EXIT}. & str\\
    \texttt{linear} & \texttt{yes} & Linear option for X-axis of the histogram\tnote{2}. & str\\
    \texttt{scatnum} & -1 & Scattering number to be counted\tnote{3}.   & 1 \\
    
    \bottomrule
    \end{tabular}
    \begin{tablenotes}
        \item [1] The default value $0$ implies momentum transfer of inelastic scattering, while a value of $1$ means momentum transfer of elastic scattering.
        \item [2] The default, \texttt{yes}, implies the linear X-axis, while \texttt{no} changes the X-axis to log scale. 
        \item [3] Should be an integer greater than or equal to -1. The default value implies the histogram accumulates event regardless of the scattering number. Other values enables the accumulation on the specified scattering number. The scattering number is counted by a \texttt{MultiScat} scorer, hence such a scorer should be attached to the volume of interest if \texttt{scatnum} is other than -1.
    \end{tablenotes}
    \end{threeparttable}

\subsection{\rm\texttt{Angular}}
\label{ssAngular}

\begin{table}[htb]
\begin{threeparttable}[c]
	\centering                       
	\caption{Parameters available in \texttt{Angular} configuration strings. Parameters without default values are required, the others are optional. }     
	\label{tableAngular} 
	
    \begin{tabular}{L{5em}L{5em}L{17em}L{3em}} 
    \toprule
    Parameter & Default & Description & Unit   \\ 

    \midrule
    \texttt{name} &  & Name of the histogram. & str \\
    \texttt{sample\_pos} &  & Position vector of sample in global reference.
 & \si{\milli\meter} \\
    \texttt{beam\_dir} &  & Direction vector of beam. & \si{\milli\meter} \\
    \texttt{dist} &  & Distance between particle source and sample. & \si{\milli\meter} \\
    \texttt{min} &  & Minimum scattering angle of the histogram, range of which should be $[0,180]^\circ$.   & \si{\degree} \\ 
    \texttt{max} &  & Maximum scattering angle of the histogram, range of which should be $[0,180]^\circ$.   & \si{\degree}  \\
    \texttt{numbin} & 100 & Bin number of the histogram. & 1 \\
    \texttt{ptstate} & \texttt{ENTRY} & Particle tracing state, available options are: \texttt{SURFACE}, \texttt{ENTRY}, \texttt{ABSORB}, \texttt{PROPAGATE}, \texttt{EXIT}. & str \\
    \texttt{linear} & \texttt{yes} & Linear option for X-axis of the histogram\tnote{1}.  & str\\
    
    \bottomrule
    \end{tabular}
    \begin{tablenotes}
        \item [1] The default, \texttt{yes}, implies the linear X-axis, while \texttt{no} changes the X-axis to log scale.
    \end{tablenotes}
\end{threeparttable}
\end{table}

\subsection{\rm\texttt{VolFluence}}
\label{ssVolFluence}

\begin{threeparttable}[c]

	\centering                       
	\caption{Parameters available in \texttt{VolFluence} configuration strings. Parameters without default values are required, the others are optional. }     
	\label{tableVolFluence} 
	
    \begin{tabular}{p{5em}p{5em}p{17em}p{3em}} 
    \toprule
    Parameter & Default & Description & Unit   \\ 

    \midrule
    \texttt{name} &  & Name of the histogram. & str \\
    \texttt{min} &  & Minimum energy of the histogram. &  \si{\eV}\\ 
    \texttt{max} &  & Maximum energy of the histogram. &  \si{\eV}\\ 
    \texttt{numbin} & 100 & Bin number of the histogram. & 1 \\
    \texttt{ptstate} & \texttt{PROPAGATE} & Particle tracing state, available options are: \texttt{SURFACE}, \texttt{ENTRY}, \texttt{ABSORB}, \texttt{PROPAGATE}, \texttt{EXIT}. & str \\
    \texttt{linear} & \texttt{yes} & Linear option for X-axis of the histogram\tnote{1}. & str\\
    
    \bottomrule
    \end{tabular}
    \begin{tablenotes}
        \item [1] The default, \texttt{yes}, implies the linear X-axis, while \texttt{no} changes the X-axis to log scale.
    \end{tablenotes}
\end{threeparttable}

\end{document}